\documentclass[%
aps,prl,twocolumn,
superscriptaddress,
amsmath,amssymb,
]{revtex4-2}

\usepackage{graphicx}
\usepackage{dcolumn}
\usepackage{bm}
\usepackage{hyperref}

\usepackage{physics}
\usepackage{dsfont}
\usepackage{tikz,pgfplots}

\usepackage{float}

\newcommand{\x}{\mathbf{x}}
\newcommand{\y}{\mathbf{y}}

\newcommand{\p}{\mathbf{p}}
\newcommand{\q}{\mathbf{q}}

\newcommand{\mT}{\mathcal{T}}

\newcommand{\mH}{\mathcal{H}}

\makeatletter
\newcommand*{\rom}[1]{\expandafter\@slowromancap\romannumeral #1@}
\makeatother

\usepackage{amsthm}
\newtheorem*{thm*}{Theorem}
\newtheorem{thm}{Theorem}
\newtheorem{lem}{Lemma}

\begin{document}


\title{Quantum Uncertainty Principles for Measurements with Interventions}

\author{Yunlong~Xiao}
\email{mathxiao123@gmail.com}
\address{Institute of High Performance Computing (IHPC), Agency for Science Technology and Research (A*STAR), 1 Fusionopolis Way, \#16-16 Connexis, Singapore 138632, Republic of Singapore}
\address{Nanyang Quantum Hub, School of Physical and Mathematical Sciences, Nanyang Technological University, Singapore}
\author{Yuxiang~Yang}
\email{yuxiang@cs.hku.hk}
\address{QICI Quantum Information and Computation Initiative, Department of Computer Science, The University of Hong Kong, Pokfulam Road, Hong Kong}
\address{Institute for Theoretical Physics, ETH Z\"urich, 8093 Z\"urich, Switzerland}
\author{Ximing~Wang}
\address{Nanyang Quantum Hub, School of Physical and Mathematical Sciences, Nanyang Technological University, Singapore} 
\author{Qing~Liu}
\address{Key Laboratory for Information Science of Electromagnetic Waves (Ministry of Education), Fudan University, Shanghai 200433, China}
\address{Nanyang Quantum Hub, School of Physical and Mathematical Sciences, Nanyang Technological University, Singapore}
\author{Mile~Gu}
\email{mgu@quantumcomplexity.org}
\address{Nanyang Quantum Hub, School of Physical and Mathematical Sciences, Nanyang Technological University, Singapore}
\address{Centre for Quantum Technologies, National University of Singapore, Singapore}
\address{MajuLab, CNRS-UNS-NUS-NTU International Joint Research Unit, UMI 3654, Singapore}

\date{\today}
             

\begin{abstract} 
Heisenberg's uncertainty principle implies fundamental constraints on what properties of a quantum system can we simultaneously learn. However, it typically assumes that we probe these properties via measurements at a single point in time. In contrast, inferring causal dependencies in complex processes often requires interactive experimentation - multiple rounds of interventions where
we adaptively probe the process with different inputs to observe how they affect outputs. Here we demonstrate universal uncertainty principles for general interactive measurements involving arbitrary rounds of interventions. As a case study, we show that they imply an uncertainty trade-off between measurements compatible with different causal dependencies.
\end{abstract}

\maketitle


\noindent\textbf{Introduction --} We learn about physical systems through measurement, and the uncertainty principle fundamentally limits what we can simultaneously learn~\cite{Heisenberg1927}. Quantum mechanics states the existence of incompatible measurements (e.g., position and momentum of a free particle), such that predicting both outcomes to absolute precision is impossible ~\cite{Kennard1927,weyl1928gruppentheorie,PhysRev.34.163}. Subsequent use of information theory then led to various entropic uncertainty relations that quantified uncertainty using entropic measures~\cite{PhysRevLett.50.631}, culminating 
with universal uncertainty relations that provide general constraints of the joint probabilities of incompatible  measurements~\cite{PhysRevLett.111.230401,Pucha_a_2013,PhysRevA.89.052115,Pucha_a_2018}. 

\begin{figure}[H]
\centering
\includegraphics[width=0.425\textwidth]{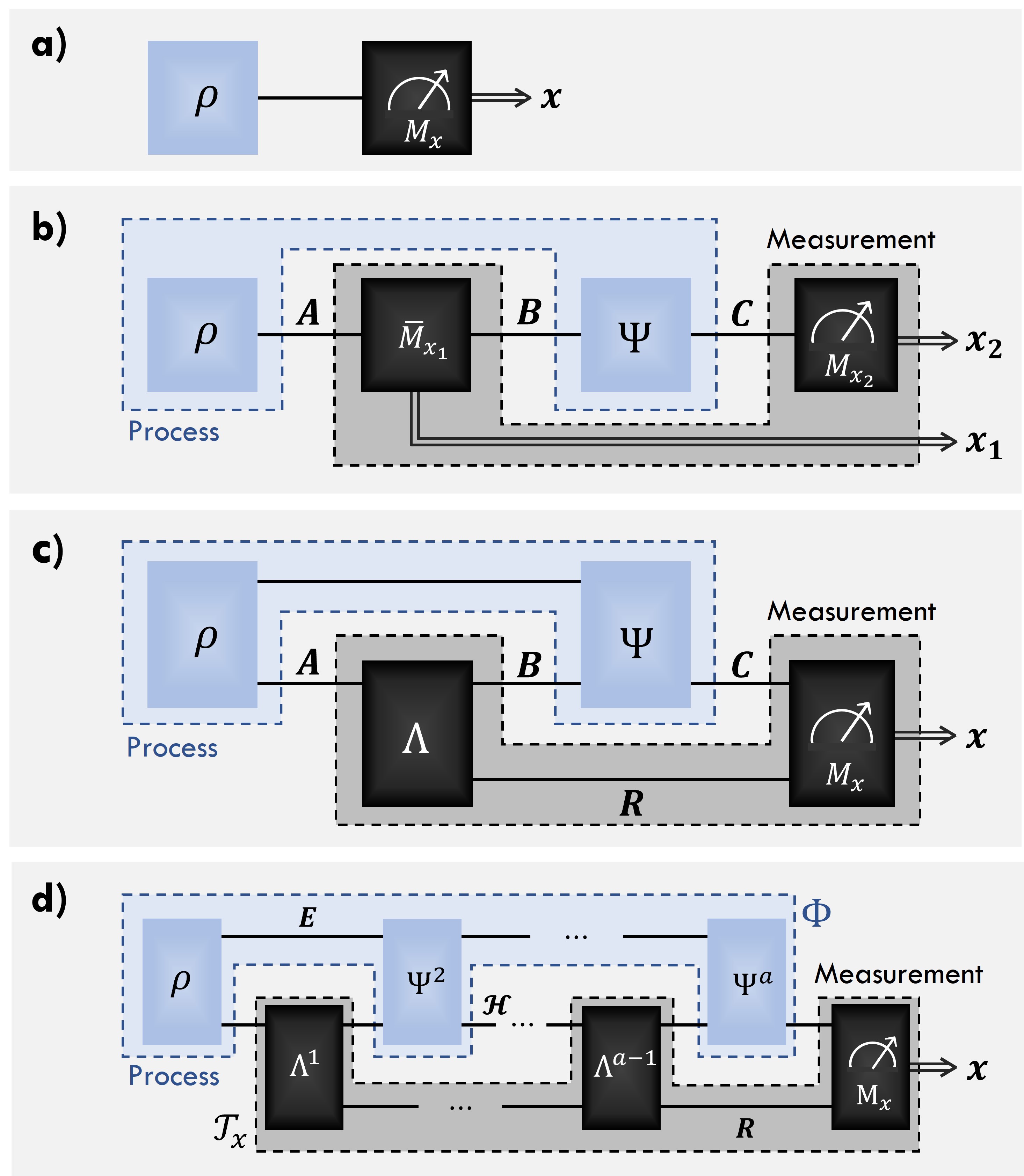}
\caption{{\bf Interactive Measurements:} Our uncertainty relations apply to all interactive measurements, including (a) passive measurements (framed by standard uncertainty relations) and (b) two-time measurements, where a quantum system first passes a quantum instrument that incorporates both a measurement outcome and the output state and later gets measured, as described by the framework of pseudo-density matrix~\cite{fitzsimons2015quantum,Marletto2019}. Our results also pertain (c) non-Markovian interactive measurements that involve coherently interacting the system with a quantum register $R$ and doing some joint measurement at a subsequent time-step and, most generally, (d) any interactive measurement $\mT_x$ with interventions at $a-1$ different time-steps.}
\label{fig:IM}
\end{figure}

Yet these relations pertain to only passive measurements, where a system is left to evolve freely before observation (see Fig.~\ref{fig:IM}a). In contrast, the most powerful means of learning involve intervention. When toddlers learn of their environment, they do not merely observe. Instead, they actively intervene -- performing various actions, observing resulting reactions and adapting future actions based on observations. Such \emph{interactive measurements} are essential to fully infer causation, so we may know if one event caused another or if both emerged from some common-causes~\cite{pearl_2009}. Indeed, interactive measurements permeate diverse sciences. Whether using reinforcement learning to identify optimal strategic behaviour or sending data packets to probe the characteristics of a network~\cite{lample2017playing,paparo2014quantum,nguyen2004active}. Such interactive measurement process also describe many quantum protocols, including quantum illumination, quantum-enhanced agents and non-Markovian open systems~\cite{PhysRevX.12.011007,doi:10.1126/science.1160627,PhysRevE.99.042103,PhysRevA.97.012127}.

Could uncertainty principles also fundamentally constrain such interactive measurements (see Fig.~\ref{fig:IM}b-d)? How would such principle interplay with interventions 
\noindent
aimed to discern causal structure? Here, we explore these questions by deriving a universal uncertainty principle that constrains the joint measurement probabilities of interactive measurements. This principle then pinpoints when two interactive measurements are non-compatible -- and quantifies the necessary trade-offs in the certainty of their measurement outcomes. 
Our results make no assumptions on the number of interventions or the causal structure of processes we probe, and encompass previous uncertainty relations for states and channels as special cases~\cite{kraus1983states,PhysRevA.77.062112,PhysRevResearch.3.023077}. We apply them to interactive measurements compatible with direct-cause vs common-cause, showing that they satisfy an uncertainty trade-off analogous to position and momentum. 


\noindent\textbf{Framework --} The premise of an interactive measurement consists of an agent that wishes to probe the dynamics of some unknown quantum process $\Phi$. Here $\Phi$ can be modelled as an open quantum system, consisting of a system accessible to the agent with $\mH$ coupled with some generally non-Markovian environment $E$ (see blue shaded region in Fig.~\ref{fig:IM}d). Initially, the $\mH$-$E$ system is in some joint state $\rho$. At each time-step $k$, the system and environment jointly evolve under $\Psi^k$. $\Phi$ is then completely defined by the set $\{\Psi^k\}^a_{k=2}$ and the initial state $\rho$, where $a$ represent to the number of time-steps. In literature, $\Phi$
offers the most general representations of non-Markovian quantum stochastic processes~\cite{PRXQuantum.2.030201} and is also closely related to concepts of higher-order quantum maps, adaptive agents, and causal networks~\cite{2008Chiribella,PhysRevLett.101.060401,PhysRevA.80.022339,doi:10.1098/rspa.2018.0706,8678741,PRXQuantum.2.030335}.

Interactive measurements then represent the most general means for an agent to determine properties of $\Phi$ (see black shaded region in Fig.~\ref{fig:IM}d): the agent initializes some internal memory register $R$; between time-steps (i.e., before $\Psi^k$ with $2\leqslant k\leqslant a$), the agent performs an \emph{intervention} -- some general quantum operation $\Lambda^k$ that interacts her memory $R$ with the accessible system $\mH$; after $a-1$ such interventions, the agent finally makes a joint measurement with respect to some positive operator valued measure (POVM) $M:= \{M_x\}$ on the joint $\mH$-$R$ system to obtain some outcome $x$. Thus, each interactive measurement $\mathcal{T}$ is completely defined by set of interventions $\{\Lambda^k\}^{a-1}_{k=1}$ and POVM $M$. Just as a conventional POVM measurement on a quantum state induces some probability distribution over measurement outcomes, so does an interactive measurement on a quantum process. Analogous to eigenstates, we say $\Phi$ is an \emph{eigencircuit} of $\mathcal{T}$ if $\Phi$ always yields a definite outcome when measured by $\mathcal{T}$. 

We make two remarks. (1) The interactive measurements encompass \emph{everything} an agent can possibly do causally. Notably, $R$ can also store classical information. For example, making a projective measurement and conditioning future action on the system based on the result of these measurements. (2) Both $\Phi$ and $\mathcal{T}$ have succinct representations using Choi-Jamio\l kowski operators, often referred to as quantum combs~\cite{PhysRevLett.101.060401,PhysRevA.80.022339} or process tensors~\cite{PhysRevA.97.012127}. We provide a rigorous mathematical treatment in supplemental material~\cite[Sec.~\rom{1}B and \rom{1}C]{SM}.

\begin{figure*}[ht]
\centering
\includegraphics[width=0.9\textwidth]{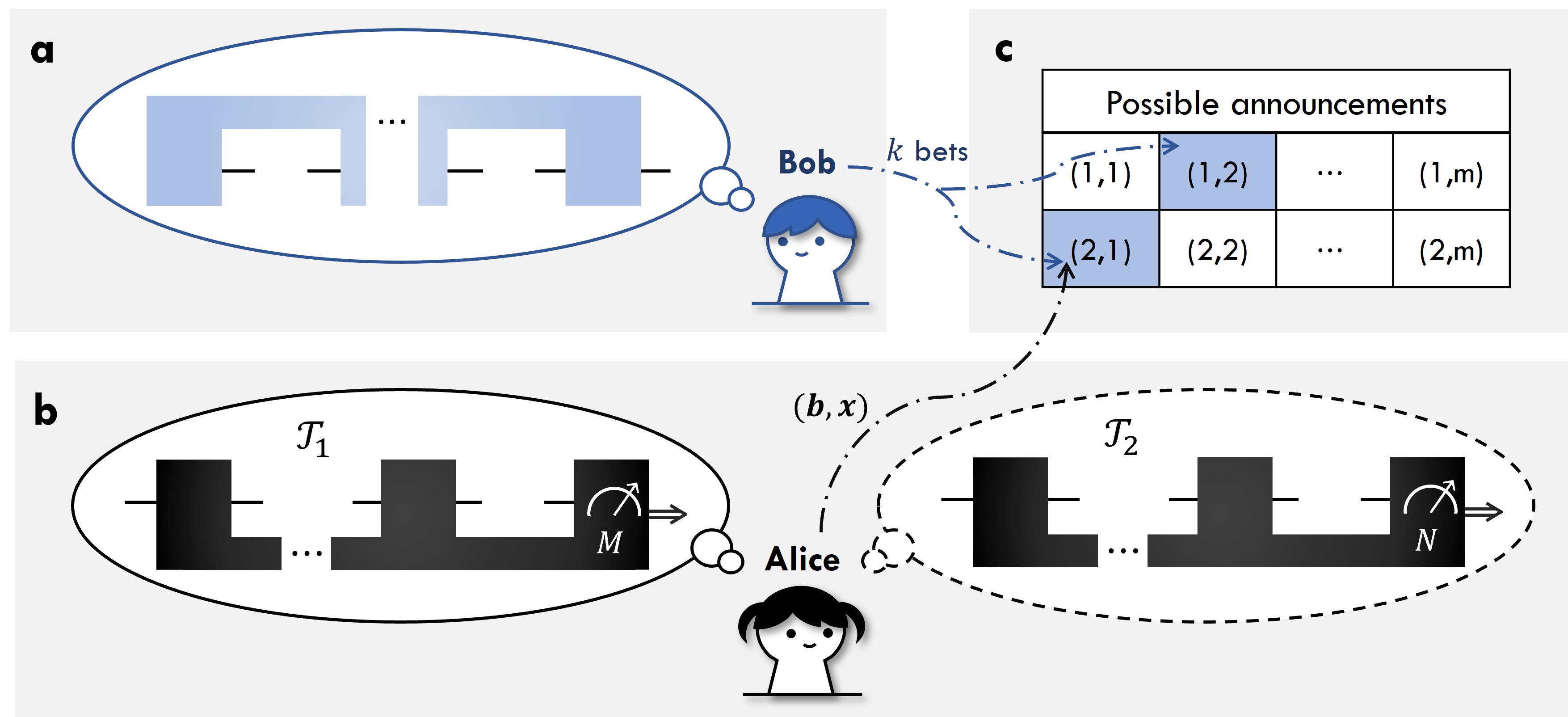}
\caption{{\bf The Quantum Roulette:} The quantum roulette is a game that aids in interpreting lower bounds for the combined uncertainty of two general interactive measurements $\{\mathcal{T}_{b}\}_{b = 1,2}$. $\mathcal{T}_{1}$ and $\mathcal{T}_{2}$ -- picture in (b). Now introduce a quantum  `roulette table' with $2 \times m$ grid of cells (c), labelled $(b,x)$ with $x = 1,\ldots,m$. In the $k^{th}$ order game, Bob begins with $k$ chips, of which he can allocate to $k$ of these cells. Bob then supplies Alice with a dynamical process $\Phi$ (a). Alice selects a $b$ at random, and measures $\Phi$ with $\mathcal{T}_{b}$ to obtain outcome $x$. Bob wins if he has a chip on the cell $(b,x)$. Lem.~\ref{lem:uur} and Thm.~\ref{thm:eur} then relate Bob's winning probabilities with the incompatibility between $\mathcal{T}_{1}$ and $\mathcal{T}_{2}$.
}
\label{fig:multiple}
\end{figure*}



\noindent\textbf{Uncertainty Principles -- }  
In conventional quantum theory, certain observables are mutually incompatible. Given an observable $\mathcal{O}$ whose outcomes $o_k$ occurs with probability $p_k$, we can quantify the uncertainty by the Shannon entropy $H(\mathcal{O}) := - \sum_k p_k \log p_k$. The entropic uncertainty principle then states that there exists mutually non-compatible observables $\mathcal{O}_1$ and $\mathcal{O}_2$, such that the joint uncertainty $H(\mathcal{O}_1) + H(\mathcal{O}_2)$ is always lower-bounded by some state-independent constant $C > 0$~\cite{PhysRevD.35.3070,PhysRevLett.60.1103,2010Berta-NP}.


Can we identify similar uncertainty relations for general interactive measurements? We answer this question by employing majorization~\cite{marshall2010inequalities}. Consider two probability vectors $\x$ and $\y$, whose elements $x_k$ and $y_k$ are arranged in non-increasing order. We say $\x$ is majorized by $\y$, written as $\x\prec\y$, if $\sum_{k=1}^{i}x_k\leqslant\sum_{k=1}^{i}y_k$ holds for all index $i$. The rationale is that majorization maintains significant connections with entropy since $\x\prec\y$ implies that $H(\x) \geqslant H(\y)$. In fact, $\x\prec\y$ implies $f(\x) \geqslant f(\y)$ for a large class of functions known as \emph{Schur-concave functions}. Such functions align with those that remain non-decreasing under random relabeling of measurement outcomes, and have been proposed as the most general class of uncertainty quantifiers~\cite{PhysRevLett.111.230401}. Thus, majorization constraints on outcome probabilities for conventional quantum measurements are referred to as universal uncertainty relations ~\cite{PhysRevLett.111.230401,Pucha_a_2013,PhysRevA.89.052115,Pucha_a_2018}. Here, we establish such a \emph{universal uncertainty relation for general interactive measurements} (See supplemental material~\cite[Sec.~\rom{2}D]{SM} for the proof):

\begin{lem}\label{lem:uur}  
Consider two distinct interactive measurements $\mathcal{T}_{1}$ and $\mathcal{T}_{2}$ on some dynamical process $\Phi$, with outcomes described by probability distributions $\p$ and $\q$. There then exists a probability vector $\mathbf{v}(\mathcal{T}_{1}, \mathcal{T}_{2})$ such that
\begin{align}\label{eq:eur}
\frac{1}{2}\p\oplus\frac{1}{2}\q
\prec
\mathbf{v}(\mathcal{T}_{1}, \mathcal{T}_{2}).
\end{align}
Here the vector-type bound $\mathbf{v}(\mathcal{T}_{1}, \mathcal{T}_{2})$ is independent of $\Phi$, and hence captures the essential incompatibility between 
$\mathcal{T}_{1}$ and $\mathcal{T}_{2}$. Meanwhile, $\oplus$ represents the concatenation of vectors. For example, $(1, 0)\oplus(1/2, 1/2) = (1, 0, 1/2, 1/2)$.  
\end{lem}

Our result for interactive measurement is also universal in this sense. In particular, they imply an infinite family of uncertainty relations, namely $f(\p/2\oplus\q/2)\geqslant f(\mathbf{v}(\mathcal{T}_{1}, \mathcal{T}_{2}))$ for any Schur-concave function $f$ (including R\'enyi entropies). 
Choosing $f$ as the Shannon entropy, Lem.~\ref{lem:uur} then results in entropic bounds for general interactive measurements (see~\cite[Sec.~\rom{2}D]{SM} for details):

\begin{thm}\label{thm:eur}  
Given two interactive measurements $\mathcal{T}_{1}$ and $\mathcal{T}_{2}$ acting on some dynamical process $\Phi$. The entropies of their measurement outcomes~\footnote{Denote the probability distribution of outcomes when $\mathcal{T}_{1}$ is measured as $\p$, then the uncertainty of $\mathcal{T}_{1}$ can be quantified by Shannon entropy, i.e. $H(\mathcal{T}_{1})_{\Phi}:= H(\p)$.} satisfy
\begin{align}\label{eq:xeur}
H(\mathcal{T}_{1})_{\Phi} + H(\mathcal{T}_{2})_{\Phi} \geqslant C(\mathcal{T}_{1}, \mathcal{T}_{2}),
\end{align}
where $C(\mathcal{T}_{1}, \mathcal{T}_{2})$ -- measuring incompatibility between $\mathcal{T}_{1}$ and $\mathcal{T}_{2}$ -- is non-negative and independent of $\Phi$. $C(\mathcal{T}_{1}, \mathcal{T}_{2})$ can be explicitly computed. It is strictly non-zero whenever $\mathcal{T}_{1}$ and $\mathcal{T}_{2}$ have no common eigencircuit. 
\end{thm}

In~\cite[Sec.~\rom{2}D]{SM}, we illustrate a choice of $C(\mathcal{T}_{1}, \mathcal{T}_{2})$ that reduces to $\log(1/c)$ when $\mathcal{T}_{1}$ and $\mathcal{T}_{2}$ are standard quantum measurements. Here $c$ stands for the maximal overlap between measurements~\cite{PhysRevLett.50.631}. Meanwhile, just as there exist many alternative bounds beyond $\log(1/c)$~\cite{SANCHESRUIZ1998189,GHIRARDI200332,PhysRevA.77.042110,PhysRevLett.106.110506,PhysRevLett.108.210405,PhysRevA.89.022112,PhysRevA.91.032123,Xiao_2016,RevModPhys.89.015002,XiaoPhD,PhysRevLett.122.100401,Xiao_2020}, there are many other valid bounds for $H(\mathcal{T}_{1})_{\Phi} + H(\mathcal{T}_{2})_{\Phi}$ (See~\cite[Sec.~\rom{2}D]{SM}). Here we focus on a choice of $C(\mathcal{T}_{1}, \mathcal{T}_{2})$ that can give tighter bounds in causal inference settings. More results are presented in~\cite[Sec.~\rom{2}C]{SM}.

Our formulations of $\mathbf{v}(\mathcal{T}_{1}, \mathcal{T}_{2})$ and $C(\mathcal{T}_{1}, \mathcal{T}_{2})$ carry direct operational meaning in a guessing game which we refer to as the \emph{quantum roulette}. The two-party game consists of (1) Alice, an agent that probes any supplied dynamical process using one of two possible interactive measurements, $\mathcal{T}_{1}$ and $\mathcal{T}_{2}$, and (2) Bob, who can engineer various dynamical processes for Alice to probe (see Fig. ~\ref{fig:multiple}).
In each round, Alice and Bob begin with a `roulette table', whose layout consists of all tuples $(b, x)$, where $b \in \{1,2\}$ and $x$ are all possible measurement outcomes of $\mathcal{T}_{1}$ and $\mathcal{T}_{2}$. Bob begins with $k$ chips, which he can use to place bets on $k$ of the possible tuples and supplies Alice with any $\Phi$ of his choosing. Alice will then select some $b \in \{1,2\}$ at random and probe $\Phi$ with $\mathcal{T}_{b}$. She finally announces both $b$ and the resulting measurement outcome $x$. Bob wins if one of his chips is on $(b, x)$.

Let $p_k$ denote Bob's maximum winning probability. Naturally $p_0 = 0$ and $p_k$ increases monotonically with $k$, tending to $1$. We define a probability vector $\mathbf{w}$ with elements $w_k = p_k - p_{k-1}$, $k = 1,2,\ldots$, representing the increase in Bob's probability of winning with $k$ rather than $k-1$ chips. In~\cite[Sec.~\rom{2}D]{SM}, we show that
$\mathbf{v}(\mathcal{T}_{1}, \mathcal{T}_{2}):= \mathbf{w}$ and $C(\mathcal{T}_{1}, \mathcal{T}_{2}):= 2 H(\mathbf{w}) - 2$ are bounds for $\p/2\oplus\q/2$ and $H(\mathcal{T}_{1})_{\Phi} + H(\mathcal{T}_{2})_{\Phi}$ respectively.

This game gives an operational criterion of non-compatibility for interactive measurements. When two observables are compatible, $H(\mathbf{w}) = 1$. This aligns with the scenario that $\mathbf{w} = (0.5,0.5,0,\ldots,0)$, which occurs when Bob's success rate is limited only by his uncertainty of which measurement Alice makes. That is, placing one counter ensures Bob can correctly predict the outcome of $\mathcal{T}_{1}$ and two counters gives him perfect prediction regardless of $b$. We see this is only possible if $\mathcal{T}_{1}$ and $\mathcal{T}_{2}$ share at least one common eigencircuit. Thus, $H(\mathcal{T}_{1})_{\Phi} + H(\mathcal{T}_{2})_{\Phi}$ is strictly greater than $0$ whenever $\mathcal{T}_{1}$ and $\mathcal{T}_{2}$ share no common eigencircuit.


\begin{figure}[t]
\centering
\includegraphics[width=0.48\textwidth]{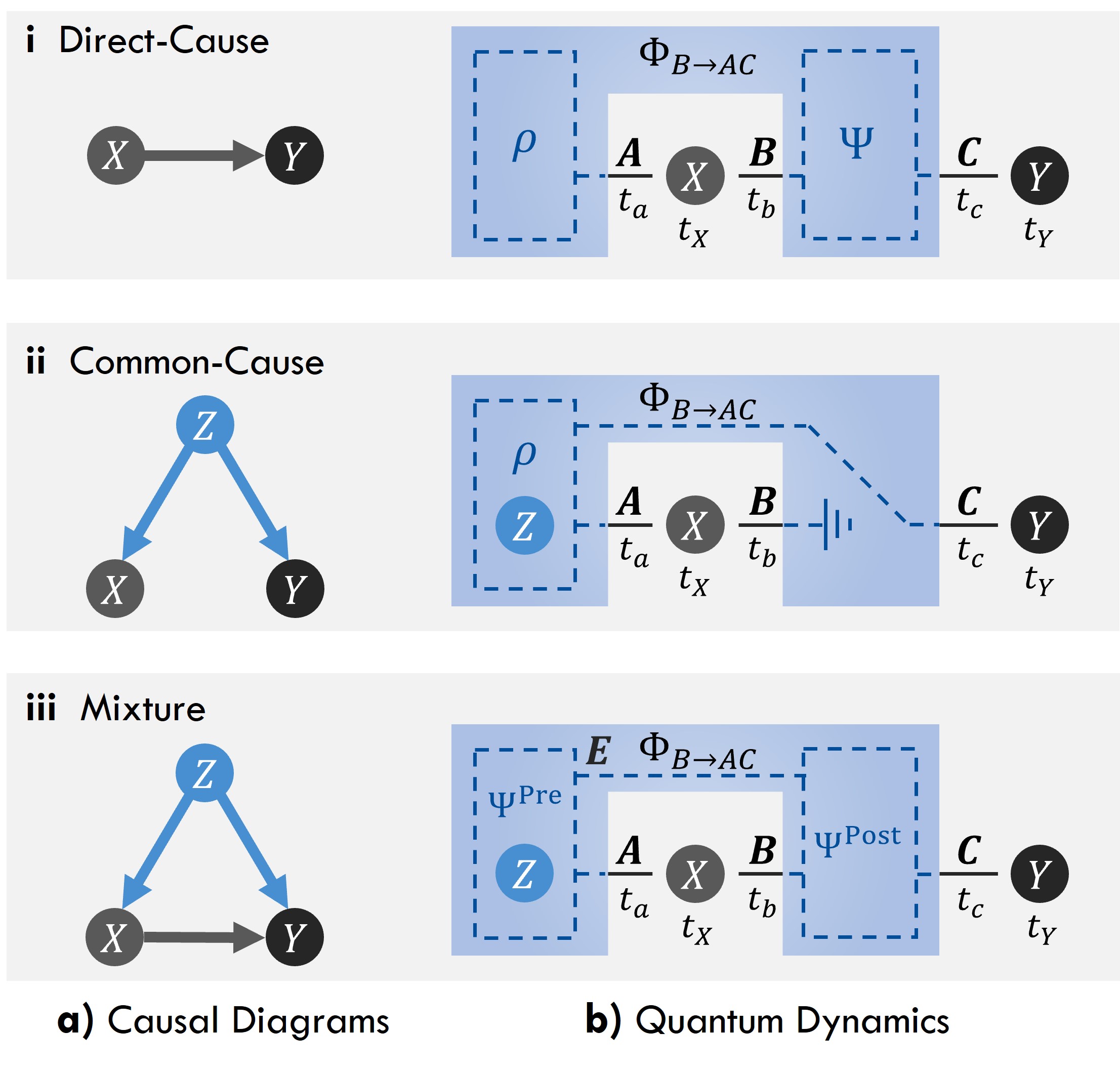}
\caption{{\bf } \textbf{Quantum Description of Causal Structures:} There are three possible causal structures for two events $X$ and $Y$, all of which can be expressed by a quantum dynamic process $\Phi_{B\rightarrow AC}$. In (i) direct-cause, $\Phi_{B\rightarrow AC}$ involves preparing a state $A$ to be observed at $X$, whose output is sent directly to $Y$ via quantum channel from $B$ to $C$. In (ii) common-cause, correlations between $X$ and $Y$ can be attributed to measurements on some pre-prepared correlated state $\rho_{AC}$ (event $Z$). Most generally (iii), $\Phi_{B\rightarrow AC}$ consists of a state-preparation process $\Psi_{\mathbb{C}\rightarrow AE}^{\text{Pre}}$ and a post-processing quantum channel $\Psi_{BE\rightarrow C}^{\text{Post}}$ (b-\romannumeral3; $E$ is an ancillary system). This then corresponds to a (possibly coherent) mixture of direct and common cause.}
\label{fig:cm}
\end{figure}

\noindent\textbf{Causal Uncertainty Relations --}  
The central relevance of interventions in causal inference makes it an appropriate illustrative example~\cite{Reichenbach1956-REITDO-2}. Consider the case where $\Phi$ represents a $d$-level system (the accessible qudit) that evolves while in possible contact with other systems (e.g. a non-Markovian environment $E$). Now suppose an agent, Alice, can access this qudit at two different points in time, say $t_X$ and $t_Y$. In general the quantum process $\Phi$ can fall under three scenarios~\cite{ried2015quantum}:
\begin{itemize}
 \setlength{\itemsep}{3pt}
  \setlength{\parskip}{0pt}
  \setlength{\parsep}{0pt}
    \item[(i)] The system at $t_X$ is a \emph{direct cause} of the system at $t_Y$: the qudit at $t_Y$ is the output of some quantum map acting on the qubit at $t_X$ (Fig.~\ref{fig:cm}b-\romannumeral1).
   \item[(ii)] The system at $t_X$ and $t_Y$ share a \emph{common cause}: the qudit at $t_X$ is correlated with an environmental qudit $E$. $E$ is measured at time $t_Y$ (Fig.~\ref{fig:cm}b-\romannumeral2).
    \item[(iii)] A mixture of both, corresponding to a general non-Markovian quantum process (Fig.~\ref{fig:cm}b-\romannumeral3). 
\end{itemize}

We now introduce two families of interactive measurements: $\mathcal{M}_{\text{CC}}$ and $\mathcal{M}_{\text{DC}}$, as depicted in Fig.~\ref{fig:unitary}. Each $\mathcal{T}_1 \in \mathcal{M}_{\text{CC}}$ is a \emph{maximal common-cause indicator}, such that its eigencircuits imply that $X$ and $Y$ are actually two arms of some maximally entangled state (Fig.~\ref{fig:cm}b-{\romannumeral2}). Meanwhile, each $\mathcal{T}_2 \in \mathcal{M}_{\text{DC}}$ is a \emph{maximal direct-cause indicator}, whose eigencircuit involve a lossless channel from $X$ to $Y$ (i.e., Fig.~\ref{fig:cm}b-{\romannumeral1} where $\Psi$ is unitary). In~\cite[Sec.~\rom{3}A]{SM}, we establish the following \emph{causal uncertainty relation}:

\begin{figure}[tb]
\centering
\includegraphics[width=0.48\textwidth]{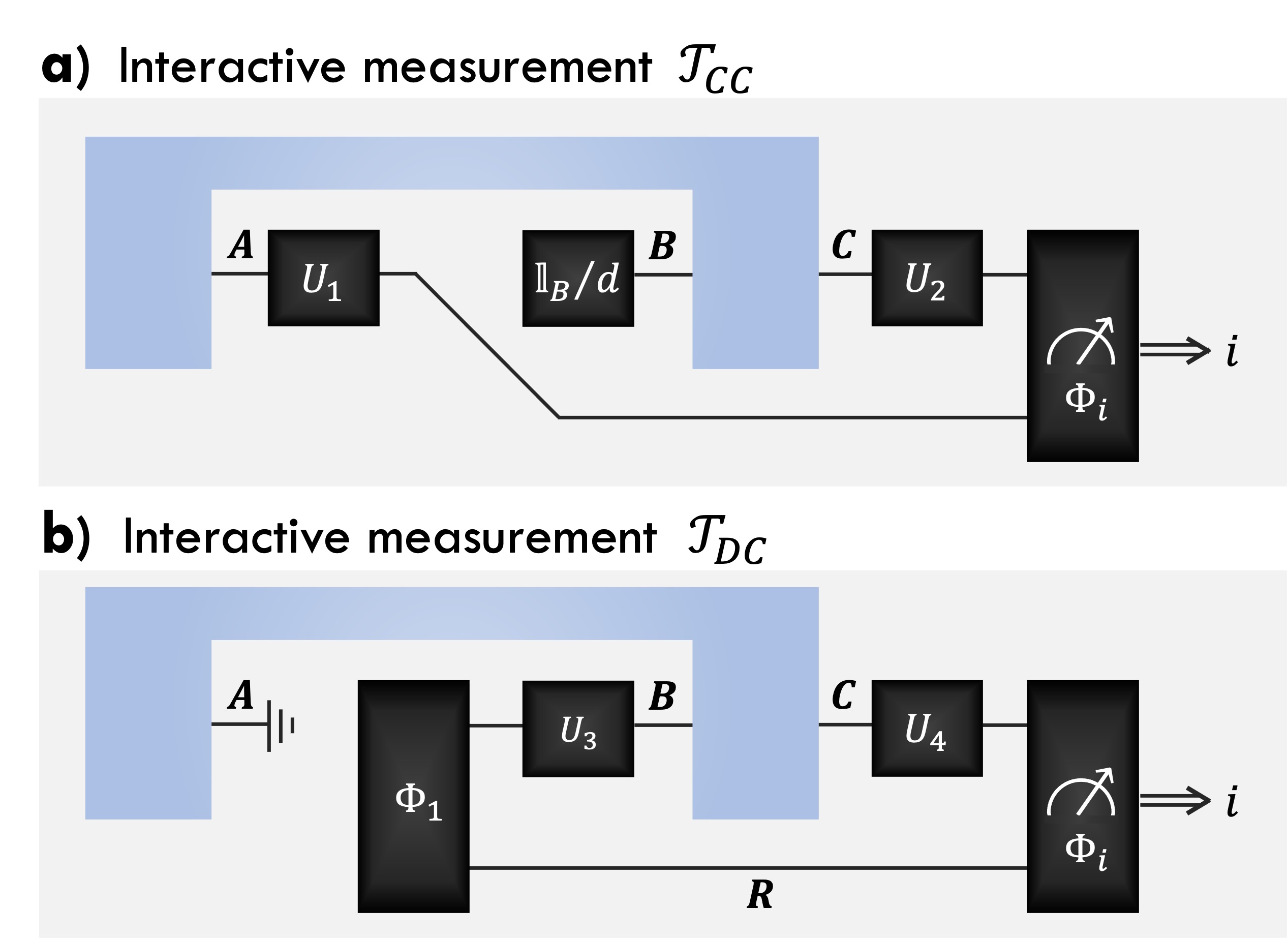}
\caption{{\bf Maximal Common-Cause and Direct-Cause Indicators:} We introduce  (a) $\mathcal{M}_{\text{CC}} = \{\mathcal{T}_{\text{CC}} (U_1, U_2)\}$ and (b) $\mathcal{M}_{\text{DC}} = \{\mathcal{T}_{\text{DC}} (U_3, U_4)\}$ as two respective families of interactive measurements with a single intervention. Here, system $A$, $B$ and $C$ are d-level quantum systems (qudits), and each $U_k$, $k =1,2,3,4$ is some single-qudit unitary, and  $\ket{\Phi_1} := \sum_{k=0}^{d-1}\ket{kk}/\sqrt{d}$. Measurements are done with respect to a maximally entangling basis $\{\Phi_i\}_i$ with $d^2$ possible outcomes. The two measurement families are incompatible, and satisfy the causal uncertainty relation in Eq.~\ref{eq:cur}.
}
\label{fig:unitary}
\end{figure} 

\begin{align}\label{eq:cur}
H(\mathcal{T}_{1})
+
H(\mathcal{T}_{2})
\geqslant
2\log d,
\end{align}
for any $\mathcal{T}_{1}\in \mathcal{M}_{\text{CC}}$ and $\mathcal{T}_{2}\in \mathcal{M}_{\text{CC}}$.
Here $H(\mathcal{T}_{i})$ ($i= 1, 2$) is the Shannon entropy of the probability distribution associated with outcomes when $\mathcal{T}_{i}$ is measured.
Furthermore, this bound can be saturated. 

\begin{figure}[tb]
\centering
\includegraphics[width=0.48\textwidth]{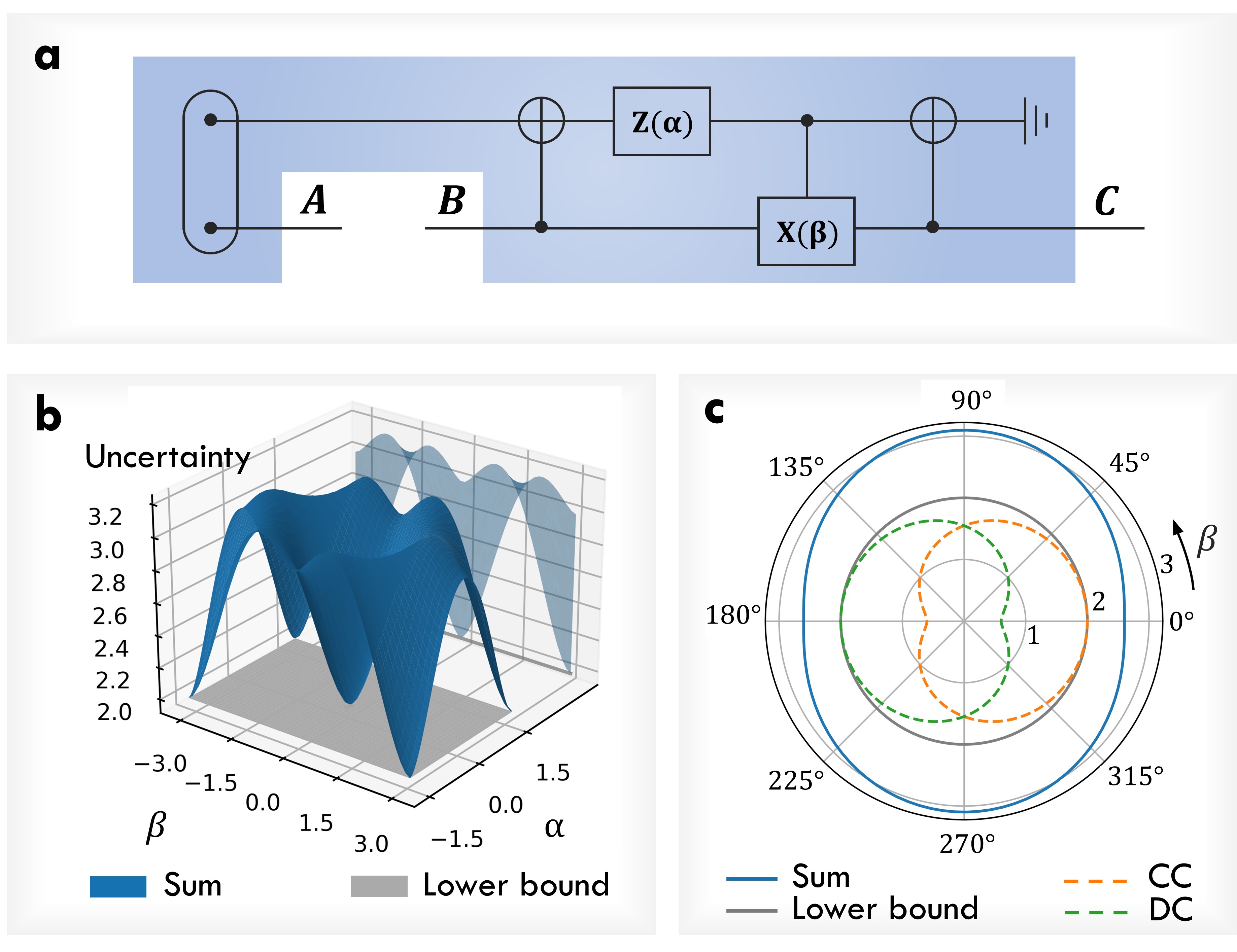}
\caption{{\bf Causal Uncertainty Relations on Non-Markovian Dynamics:} Consider a single qubit -- bottom rail of the circuit in (a) -- undergoing non-Markovian evolution. Here the systems are initialized in maximally entangled state, $Z(\theta)$ and $X(\theta)$ represent single-qubit rotation gates in $X$ and $Z$ axis. (b) illustrates the combined uncertainty $H(\mathcal{T}_{1})+H(\mathcal{T}_{2})$, where $\mathcal{T}_{1} \in {M}_{\text{CC}}$ and $\mathcal{T}_{2} \in {M}_{\text{DC}}$ are respectively common-cause and direct-cause indicators in Fig.~\ref{fig:unitary} with all $U_k$ set to the identity. Observe this never goes below the fundamental lower bound of $2$ (gray plane). (c) illustrates $H(\mathcal{T}_{1})$ (green dashed), $H(\mathcal{T}_{2})$ (red dashed) and their sum (blue solid) for $\alpha = -\pi/4$ and various values of $\beta$, corresponding to various coherent superpositions of common-cause and direct-cause circuits.}
\label{fig:circuit}
\end{figure}

Consider the application of this uncertainty to a specific parameterized quantum circuit $\Phi_{\alpha, \beta}$ (Fig.~\ref{fig:circuit}a) describing a single qubit undergoing non-Markovian evolution. Fig.~\ref{fig:circuit}b then demonstrates the combined uncertainty $H(\mathcal{T}_{1})+H(\mathcal{T}_{2})$ for various values of $\alpha$ and $\beta$, including cases where they 
saturate the lower bound of $2$. We also note that unlike classical processes, which must be either purely common-cause, or purely direct-cause, or a probabilistic mixture of both -- quantum processes can feature richer causal dependencies~\cite{causal-NC}. Fig.~\ref{fig:circuit}c depicts this for the cross-section of $\alpha = \pi/4$. Such circuits include the coherent superposition of direct and common cause as a special case. Our causal uncertainty relation also applies to these uniquely quantum causal structures.

\noindent\textbf{Discussion --} 
The most powerful means of learning involves interactive measurement -- a procedure in which we can intervene by injecting (possible entangled) quantum states into the process over multiple time-steps before observing the final output. Here, we derive entropic uncertainty relations that govern all interactive measurements, bounding their joint uncertainty whenever such measurement outcomes are non-compatible. In the context of causal inference, they predict a uniquely quantum entropic trade-off between measurements that probe for direct and common cause. More generally, our relations encompass all possible means for an agent to interact and learn about a target quantum system and thus include previously studied uncertainty relations on states and channels as special cases.

One potential application of such relations is the metrology of unknown quantum processes with memory~\cite{PhysRevLett.123.110501,altherr2021quantum,PhysRevLett.129.240503}. In practice, full tomography of a general quantum process can be extremely costly. Even a single non-Markovian qubit measured at two different times requires $54$ different interactive measurements~\cite{Feix_2017}. Our result may help us ascertain specific properties of a process while avoiding this costly procedure. In~\cite[Sec.~\rom{4}B]{SM}, we illustrate how our causal uncertainty relations imply that a single interactive measurement can rule out specific causal structures.  Indeed, quantum illumination and adaptive sensing can both cast as measuring desired properties of a candidate quantum process, and thus could benefit from such an approach. 

Interactive measurements through repeated interventions also emerge in other settings~\cite{PhysRevA.81.022121,Mile2012,PhysRevLett.114.090403}. In quantum open systems, sequential intervention provides a crucial toolkit for characterizing non-Markovian noise~\cite{LI20181,PhysRevA.100.052104,White2020,PhysRevLett.129.030401}. Meanwhile, in reinforcement learning, quantum agents that continuously probe an environment show enhancements in enacting or learning complex adaptive behaviour~\cite{paparo2014quantum,thompson2017using,elliott2022quantum}. Investigating uncertainty relations specific to such contexts has exciting potential, perhaps revealing new means of probing non-Markovian dynamics, or fundamental constraints on how well an agent can simultaneously optimize two different rewards.


\begin{acknowledgements}
\section{Acknowledgments} 
We would like to thank Varun Narasimhachar, Jayne Thompson, and Bartosz Regula for fruitful discussions. This work is supported by the Singapore Ministry of Education Tier 2 Grant MOE-T2EP50221-0005, the National Research Foundation, Singapore, and Agency for Science, Technology and Research (A*STAR) under its QEP2.0 programme (NRF2021-QEP2-02-P06), the National Research Foundation and   under the NRF-QEP program (NRF2021-QEP2-02-P06), The Singapore Ministry of Education Tier 1 Grant RG146/20, FQXi-RFP-1809 (The Role of Quantum Effects in Simplifying Quantum Agents) from the Foundational Questions Institute and Fetzer Franklin Fund (a donor-advised fund of Silicon Valley Community Foundation). Y. X. is supported by A*STAR's Central Research Fund (CRF UIBR). Y. Y. acknowledges the support from the Swiss National Science Foundation via the National Center for Competence in Research ``QSIT" as well as via project No.\ 200020\_165843, the support from Guangdong Basic and Applied Basic Research Foundation (Project No. 2022A1515010340), and the support from the Hong Kong Research Grant Council (RGC) through the Early Career Scheme (ECS) grant 27310822. Any opinions, findings and conclusions or recommendations expressed in this material are those of the author(s) and do not reflect the views of National Research Foundation or the Ministry of Education, Singapore.
\end{acknowledgements}


\bibliography{Bib}
\end{document}



\title{Quantum Uncertainty Principles for Measurements with Interventions\\Supplemental Material}

\author{Yunlong~Xiao}
\email{mathxiao123@gmail.com}
\address{Institute of High Performance Computing (IHPC), Agency for Science Technology and Research (A*STAR), 1 Fusionopolis Way, \#16-16 Connexis, Singapore 138632, Republic of Singapore}
\address{Nanyang Quantum Hub, School of Physical and Mathematical Sciences, Nanyang Technological University, Singapore 637371, Singapore}
\author{Yuxiang~Yang}
\email{yuxiang@cs.hku.hk}
\address{QICI Quantum Information and Computation Initiative, Department of Computer Science, The University of Hong Kong, Pokfulam Road, Hong Kong}
\address{Institute for Theoretical Physics, ETH Z\"urich, 8093 Z\"urich, Switzerland}
\author{Ximing~Wang}
\address{Nanyang Quantum Hub, School of Physical and Mathematical Sciences, Nanyang Technological University, Singapore 637371, Singapore}
\author{Qing~Liu}
\address{Nanyang Quantum Hub, School of Physical and Mathematical Sciences, Nanyang Technological University, Singapore 637371, Singapore}
\author{Mile~Gu}
\email{mgu@quantumcomplexity.org}
\address{Nanyang Quantum Hub, School of Physical and Mathematical Sciences, Nanyang Technological University, Singapore 637371, Singapore}
\address{Complexity Institute, Nanyang Technological University, Singapore 639673, Singapore}

\date{\today}
             
\begin{abstract} 
In this supplemental material, we formulate several uncertainty relations for measurements with interventions, extending the results presented in the main text of our work. As a by-product, we have derived causal uncertainty relations for quantum dynamics, establishing a trade-off between common-cause and direct-cause in quantum causal inference. Such a fundamental trade-off have been further utilized to infer the causality associated with parameterized quantum circuits, which are the essential building blocks for Noisy Intermediate-Scale Quantum (NISQ) technologies. Detailed analyses and proofs of the results presented in the main text have also been included. More precisely, to systemically investigate the most general quantum dynamics with definite causal orders and the corresponding measuring processes with interventions, we introduce the framework of \emph{quantum circuit fragments} and \emph{interactive measurements} in Sec.~\ref{sec:QCF}. The uncertainty principle for multiple interactive measurements has been demonstrated in Sec.~\ref{sec:UUR}, which works for any quantum circuit fragments. We have further developed the causal uncertainty relation and detailed its application to causal inference in Sec.~\ref{sec:CUR}. Finally, numerical experiments for our results have been provided in Sec.~\ref{sec:NE}. It is worth noting that we may reiterate some of the steps in the main text to make this supplemental material more explicit and self-contained.

\end{abstract}

\maketitle
\tableofcontents

\newpage


\section{\label{sec:QCF}Quantum Circuit Fragment}
In this section, we introduce our notations and prepare the groundwork for results presented in the main text of our work. Additionally, the general quantum dynamics and the corresponding measuring processes with interventions have been rigorously formulated as \emph{quantum circuit fragments} and \emph{interactive measurements}.
This preparatory section contains four subsections: In Subsec.~\ref{subsec:Channel-Superchannel}, we give a brief introduction to the concept of quantum channels, superchannels, and their Choi-Jamio\l kowski operators. Subsequently, in Subsec.~\ref{subsec:Fragment}, we move on to quantum dynamics with definite causal order, and formally introduce the concept of quantum circuit fragments. The most general measuring processes for quantum circuit fragments -- interactive measurements -- have been constructed in Subsec.~\ref{subsec:Interactive-Measurement}. Finally, a special type of quantum circuit fragments, known as causal maps in quantum causal inference, has been discussed in the last subsection, i.e. Subsec.~\ref{subsec:CMaps}.


\subsection{\label{subsec:Channel-Superchannel}Quantum Channels and Superchannels}

The fundamental building blocks that makeup quantum technologies and quantum information processing are quantum channels. Physically, the preparation of quantum states, the implementation of quantum measurements, the noise arising from the system-environment interactions, and the operation (including quantum gates) carried out on quantum devices are all characterized by the concept of quantum channel. In this subsection, we will give a brief introduction to the concept of the quantum channel and the corresponding mathematical toolkit. For detailed consideration of the quantum channels and their mathematical properties, we refer the reader to Refs.~\cite{nielsen_chuang_2010,wilde_2013,watrous_2018}.

A linear map $\mE$ from system $A$ to system $B$ is called a quantum channel if it is both completely positive (CP) and trace-preserving (TP). The property of completely positive implies that by applying quantum channels to part of a quantum system, the resultant system is still a well-defined quantum system. Meanwhile, as indicated by the name, the trace-preserving property guarantees that any channel's output state still has trace $1$, when the input is a quantum state.

Instead of dealing with a map (i.e. quantum channel) directly, usually we prefer to investigate the properties of its matrix representation. To do so, let us start with the concept of Choi-Jamio\l kowski (CJ) operators \cite{JAMIOLKOWSKI1972275,CHOI1975285}. Formally, it is defined as

\begin{figure}[h]
\includegraphics[width=0.4\textwidth]{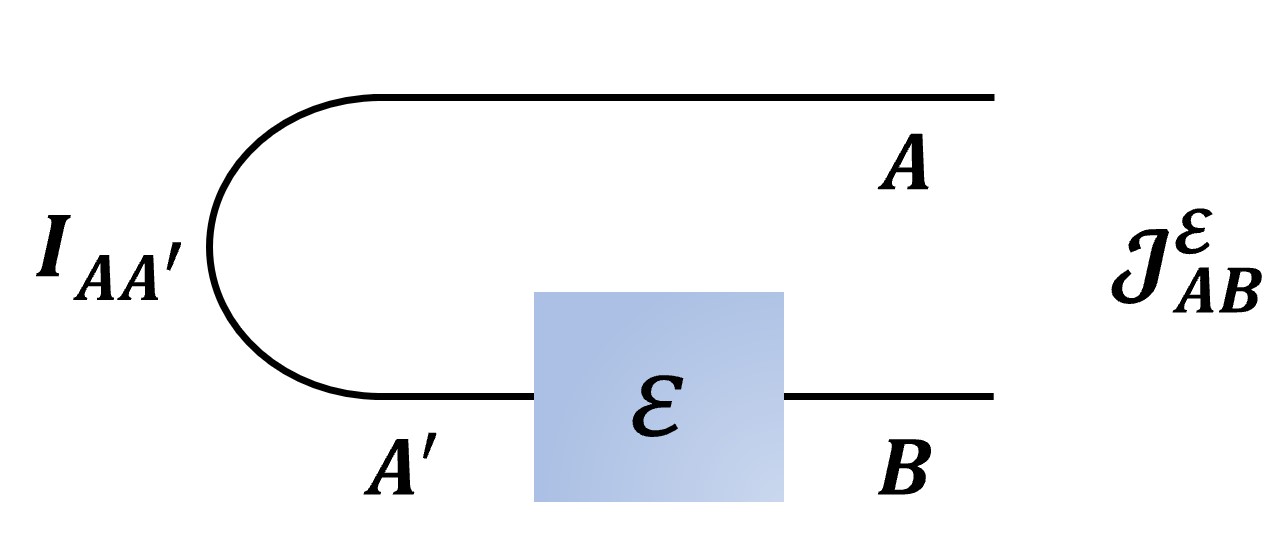}
\caption{(color online) Circuit realization of generating the CJ operator $J^{\mE}_{AB}$ (see Eq.~\ref{eq:cj}). Here $I_{AA^{'}}$ stands for $\ketbra{I}{I}_{AA^{'}}$, and $\ket{I}_{AA^{'}}$ represents the unnormalized maximally entangled state $\sum_{i}\ket{i}_{A}\ket{i}_{A^{'}}$. 
}
\label{fig:cj}
\end{figure}

\begin{mydef}
{Choi-Jamio\l kowski (CJ) Operator~\cite{JAMIOLKOWSKI1972275,CHOI1975285}}{CJ}
For a quantum channel $\mathcal{E}: A\to B$, its Choi-Jamio\l kowski operator $J^{\mathcal{E}}_{AB}$ is defined as
\begin{align}\label{eq:cj}
J^{\mathcal{E}}_{AB}
:=
\id_{A}\otimes\mathcal{E}_{A^{'}\to B} (\ketbra{I}{I}_{AA^{'}}),
\end{align}
where $\ket{I}_{AA^{'}}:= \sum_{i}\ket{i}_{A}\ket{i}_{A^{'}}$ is the unnormalized maximally entangled state with $A^{'}$ being a replica of system $A$, and $\{\ket{i}\}$ being an orthonormal basis on $A$. See Fig.~\ref{fig:cj} for an illustration of Eq.~\ref{eq:cj}.
\end{mydef}

Here the correspondence between a quantum channel $\mathcal{E}$ and its CJ operator $J^{\mathcal{E}}$ is referred as the Choi-Jamio\l kowski isomorphism in quantum information theory \cite{nielsen_chuang_2010,wilde_2013,watrous_2018}. For any quantum state $\rho$ acting on system $A$, and a quantum channel $\mathcal{E}: A\to B$, the output state is fully characterized by the CJ operator $J^{\mE}$. More precisely, the output on system $b$ can be written as
\begin{align}\label{eq:state-under-channel}
    \mE(\rho)= \Tr_{A}[J^{\mE}_{AB}\cdot \rho^{\T}_{A}\otimes \1_{B}].
\end{align}
Using the language of CJ operators, we know that $\mathcal{E}$ is (\romannumeral1) completely positive (CP) if and only if its CJ operator satisfies $J^{\mathcal{E}}\geqslant0$ and (\romannumeral2) trace-preserving (TP) if and only if its CJ operator satisfies $\Tr_{B}[J^{\mathcal{E}}_{AB}]=\1_{A}$. Here $A$ and $B$ represent the input and output systems of $\mE$ respectively. The notation  $\Tr_{A}$ stands for the partial trace over system $A$.

The development of quantum telecommunications and networks requires the investigations of quantum channels with multiple inputs and outputs at different time ticks, which comply with the causality in the theory of relativity. More precisely, in the theory of quantum information, such a causality is restricted by the concept of non-signaling (NS), which is formally defined as

\begin{mydef}
{Non-Signaling (NS)}{NS}
Given a bipartite channel $\mathcal{E}: AC\to BD$, it is called non-signaling from $C\to D$ to $A\to B$ if the following condition is satisfied
\begin{align}\label{eq:ns1}
    \Tr_{D}\circ\,\mathcal{E}_{AC\to BD}
    =
    \mathcal{F}_{A\to B}\otimes\Tr_{C},
\end{align}
for some quantum channel $\mathcal{F}$ from $A$ to $B$. On the other hand, we say $\mE$ is non-signaling from $A\to B$ to $C\to D$, if there exists a quantum channel $\mathcal{G}$ from $C$ to $D$ such that
\begin{align}\label{eq:ns2}
    \Tr_{B}\circ\,\mathcal{E}_{AC\to BD}
    =
    \mathcal{G}_{C\to D}\otimes\Tr_{A}.
\end{align}
Finally, the bipartite channel $\mE$ is said to be non-signaling if it meets the condition of both Eq.~\ref{eq:ns1} and Eq.~\ref{eq:ns2}.
\end{mydef}

\begin{figure}[h]
\includegraphics[width=0.7\textwidth]{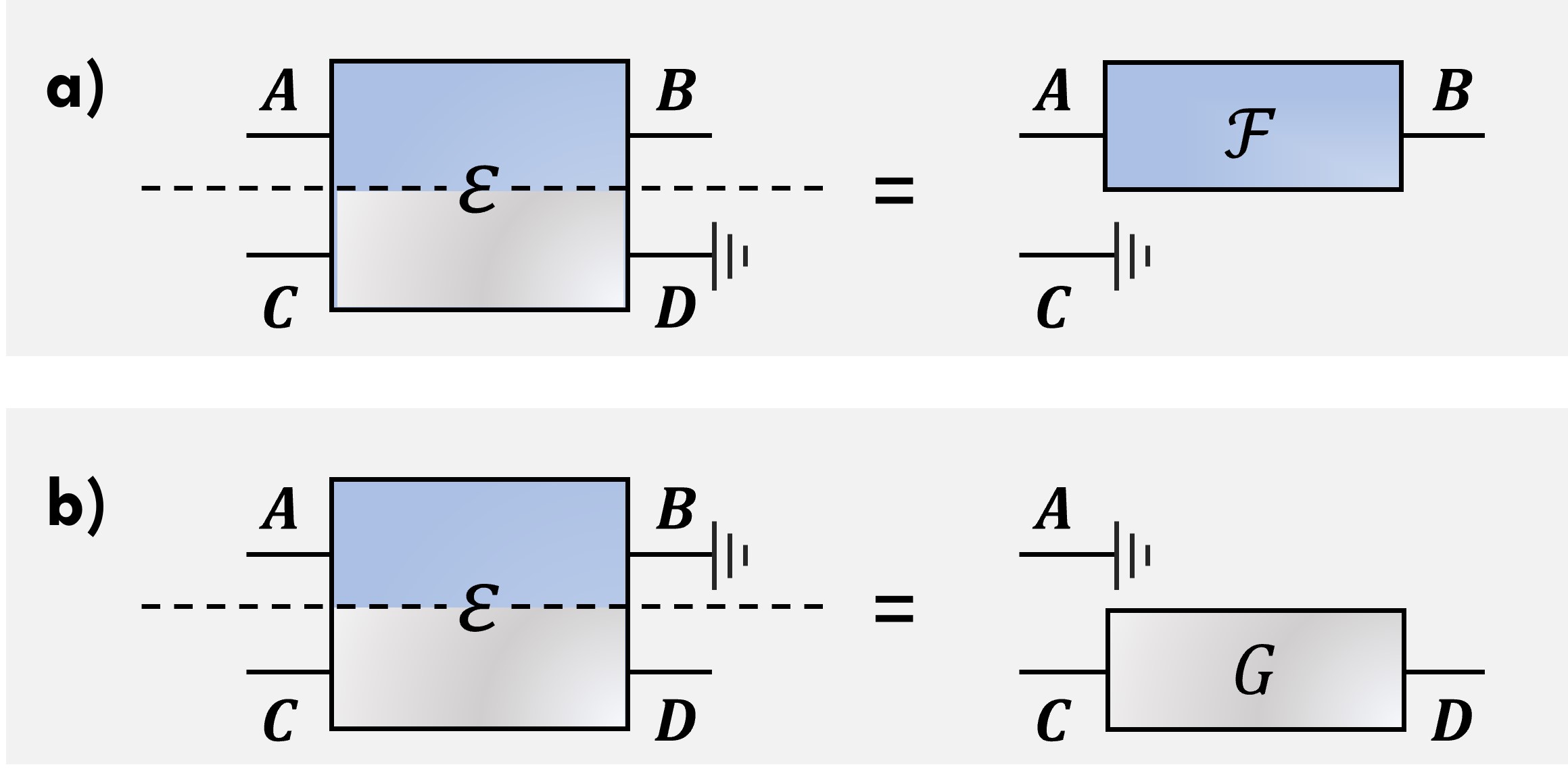}
\caption{(color online) Pictorial demonstrations of non-signaling (NS) bipartite quantum channel $\mE$ from systems $AC$ to $BD$: (a) NS from $C\to D$ to $A\to B$ (see Eq.~\ref{eq:ns1}); (b) NS from $A\to B$ to $C\to D$ (see Eq.~\ref{eq:ns2}).
}
\label{fig:ns}
\end{figure}

\noindent
Noted that in Def.~\ref{def:NS}, Eq.~\ref{eq:ns1} means that for any bipartite input state $\rho_{AC}$ acting on systems $AC$, we have 
\begin{align}
    \Tr_{D}\circ\,\mathcal{E}_{AC\to BD}(\rho_{AC})
    =
    \Tr_{D}
    [
    \mE_{AC\to BD}(\rho_{AC})
    ]
    =
    \mathcal{F}_{A\to B}\otimes\Tr_{C}
    (\rho_{AC})
    =
    \mF_{A\to B}
    (\rho_{A}),
\end{align}
where $\rho_{A}:= \Tr_{C}[\rho_{AC}]$ is the reduced state of $\rho_{AC}$ on system $A$. A similar situation applies to Eq.~\ref{eq:ns2}, as shown in Fig.~\ref{fig:ns}.

Physically, Eq.~\ref{eq:ns1} describes a situation where $C\to D$ is a quantum dynamical process that happened after the process of $A\to B$. Hence, information cannot be transmitted from the future (i.e. $C\to D$) to the past (i.e. $A\to B$). A similar statement holds for Eq.~\ref{eq:ns2}, where the temporal order for $C\to D$ and $A\to B$ has been switched. 

\begin{figure}[h]
\includegraphics[width=0.35\textwidth]{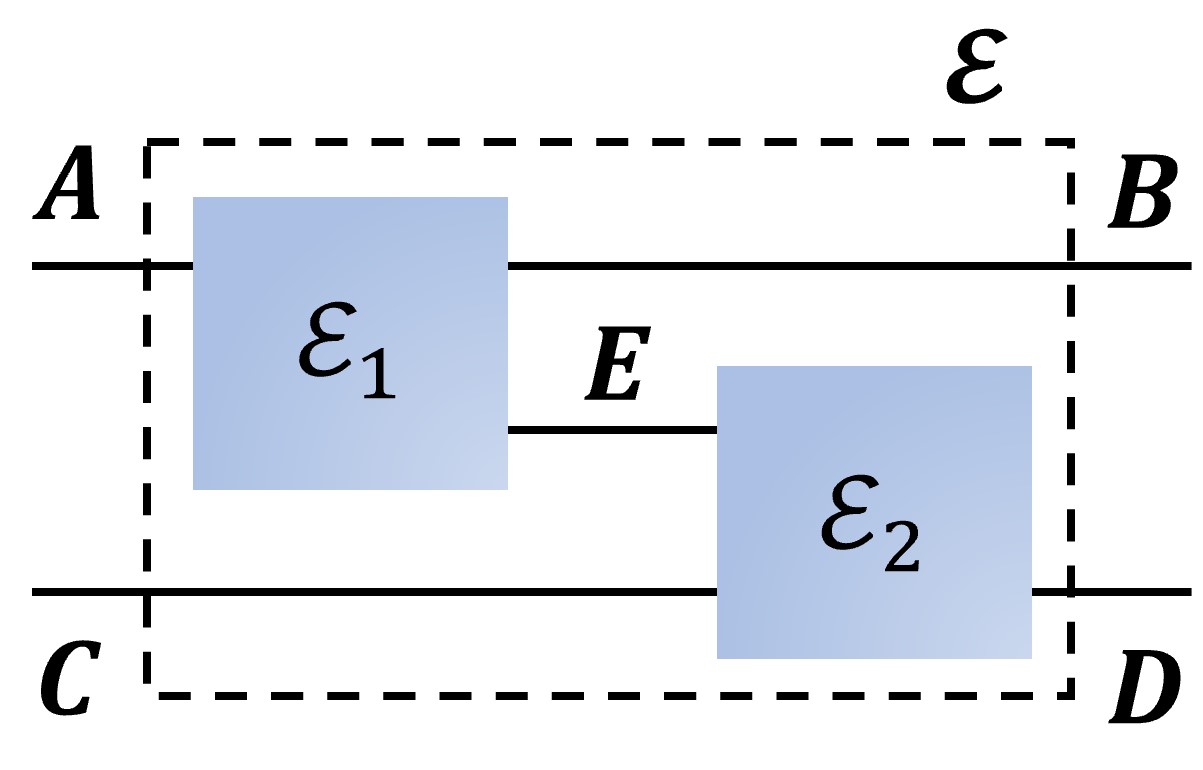}
\caption{(color online) Sequential composition of quantum channels $\mE_1$ and $\mE_2$, i.e. $\mathcal{E}:= \mathcal{E}_{2}\circ\mathcal{E}_{1}$, where $\mE_1$ is a channel from system $A$ to $BE$, and meanwhile $\mE_2$ is a channel from systems $CE$ to $D$. System $E$ is the link that connects them together.
}
\label{fig:link}
\end{figure}

The non-signaling condition also plays an important role in the composition of dynamical processes. Let us consider a concrete example. Assume a quantum channel $\mathcal{E}_{1}: A\to BE$ is followed by another channel $\mathcal{E}_{2}: CE \to D$, where the two channels are connected by a memory system $E$, as illustrated in Fig.~\ref{fig:link}. In this case, the whole quantum dynamics $\mathcal{E}:= \mathcal{E}_{2}\circ\mathcal{E}_{1}$ is a linear map from $AC$ to $BD$, satisfying the following conditions: 
\begin{enumerate}[(i)]
\item Completely Positive (CP),
\item Trace-Preserving (TP),
\item Non-Signaling (NS) from $C\to D$ to $A\to B$.
\end{enumerate}

Denote the CJ operators of quantum channels $\mathcal{E}_{1}$ and $\mathcal{E}_{2}$ as $J^{1}_{ABE}$ and $J^{2}_{CDE}$ (or simply $J^1$ and $J^2$) respectively. It now follows immediately that the CJ operator of $\mE= \mathcal{E}_{2}\circ\mathcal{E}_{1}$ is given by~\cite{PhysRevA.80.022339}
\begin{align}\label{eq:e1-e2}
J^{\mE}=
\Tr_{E}[(J^{1})^{\T_{E}}\cdot J^{2}],
\end{align}
where $\T_{E}$ stands for the partial transpose over system $E$. Here the non-signaling condition indicates the temporal order between processes $\mE_1$ and $\mE_2$. As a by-product, Eq.~\ref{eq:e1-e2} motivates the abstract definition of {\it link product} $\star$ \cite{PhysRevLett.101.060401,PhysRevA.80.022339}.

\begin{mydef}
{Link Product~\cite{PhysRevLett.101.060401,PhysRevA.80.022339}}{LP}
Given two operators $M$ and $N$ acting on systems $XY$ and $YZ$ respectively, we define their link product $M\star N$ as 
\begin{align}
    M\star N:= \Tr_{Y}[M^{\T_{Y}}\cdot N],
\end{align}  
where the common space $Y$ is appeared as the link between operators $M$ and $N$, and has been ``swallowed'' by the product $\star$. Here $\Tr_{Y}$ and $\T_{Y}$ represent the partial trace and partial transpose over the system $Y$ respectively.
\end{mydef}
\noindent
Thanks to link product, now Eq.~\ref{eq:state-under-channel} can be simplified as 
\begin{align}
\mE(\rho)= J^{\mE}\star\rho.
\end{align}

\noindent
Equipped with the link product, now we can rewrite the CJ operator of $\mathcal{E}= \mathcal{E}_{2}\circ\mathcal{E}_{1}$ as 
\begin{align}
J^{\mathcal{E}}= J^1\star J^2. 
\end{align}
Here $\mathcal{E}= \mathcal{E}_{2}\circ\mathcal{E}_{1}$ is a typical example of quantum superchannel, where channels $\mathcal{E}_{1}$ and $\mathcal{E}_{2}$ are known as the pre-processing and post-processing of $\mathcal{E}$ respectively. Similar to the case of quantum channels, all above restrictions on $\mE$ can be translated into the language of CJ operators; a map $\mathcal{E}$ from $AC$ to $BD$ is (\romannumeral1) completely positive (CP) if and only if $J^{\mathcal{E}}\geqslant0$, (\romannumeral2) trace-preserving (TP) if and only if $\Tr_{BD}[J^{\mathcal{E}}]= \mathds{1}_{AC}$, and (\romannumeral3) non-signaling (NS) from $C\to D$ to $A\to B$ if and only if $\Tr_{D}[J^{\mathcal{E}}]= \Tr_{CD}[J^{\mathcal{E}}]\otimes\mathds{1}_{C}/d_{C}$ with $d_{C}:=\dim C$. Note that here $\Tr_{CD}[J^{\mathcal{E}}]/d_{C}$ forms a CJ operator for some quantum channel, as it is both completely positive (CP) and trace-preserving (TP).

\begin{figure}[h]
\includegraphics[width=0.4\textwidth]{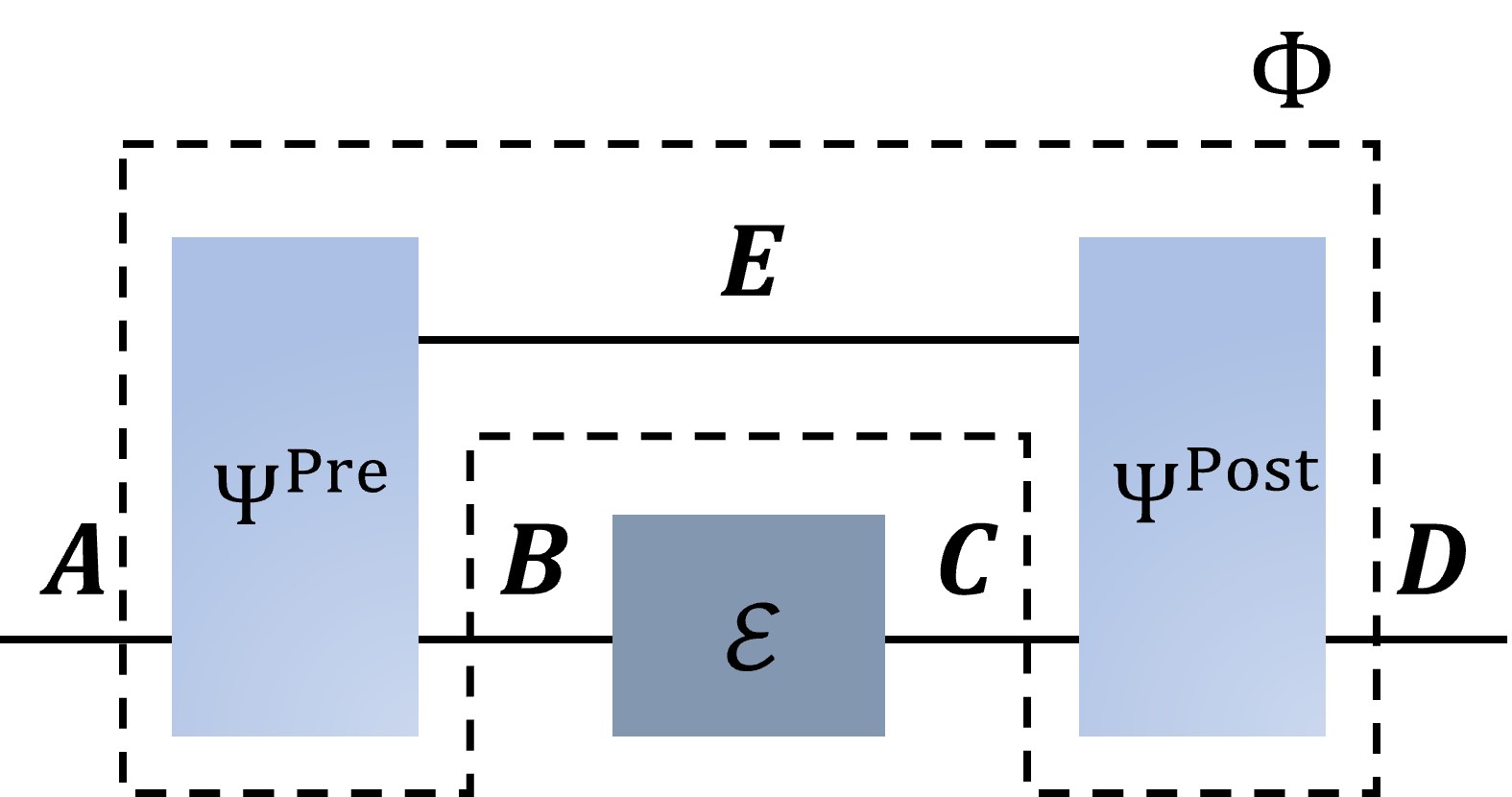}
\caption{(color online) Physical realization of the quantum superchannel $\Phi= \Psi^{\text{Post}}\circ\Psi^{\text{Pre}}$, whose action on input channel is described by Eq.~\ref{eq:sup}. Here channel $\Psi^{\text{Pre}}$ stands for the pre-processing of superchannel $\Phi$, channel $\Psi^{\text{Post}}$ represents the post-processing of superchannel $\Phi$, and they are connected through the memory system $E$.
}
\label{fig:sup}
\end{figure}

In fact, all quantum superchannels satisfy above conditions, namely completely positive (CP), trace-preserving (TP), and non-signaling (NS) from one process to another. Furthermore, the inverse statement -- a linear map satisfying completely positive (CP), trace-preserving (TP), and non-signaling (NS) from one process to another forms a quantum superchannel -- is also true~\cite{2008Chiribella}. Mathematically, given a bipartite quantum channel $\Phi: AC\to BD$ and non-signaling (NS) from $C\to D$ to $A\to B$, there exist two quantum channels -- $\Psi^{\text{Pre}}$ (known as pre-processing) and $\Psi^{\text{Post}}$ (known as post-processing) -- such that for any quantum channel $\mE$ from system $B$ to system $C$ and quantum state $\rho$ acting on system $A$, we have
\begin{align}\label{eq:sup}
    \Phi(\mE)(\rho)= \Psi^{\text{Post}}\circ \mE\circ \Psi^{\text{Pre}} (\rho).
\end{align}
Here $\Phi$ represents a general manipulation of quantum channels with definite causal order, which maps a quantum channel of form $B\to C$ to a resultant channel of form $A\to D$, visualizing in Fig.~\ref{fig:sup}.

Remark that, on the one hand, a quantum channel can be viewed as a special kind of superchannel with only one process (either pre- or post-), and no quantum memory system. On the other hand, quantum superchannels can also be recognised as bipartite quantum channels with additional conditions, namely NS from a later process to early one. From this perspective, it is straightforward to formulate the CJ operator of a superchannel, such as $\Phi: AC\to BD$.
\begin{align}
    J^{\Phi}_{ABCD}
    :=
    \id_{AC}\otimes\Phi_{A^{'}C^{'}\to BD} (\ketbra{I}{I}_{AA^{'}}
    \otimes
    \ketbra{I}{I}_{CC^{'}}),
\end{align}
where $\ket{I}_{AA^{'}}$ and $\ket{I}_{CC^{'}}$ are unnormalized maximally entangled states acting on systems $AA^{'}$ and $CC^{'}$ respectively. For a quantum channel $\mE: B\to C$, its resultant channel under $\Phi$ has the following CJ operator,
\begin{align}
    J^{\Phi(\mE)}
    =
    J^{\Phi}\star J^{\mE},
\end{align}
where $J^{\mE}$ stands for the CJ operator of the channel $\mE$.


\subsection{\label{subsec:Fragment}Quantum Circuit Fragments: Multiple Quantum Processes with Definite Causal Order}

\begin{figure}[h]
\includegraphics[width=0.75\textwidth]{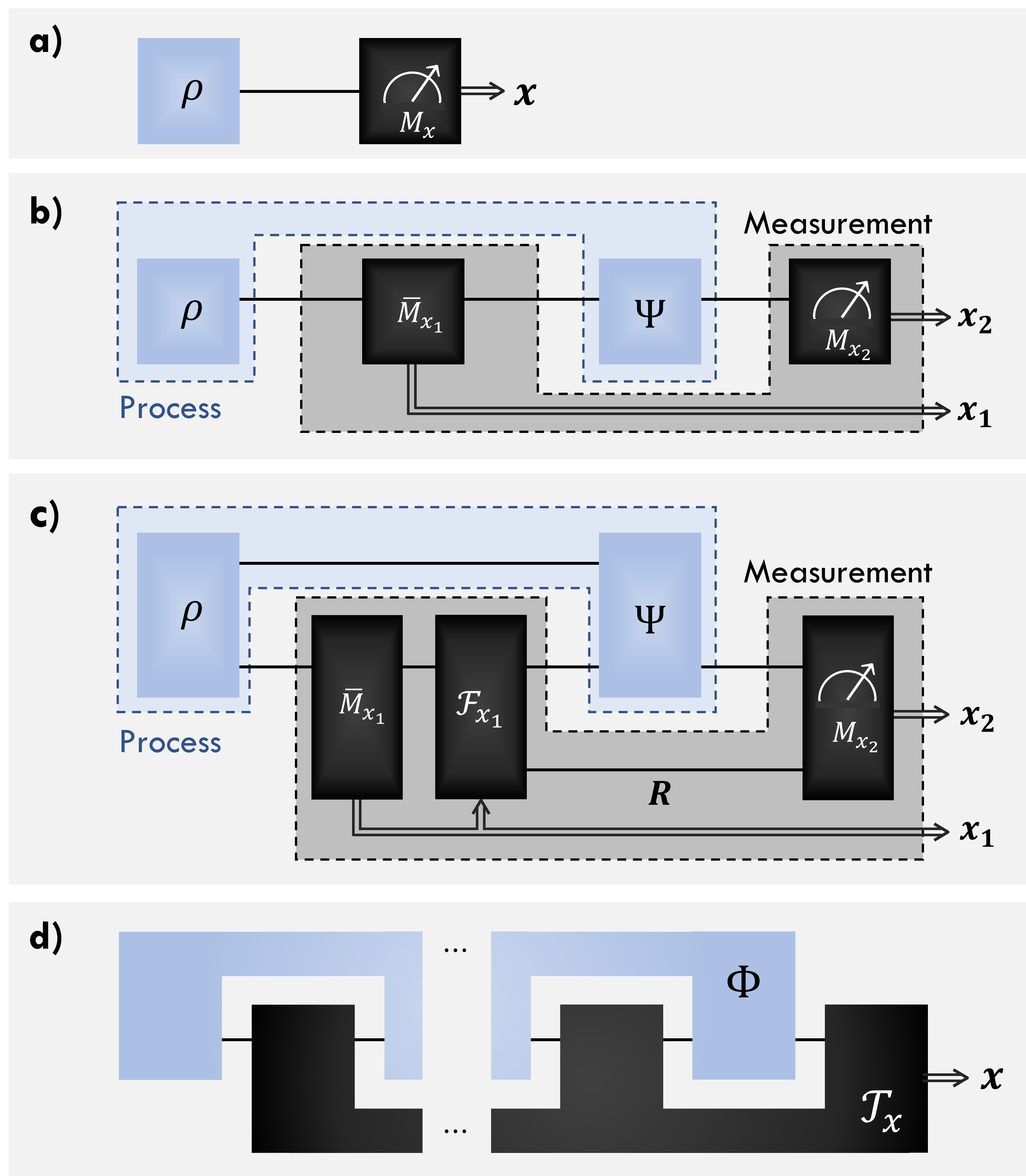}
\caption{(color online) Quantum circuit fragments and interactive measurements: (a) Passive measurement, where the system described by $\rho$ is left to evolve before feeding into the measuring process $M:= \{M_x\}_x$. (b) Measuring quantum dynamics at two different time points. To investigate the system of $\rho$ and its dynamics $\Psi$, we have executed a quantum instrument $\overline{M}$ at time point $t_1$, followed by a quantum measurements $M$ taken at time point $t_2$, where $t_1< t_2$. The measurement outcome obtained at time $t_1$ and $t_2$ are denoted as $x_1$ and $x_2$ respectively. In this example, the entire quantum dynamics, including the preparation of initial state $\rho$ and quantum evolution $\Psi$, is marked with blue color. Meanwhile, the adaptive measuring protocol forms a simple example of interactive measurement, which is colored black. (c) Correlated quantum dynamics and measurement with interventions. Based on previous measurement outcome $x_1$, an intervention $\mF_{x_1}$, characterized by a quantum channel, has been applied to the quantum dynamics. In this case, quantum information has been transmitted to later time point by using $\mF_{x_1}$. (d) Quantum circuit fragment $\Phi$ and interactive measurement $\mT:= \{\mT_x\}_x$. The most general picture of quantum dynamics $\Phi$ is described by quantum circuit fragment (colored blue). Here $\mT_x$ (colored black) represents an interactive measurement. Before obtaining a classical outcome $x$, $\Phi$ and $\mT_x$ have interacted multiple times.
}
\label{fig:motivation}
\end{figure}

Current quantum theories typically concern passive measurements, where a system is left to evolve freely before observation. A visual illustration of passive measurement has been provided in Fig.~\ref{fig:motivation}a, where the blue box represents state-preparational channel, emitting independent and identically distributed (i.i.d.) copies of state $\rho$. On the right side of Fig.~\ref{fig:motivation}a, the black box, waiting to receive an input, stands for the general positive operator valued measure (POVM) $M:= \{M_x\}_x$. To further investigate the system of interest (characterized by $\rho$) and its evolution (illustrated by $\Psi$), the observer might
implement measurement $M:= \{M_x\}_x$ at two different time points $t_1$ and $t_2$ ($t_1< t_2$) and collect the corresponding outcome $x_1$ and $x_2$, as demonstrated in Fig.~\ref{fig:motivation}b. For simplicity, here we assume that the measurements executed in $t_1$ and $t_2$ are the same. However, generally they do not have to be the same. In this case, the first measurement occurred at time $t_1$ interacts with the system, influencing the second measurement. Such a measuring protocol (colored black in Fig.~\ref{fig:motivation}b) forms a simple example of interactive measurement. In addition, systems can have prior correlations. For example, the quantum dynamics inside blue dashed box of Fig.~\ref{fig:motivation}c is what it looks like. Now based on the measurement outcome $x_1$ obtained at $t_1$, we can then apply an intervention, described by a quantum channel $\mF_{x_{1}}$ (depending on the outcome $x_1$), and finalize the measuring process by carrying out a joint measurement. An illustration is given in the black dashed box of Fig.~\ref{fig:motivation}c. Most generally, systems can have prior correlations over different time points, and measuring processes can contain multiple interventions, limned by Fig.~\ref{fig:motivation}d. The quantum dynamics and the measurement with interventions are colored blue and black respectively. In this work, the entire class of quantum dynamics with definite causal order and measurement with interventions mentioned above are characterized by the framework of \emph{quantum circuit fragments} (see Def.~\ref{def:QCF} of Subsec.~\ref{subsec:Fragment}) and \emph{interactive measurements} (see Def.~\ref{def:IM} of Subsec.~\ref{subsec:Interactive-Measurement}) respectively, and our results involves the most general picture of them.

In this subsection, we focus on the idea of quantum circuit fragments. Roughly speaking, quantum channels have been employed to characterize the quantum dynamics with a single process. Meanwhile, quantum superchannels offer us the most general way of manipulating quantum channels~\cite{2008Chiribella}, constituting two processes that occur at different time ticks. These specific quantum dynamics give us a glimpse of a more general framework -- quantum circuit fragments, where multiple quantum processes are connected through quantum memories with definite causal orders. Formally, the `Quantum Circuit Fragments' or briefly `Circuit Fragments' is defined as

\begin{figure}[h]
\includegraphics[width=1\textwidth]{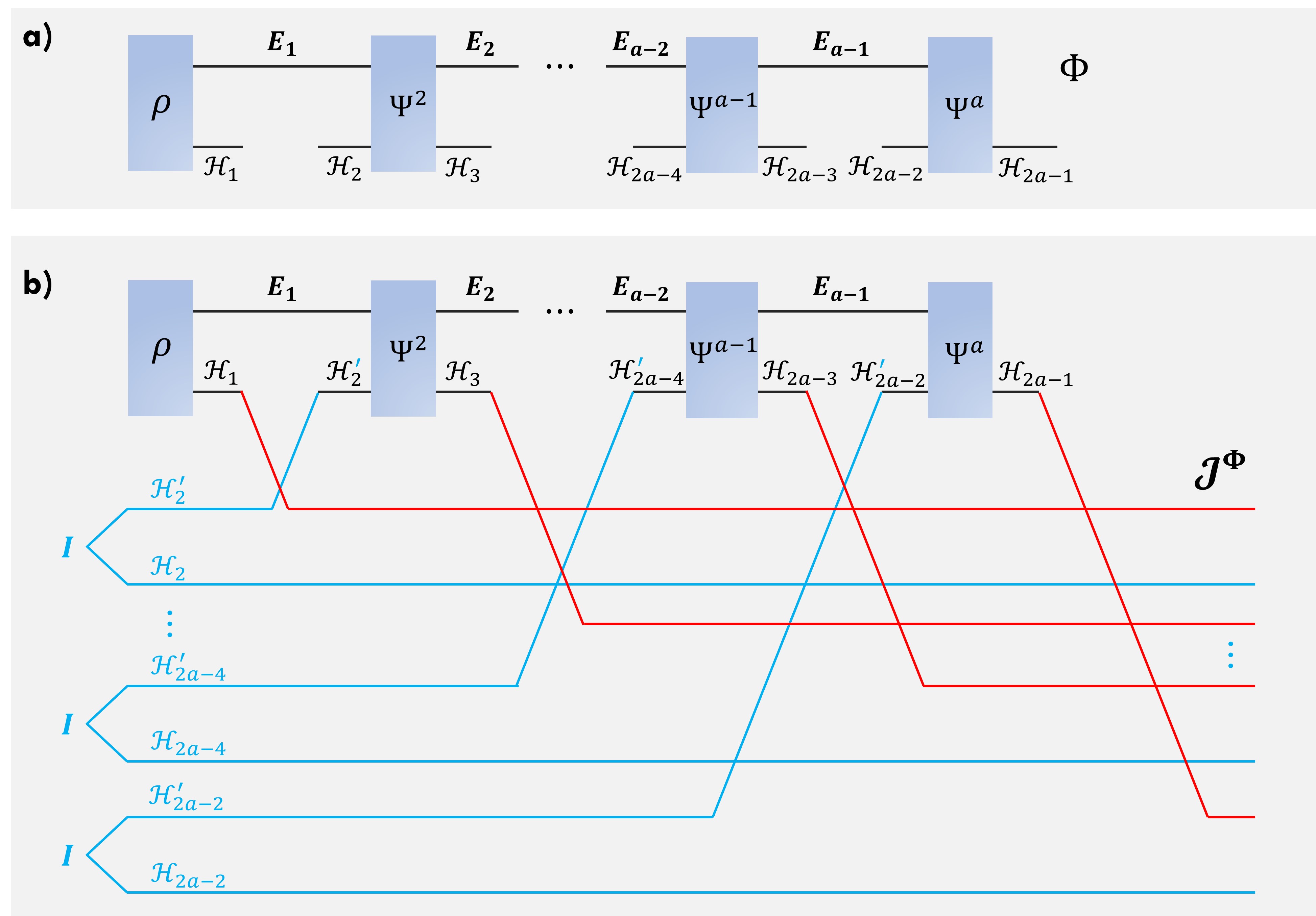}
\caption{(color online) Quantum circuit fragment $\Phi$ (a) and its CJ operator $J^{\Phi}$ (b). Here the quantum circuit fragment $\Phi$ is characterized by Eq.~\ref{eq:QCF}, including an initial state $\rho$ and $a-1$ multiple quantum processes $\{\Psi^{i}\}_{i=2}^{a}$. The corresponding CJ operator $J^{\Phi}$ is given by Eq.~\ref{eq:qcf-J}. In (b), the blue lines indicate the systems coming from $\ketbra{I}{I}_{\otimes_{i=2}^{a}\mH_{2i-2}\mH_{2i-2}^{'}}$ (see Eq.~\ref{eq:QCFI}), and the red lines represent the output systems of quantum processes $\{\Psi^{i}\}_{i=2}^{a}$.
}
\label{fig:QCF}
\end{figure}

\begin{mydef}
{Quantum Circuit Fragments}{QCF}
A bipartite quantum state $\rho$ is prepared in systems $\mH_{1}E_{1}$, and subjected to $a-1$ quantum channels $\{\Psi_{2}, \ldots, \Psi_{a}\}$ in the form of
\begin{align}
    &\Psi^{i}:\, \mH_{2i-2}E_{i-1}\to \mH_{2i-1}E_{i},\quad 2\leqslant i\leqslant a-1,\\
    &\Psi^{a}:\, \mH_{2a-2}E_{a-1}\to \mH_{2a-1}.
\end{align}
Then we call the quantum dynamics 
\begin{align}\label{eq:QCF}
    \Phi
    := 
    \Psi^{a}\circ\Psi^{a-1}
    \circ
    \cdots
    \circ
    \Psi^{2}(\rho)
\end{align} 
a quantum circuit fragment (see Fig.~\ref{fig:QCF}a). Denote the set of all quantum circuit fragments in the form of Eq.~\ref{eq:QCF} as $\mathfrak{F}_{a}$.
\end{mydef}

\noindent
Here $\Phi$ is a quantum channel with $a$ individual processes $\{\Psi^{i}\}_{i=1}^{i=a}$, where $\Psi^{1}:= \rho$ acting on systems $\mH_1 E_1$.
Its input systems and output systems are $\otimes_{i=2}^{a}\mH_{2i-2}$ and $\otimes_{i=1}^{a}\mH_{2i-1}$ respectively. More precisely, $\Phi: \otimes_{i=2}^{a}\mH_{2i-2}\to\otimes_{i=1}^{a}\mH_{2i-1}$ is a quantum channel satisfying non-signaling (NS) from $\mH_{2i-2}\to\mH_{2i-1}$ to $\mH_{2i-4}\to\mH_{2i-3}$ for all $2\leqslant i\leqslant a$, with $\mH_{0}= \mathds{C}$. Noted that here all the memory systems $\{E_{i}\}_{i=1}^{a-1}$ of $\Phi$ have been swallowed by the composition of quantum channels. Similar to the CJ operators of quantum superchannels discussed in Sec.~\ref{subsec:Channel-Superchannel}, the CJ operator of $\Phi$ can be obtained by considering the following equation
\begin{align}\label{eq:qcf-J}
    J^{\Phi}
    :=
    \id_{\otimes_{i=1}^{a}\mH_{2i-2}}
    \otimes
    \Psi^{a}_{\mH_{2a-2}^{'}E_{a-1}\to \mH_{2a-1}}
    \circ
    &\Psi^{a-1}_{\mH_{2a-4}^{'}E_{a-2}\to \mH_{2a-3}E_{a-1}}
    \circ
    \cdots\notag\\
    \cdots
    \circ
    &\Psi^{2}_{\mH_{2}^{'}E_{1}\to \mH_{3}E_{2}}
    (
    \rho_{\mH_{1}E_{1}}
    \otimes
    \ketbra{I}{I}_{\otimes_{i=2}^{a}\mH_{2i-2}\mH_{2i-2}^{'}}
    ),
\end{align}
where the multiple copies of unnormalized maximally entangled state
$\ketbra{I}{I}_{\otimes_{i=2}^{a}\mH_{2i-2}\mH_{2i-2}^{'}}$ is defined as
\begin{align}\label{eq:QCFI}
    \ketbra{I}{I}_{\otimes_{i=2}^{a}\mH_{2i-2}\mH_{2i-2}^{'}}
    :=
    \ketbra{I}{I}_{\mH_{2}\mH_{2}^{'}}
    \otimes
    \ketbra{I}{I}_{\mH_{4}\mH_{4}^{'}}
    \otimes
    \cdots
    \otimes
    \ketbra{I}{I}_{\mH_{2a-2}\mH_{2a-2}^{'}}.
\end{align}
The physical process of generating $J^{\Phi}$ is demonstrated in Fig.~\ref{fig:QCF}b. Denote the CJ operator of $\Psi^{i}$ (see Def.~\ref{def:QCF}) as $J^{i}$ ($i\in\{2, 3, \ldots, a-1\}$), then it is straightforward to check that $J^{\Phi}$ can also be written as a result of link product, namely
\begin{align}
    J^{\Phi}
    =
    J^{a}\star J^{a-1}\star
    \cdots
    \star J^{2}\star \rho.
\end{align}
In literature~\cite{PhysRevLett.101.060401,PhysRevA.80.022339}, $J^{\Phi}$ is called an $a$-comb, where $a$ indicates the number of processes contained in quantum circuit fragment $\Phi$. Analogous to the Choi-Jamio\l kowski (CJ) isomorphism between the quantum channels $\mE: A\to B$ and their CJ operators $J^{\mE}_{AB}$, the morphism between the quantum circuit fragment $\Phi$ and the corresponding $a$-comb $J^{\Phi}$ forms a bijection. This implies that the physical properties associated with $\Phi$ are completely characterized by its comb representation $J^{\Phi}$.

In open quantum systems, a special form of quantum circuit fragment -- known as process tensor~\cite{PhysRevA.97.012127,PhysRevLett.122.140401,PRXQuantum.2.030201} -- has been used to describe the system-environment unitary dynamics. In this model, it is assumed that the system-environment is initialized in a state $\rho_{\mH_{1}E_{1}}$. Without loss of generality, we can always assume it is a pure state (namely $\rho_{\mH_{1}E_{1}}= \phi_{\mH_{1}E_{1}}$) by enlarging the environment system $E_1$. After the preparation of $\phi_{\mH_{1}E_{1}}$, the system-environment dynamics undergo unitary interactions $\{\mU_{3:2}, \ldots, \mU_{2a-1:2a-2}\}$ at $a-1$ time steps. Now the whole process $\Phi$ is demonstrated by the following map
\begin{align}\label{eq:process-tensor}
    \Phi
    = 
    \Tr_{E_{a}}[
    \,
    \mU_{2a-1:2a-2}\circ\cdots\circ\mU_{3:2}(\phi_{\mH_{1}E_{1}})
    ],
\end{align}
where, for any input state $\sigma$, the output state under $\mU$ is given by $\mU(\sigma)= U \sigma U^{\dag}$. It is easy to verify that this process tensor can be viewed as a special form of quantum circuit fragment (see Def.~\ref{def:QCF}) by setting
\begin{align}
    &\rho= \phi_{\mH_{1}E_{1}}:\, \mathds{C} \to \mH_{1}E_{1},\\
    &\Psi^{i}= \mU_{2i-1: 2i-2}:\, \mH_{2i-2}E_{i-1}\to \mH_{2i-1}E_{i},\quad 2\leqslant i\leqslant a-1,\\
    &\Psi^{a}= \mU_{2a-1:2a-2}:\, \mH_{2a-2}E_{a-1}\to \mH_{2a-1}E_{a}.
\end{align}
Remark that, here the $a$-th quantum process is a unitary interaction from systems $\mH_{2a-2}E_{a-1}$ to $\mH_{2a-1}E_{a}$ followed by the partial trace over environmental system $E_{a}$. 

Besides process tensor, in the context of different theories, different concepts of quantum dynamics with multiple processes have been introduced, such as the causal maps in quantum causal inference~\cite{ried2015quantum,causal-NC}, the pseudo-density matrices (PDMs) in witnessing temporal correlations~\cite{fitzsimons2015quantum}, the collisional model of 
quantum thermometer~\cite{PhysRevLett.123.180602}, the quantum neural networks (QNNs)~\cite{QNN} and more general parameterized quantum circuits (PQCs)~\cite{RevModPhys.94.015004} in the noisy intermediate-scale quantum (NISQ) era~\cite{Preskill2018quantumcomputingin}, and so forth. Despite having  different names, there is a common point for all these concepts -- they contain multiple quantum processes with definite causal order. Finally, it is worth mentioning that all of them can be treated as special cases of our quantum circuit fragment.



\subsection{\label{subsec:Interactive-Measurement}Interactive Measurements: How to Measure the Quantum Circuit Fragments?}

Quantum circuit fragments offer us a general way of demonstrating multiple quantum processes with definite causal order. To investigate the physical properties associated with quantum circuit fragments and decoding information from such quantum circuit fragments, a measuring process is needed. Unlike state case, where a system is left to evolve freely before observation, now we can interact with the quantum dynamics and make multiple preceding interventions before final measurement. Here, the whole process is called interactive measurement. 

Let us consider the quantum circuit fragment $\Phi$ demonstrated in Eq.~\ref{eq:QCF}. The corresponding interactive measurement, denoted as $\mT$, contains $a-1$ rounds of interventions $\{\Lambda^{i}\}_{i=1}^{a-1}$ and a joint measurement $\{M_x\}_x$. Here the general interventions are described by quantum channels. In particular, the interactive measurement for quantum circuit fragment $\Phi$ is defined as

\begin{figure}[h]
\includegraphics[width=1\textwidth]{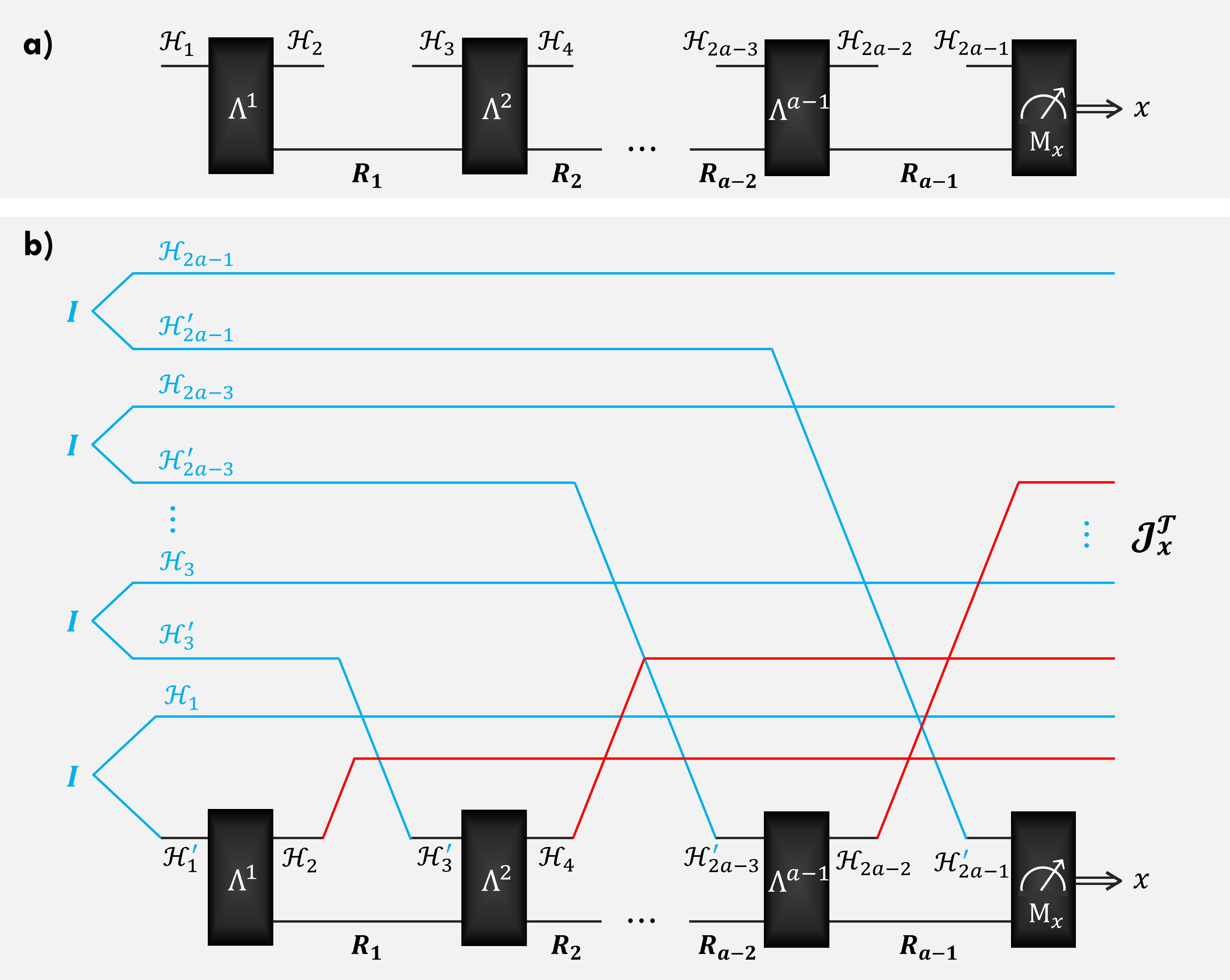}
\caption{(color online) Interactive measurement $\mT_x$ (a) and its CJ operator $J^{\mT}_{x}$ (b). Here the interactive measurement $\mT_x$ is characterized by Eq.~\ref{eq:IM}, including 
$a-1$ rounds of interventions $\{\Lambda^{i}\}_{i=1}^{a-1}$ and a joint measurement $M_x$. The corresponding CJ operator $J^{\mT}_{x}:= J^{\mT_{x}}$ is given by Eq.~\ref{eq:J_T_x}. In (b), the blue lines indicate the systems coming from $\ketbra{I}{I}_{\otimes_{i=1}^{a}\mH_{2i-1}\mH_{2i-1}^{'}}$ (see Eq.~\ref{eq:IMI}), and the red lines represent the system after interventions $\{\Lambda^{i}\}_{i=1}^{a-1}$.
}
\label{fig:IM}
\end{figure}

\begin{mydef}
{Interactive Measurements}{IM}
An interactive measurement $\mT$ designed for quantum circuit fragment $\Phi$ (See Eq.~\ref{eq:QCF}) constitutes of $a-1$ rounds of interventions $\{\Lambda^{i}\}_{i=1}^{a-1}$ in the form of
\begin{align}
    &\Lambda^{1}:\, \mH_{1}\to \mH_{2}R_{1},\\
    &\Lambda^{i}:\, \mH_{2i-1}R_{i-1}\to \mH_{2i}R_{i},\quad 2\leqslant i\leqslant a-1,
\end{align}
and a joint measurement $\{M_x\}_x$ acting on systems $\mH_{2a-1}R_{a-1}$, namely $M_{x}: \mH_{2a-1}R_{a-1}\to\mathds{C}$. Here each intervention $\Lambda^{i}$ forms a quantum channel, and different interventions are connected through memory systems $\{R_{i}\}_{i=1}^{a-1}$. Mathematically, the interactive measurement $\mT$ is characterized by the following maps $\{\mT_x\}_x$ (see Fig.~\ref{fig:IM}a); that are
\begin{align}\label{eq:IM}
    \mT_x(\cdot)
    := 
    \Tr[
    M_x\cdot\Lambda^{a-1}
    \circ
    \Lambda^{a-2}
    \circ
    \cdots
    \circ
    \Lambda^{1}(\cdot)
    ].
\end{align} 
Denote the set of all interactive measurements in the form of Eq.~\ref{eq:IM} as $\mathfrak{T}_{a}$.
\end{mydef}

\noindent
Within the framework of interactive measurement, previous interventions can be  used to improve the performance of upcoming interventions. To be more specific, take $\mT$ for instance, the interventions $\{\Lambda^{i}\}_{i=1}^{k-1}$ occurred before the implementation of $\Lambda^{k}$ strengthen our capability of gaining information from subsequent quantum interventions $\{\Lambda^{i}\}_{i=k}^{a-1}$. 
Given quantum circuit fragment $\Phi$, the probability of obtaining outcome $x$ by measuring interactive measurement $\mT$ is given by
\begin{align}\label{eq:p_x}
    p_x(\Phi, \mT)
    :=
    \Tr[
    M_x\cdot
    \Psi^{a}
    \circ
    \Lambda^{a-1}
    \circ
    \Psi^{a-1}
    \circ
    \Lambda^{a-2}
    \circ
    \cdots
    \circ
    \Psi^{2}
    \circ
    \Lambda^{1}(\rho)
    ].
\end{align}
Since the dependencies of probability distribution $p_x(\Phi, \mT)$ is usually clear from the context, we simply re-express it as $p_x$.

To characterize interactive measurements and simplify related calculations, the CJ operator of interactive measurements is a good choice. Again, let us take $\mT$ (see Eq.~\ref{eq:IM}) for an illustration. Here the input and output systems of $\mT:= \{\mT_x\}_x$ (see Eq.~\ref{eq:IM} for the definition of $\mT_x$) are $\otimes_{i=1}^{a}\mH_{2i-1}$ and $\otimes_{i=1}^{a}\mH_{2i-2}$, and the trivial system $\mathds{C}$ has been ignored. Thus, the CJ operator of each $\mT_x$ becomes 
\begin{align}\label{eq:J_T_x}
    &J^{\mT}_{x}
    :=
    J^{\mT_{x}}\notag\\
    =
    &\id_{\otimes_{i=1}^{a}\mH_{2i-1}}
    \otimes
    \Tr[
    M_x
    \cdot
    \Lambda^{a-1}
    _{\mH_{2a-3}^{'}R_{a-2}\to\mH_{2a-2}R_{a-1}}
    \circ
    \Lambda^{a-2}
    _{\mH_{2a-5}^{'}R_{a-3}\to\mH_{2a-4}R_{a-2}}
    \circ
    \cdots
    \circ
    \Lambda^{1}
    _{\mH_{1}^{'}\to\mH_{2}R_{1}}
    (\ketbra{I}{I}_{\otimes_{i=1}^{a}\mH_{2i-1}\mH_{2i-1}^{'}})
    ].
\end{align}
A circuit illustration of $J^{\mT}_{x}$ is depicted in Fig.~\ref{fig:IM}b.
Here the measurement $M_x$ is acting on systems $\mH_{2a-1}^{'}R_{a-1}$, and the operator $\ketbra{I}{I}_{\otimes_{i=1}^{a}\mH_{2i-1}\mH_{2i-1}^{'}}$ is defined as
\begin{align}\label{eq:IMI}
    \ketbra{I}{I}_{\otimes_{i=1}^{a}\mH_{2i-1}\mH_{2i-1}^{'}}
    :=
    \ketbra{I}{I}_{\mH_{1}\mH_{1}^{'}}
    \otimes
    \ketbra{I}{I}_{\mH_{3}\mH_{3}^{'}}
    \otimes
    \cdots
    \otimes
    \ketbra{I}{I}_{\mH_{2a-1}\mH_{2a-1}^{'}}.
\end{align}
Let us further denote the CJ operator of quantum circuit fragment $\Phi$ as $J^{\Phi}$, then probability $p_x$ considered in Eq.~\ref{eq:p_x} can be re-expressed as
\begin{align}
    p_x(\Phi, \mT)= J^{\Phi}\star J^{\mT}_{x},
\end{align}
where $J^{\mT}_{x}$ is defined in Eq.~\ref{eq:J_T_x}.

Another new concept that will help us demonstrate the trade-off between incompatible interactive measurements is \emph{eigencircuit}, which is formally defined as

\begin{mydef}
{Eigencircuit}{EC}
Given an interactive measurement $\mT:=\{\mT_{x}\}_{x}$, we call quantum circuit fragment $\Phi$ an eigencircuit of $\mT$ if 
\begin{align}
    J^{\Phi}\star J_{x}= 1,
\end{align}
for some $x$. Here $J^{\Phi}$ and $J_{x}$ represent the CJ operators of $\Phi$ and $\mT_{x}$ respectively.
\end{mydef}

After building the general framework of interactive measurements (see Def.~\ref{def:IM}), let us now investigate  some special forms of measurements for quantum dynamics and show that they are all special cases of our interactive measurements. First of all, consider the channel measurement, which was first introduced in Ref.~\cite{PhysRevA.77.062112} and named as process positive-operator-valued measure (PPOVM). Such a measuring process can be viewed as an interactive measurement by trivializing the input system of the first intervention, which is followed by a joint measurement directly. Second, we move on to investigating the framework of testers introduced in Ref.~\cite{PhysRevA.80.022339}. Mathematically, the testers are linear maps that take quantum causal networks as their inputs and output probability distributions. Physically, the testers generalize the concept of PPOVMs. In particular, a tester is consisted of the preparation of an initial state, finite rounds of interventions, and a joint measurement at the end of dynamical process. Analogous to relation between PPOVMs and interactive measurements, all testers can be regarded as interactive measurements with trivialized input system of the first intervention. More precisely, for general quantum testers, they can be written in the form of Eq.~\ref{eq:IM} with restricted $\Lambda^1$, where $\mH_1= \mathds{C}$, i.e. $\Lambda^1= \rho_{\mH_2 R_1}$ holds for some bipartite state acting on systems $\mH_2 R_1$.
To summarize, both PPOVMs and testers are all special cases of interactive measurements introduced in this work.

Different from previous works of investigating quantum dynamics, in this work we are not interested in the statistical properties associated with quantum circuit fragments and interactive measurements. Instead, we are caring about the physical properties that can be learned from quantum circuit fragments by implementing interactive measurements, such as their causal structures, non-Markovianity, and so on. More details will be unfolded in subsequent sections, especially the investigation of causal inference in Sec.~\ref{sec:CUR}.

\begin{figure}[h]
\includegraphics[width=1\textwidth]{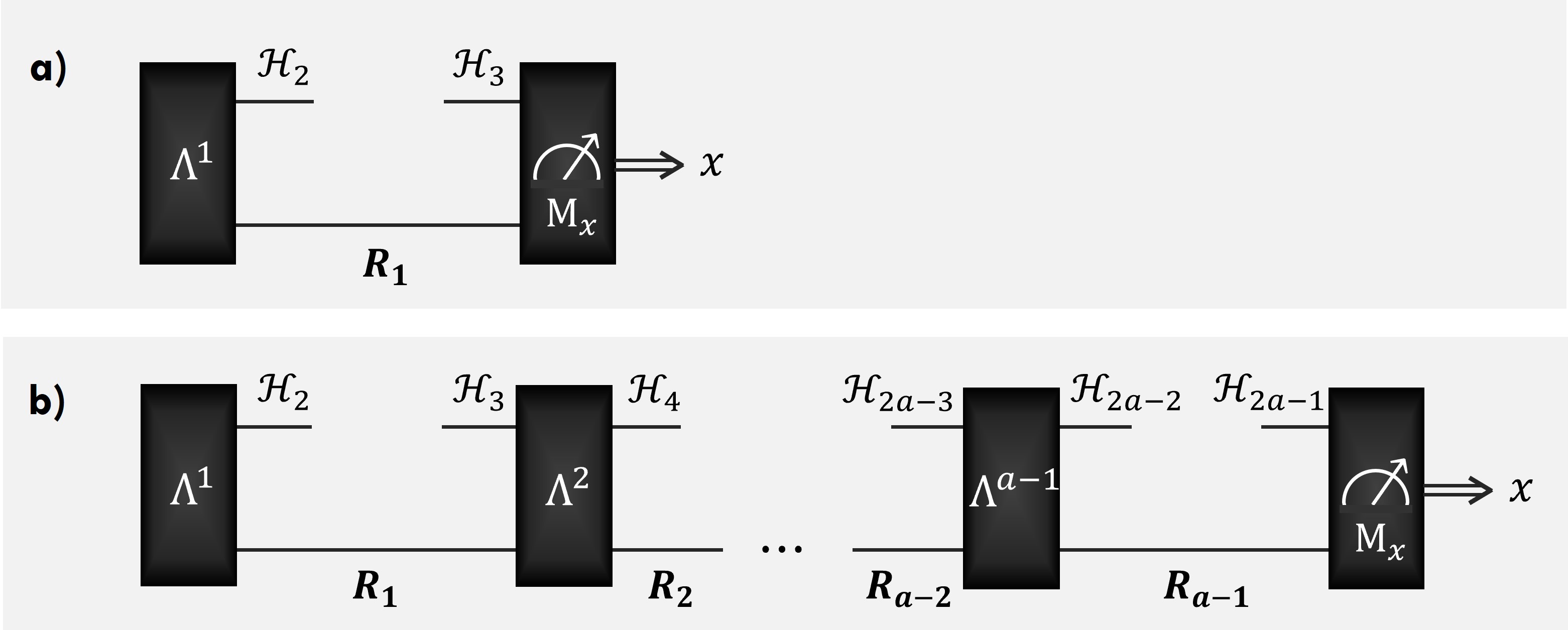}
\caption{(color online)
Circuit demonstration of process positive-operator-valued measure (PPOVM) (a) and tester (b). Both of them are obtained by trivializing the input system, i.e. $\mH_1$, of our interactive measurement model $\mT_x$ (see Eq.~\ref{eq:IMI}). In particular, for interactive measurement $\mT_x\in\mathfrak{T}_a$ depicted in Fig.~\ref{fig:IM}a, taking $\Lambda^1$ as $\rho_{\mH_2 R_1}\otimes\Tr_{\mH_1}$ leads to testers. When $a= 2$, we cover the framework of PPOVMs.
}
\label{fig:tester}
\end{figure}


\subsection{\label{subsec:CMaps}Quantum Causal Maps}

In this subsection, we turn our attention to the dynamics of quantum causal map, which was first introduced in the Ref.~\cite{ried2015quantum}, and further discuss how to obtain a causal map from process tensor in open quantum system (see Refs.~\cite{PhysRevA.97.012127,PhysRevLett.122.140401,PRXQuantum.2.030201} for more details about related topics). We start by writing down the system-environment dynamics described by process tensor $\Phi$ of the following form (see Eq.~\ref{eq:process-tensor}):
\begin{align}\label{eq:process-tensor}
    \Phi
    = 
    \Tr_{E_{a}}[
    \,
    \mU_{2a-1:2a-2}\circ\cdots\circ\mU_{3:2}(\phi_{\mH_{1}E_{1}})
    ],
\end{align}
where each $\mU_{2i-1, 2i-2}$ is a unitary map from system-environment dynamics $\mH_{2i-2}E_{i-1}$ to $\mH_{2i-1}E_{i}$ with $2\leqslant i\leqslant a$. Assume each system $\mH_{k}$ happens at time point $t_{k}$ ($1\leqslant k\leqslant 2a-1$), and we can only interact with the system-environment dynamics in two time periods -- $(t_{2i-1}, t_{2i})$ and $(t_{2j-1}, t_{2j})$ ($1\leqslant i< j\leqslant a-1$). Based on the feedback from these two time periods, our goal is to identify its causal structure.  

In this model, non-intervention, i.e. letting the system-environment dynamics evolve automatically, is equivalent to apply $\id_{\mH_{2k-1}\to\mH_{2k}}$ to system $\mH_{2k-1}$, where $1\leqslant k\leqslant a-1$ and $k\neq i, j$. To gain information from the quantum circuit fragment $\Phi$, we need to end our interactions with a measurement. Therefore, a POVM has been applied to the system $\mH_{2j-1}$, and after the time point $t_{2j}$ all systems will be discarded, including $\mH_{2j}$. Denote the state $\phi_{\mH_{2i-1}E_{i}}^{'}$ as
\begin{align}
    \phi_{\mH_{2i-1}E_{i}}^{'}
    :=
    \mU_{2i-1:2i-2}
    \circ
    \id_{\mH_{2i-3}\to\mH_{2i-2}}
    \circ
    \cdots
    \circ
    \mU_{3:2}
    \circ
    \id_{\mH_{1}\to\mH_{2}}
    (\phi_{\mH_{1}E_{1}}),
\end{align}
and define the following unitary map $\mU^{'}_{\mH_{2i}E_{i}\to\mH_{2j-1}E_{j}}$,
\begin{align}
    \mU^{'}_{\mH_{2i}E_{i}\to\mH_{2j-1}E_{j}}
    :=
    \mU_{2j-1:2j-2}
    \circ
    \id_{\mH_{2j-3}\to\mH_{2j-2}}
    \circ
    \cdots
    \circ
    \id_{\mH_{2i+1}\to\mH_{2i+2}}
    \circ
    \mU_{2i+1:2i}.
\end{align}
Now, besides the time periods $(t_{2i-1}, t_{2i})$ and $(t_{2j-1}, t_{2j})$, the whole quantum circuit fragment turns out to be $\Phi^{'}$, which is demonstrated by the following map
\begin{align}\label{eq:phi'}
    \Phi^{'}
    _{\mH_{2i}\to\mH_{2i-1}\mH_{2j-1}}
    =
    \Tr_{E_{j}}
    [\,
    \mU^{'}_{\mH_{2i}E_{i}\to\mH_{2j-1}E_{j}}(\phi_{\mH_{2i-1}E_{i}}^{'})
    ].
\end{align}

\begin{figure}[h]
\includegraphics[width=1\textwidth]{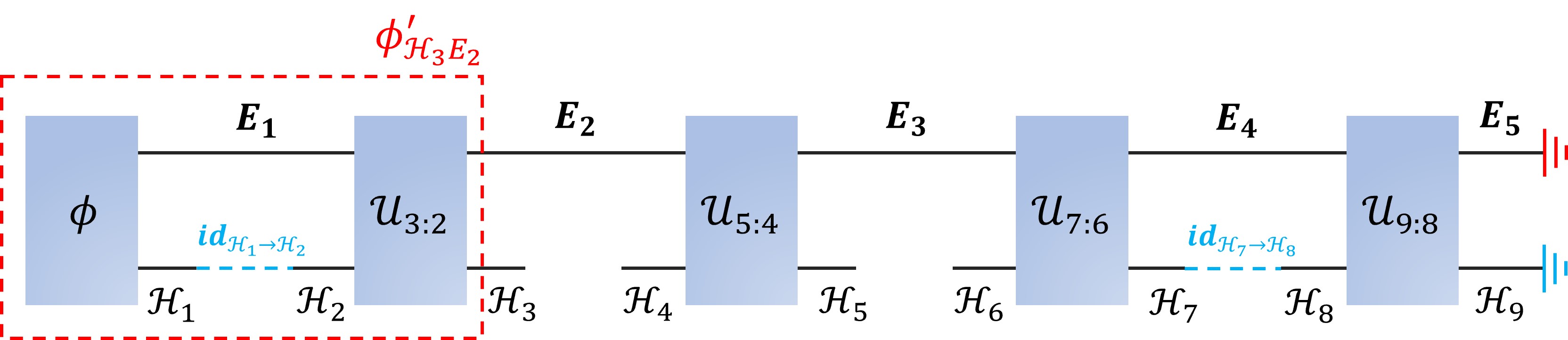}
\caption{(color online)
Investigating the causality of $\Phi$ (see Eq.~\ref{eq:cm-5}) between time periods $(t_{2i-1}, t_{2i})$ and $(t_{2j-1}, t_{2j})$. For this purpose, $\id$ channels have been applied to systems $\mH_{1}$ and $\mH_{7}$, followed by $\Tr_{E_5}$ and $\Tr_{\mH_9}$. The final quantum circuit fragment $\Phi^{'}$ belongs to the set of all causal maps, i.e. $\mathfrak{F}_{2}$, and has been characterized by Eq.~\ref{eq:cm-2}.
}
\label{fig:pt-causalmap}
\end{figure}

Take a system-environment dynamics with four rounds of unitary evaluations for instance (see Fig.~\ref{fig:pt-causalmap}); that is
\begin{align}\label{eq:cm-5}
    \Phi
    =
    \Tr_{E_{5}}[
    \,
    \mU_{9:8}
    \circ
    \mU_{7:6}
    \circ
    \mU_{5:4}
    \circ
    \mU_{3:2}
    (\phi_{\mH_{1}E_{1}})
    ]
    \in
    \mathfrak{F}_5.
\end{align}
To infer the causal models between time periods $(t_3, t_4)$ and $(t_5, t_6)$, we apply noiseless quantum channels to systems $\mH_1$ and $\mH_7$, i.e. described by $\id_{\mH_1\to\mH_2}$ and $\id_{\mH_7\to\mH_8}$ respectively. Then trace out all the systems happened after time point $t_6$, namely implementing $\Tr_{E_5}$ and $\Tr_{\mH_9}$. Thus, the resultant quantum dynamics turns out to be
\begin{align}\label{eq:cm-2}
    \Phi^{'}
    _{\mH_{4}\to\mH_{3}\mH_{5}}
    =
    \Tr_{E_{3}}
    [\,
    \mU_{5:4}(\phi_{\mH_{3}E_{2}}^{'})
    ],
\end{align}
where $\phi_{\mH_{3}E_{2}}^{'}:= \mU_{3:2}(\id_{\mH_1\to\mH_2}(\phi_{\mH_{1}E_{1}}))$.

Let us return to the model of general system-environment dynamics investigated in Eq.~\ref{eq:phi'}. To simplify our notations, we relabel the systems as
\begin{align}
    A&:= \mH_{2i-1},\\
    B&:= \mH_{2i},\\
    C&:= \mH_{2j-1},\\
    E&:= E_{i},\\
    F&:= E_{j}.
\end{align}
Equipped with new labels, the quantum circuit fragment $\Phi^{'}$ (see Eq.~\ref{eq:phi'}) can be re-expressed as
\begin{align}\label{eq:phi'decom}
    \Phi^{'}_{B\to AC}
    =
    \Tr_{F}
    [\,\mU_{BE\to CF}^{'}(\phi^{'}_{AE})].
\end{align}
Above example can be seen as a prototype of causal map in open quantum system. Generally, we call all the quantum circuit fragments $\Phi_{B\to AC}$ with $t_{A}\leqslant t_{B}\leqslant t_{C}$ quantum causal maps (see Fig.~2b of our main text), which form the set of $\mathfrak{F}_{2}$ (see Def.~\ref{def:QCF}). Such a quantum dynamics can also be viewed as a special form of superchannel with trivialized input system of pre-processing. In particular, given a causal map $\Phi_{B\to AC}\in\mathfrak{F}_{2}$, it can be divided into two quantum processes: pre-processing $\Psi^{\text{Pre}}_{\mathds{C}\to AE}$ and post-processing $\Psi^{\text{Post}}_{BE\to C}$, namely 
\begin{align}
\Phi_{B\to AC}
=
\Psi^{\text{Post}}_{BE\to C}\circ\Psi^{\text{Pre}}_{\mathds{C}\to AE}.
\end{align}
Here the subscript $\mathds{C}\to AE$ indicates that $\Psi^{\text{Pre}}_{\mathds{C}\to AE}$ is a state preparational channel. For Eq.~\ref{eq:phi'decom}, the map $\Phi^{'}_{B\to AC}$ can also be written as the composition of pre-processing $\phi^{'}_{AE}$ and post-processing $\Tr_{F}\circ\, \mU_{BE\to CF}^{'}$. Denote the CJ operators of pre-processing and post-processing as $J^{\text{Pre}}$ and $J^{\text{Post}}$ respectively, then the CJ operator of causal map $\Phi_{B\to AC}$ can be obtained from their link product; that is
\begin{align}
J^{\Phi}_{ABC}= J^{\text{Pre}}_{AE}\star J^{\text{Post}}_{BCE}. 
\end{align}
Here the CJ operator $J^{\Phi}$ of causal map $\Phi_{B\to AC}$, acting on systems $ABC$, satisfies the conditions of CPTP and NS from post-processing to pre-processing. It now follows immediately that
\begin{enumerate}[(i)]
\item CP: \quad $J^{\Phi}\geqslant0$,
\item TP: \quad $\Tr_{AC}[J^{\Phi}]= \mathds{1}_{B}$,
\item NS from post-processing to pre-processing: \quad $\Tr_{C}[J^{\Phi}]= \Tr_{BC}[J^{\Phi}]\otimes\mathds{1}_{B}/d_{B}$.
\end{enumerate}
Actually, these conditions can be further simplified as $J^{\Phi}\geqslant0$ and $\Tr_{C}[J^{\Phi}]= \rho_{A}\otimes\mathds{1}_{B}$ for some quantum state $\rho_{A}$. 

\begin{figure}[h]
\includegraphics[width=0.5\textwidth]{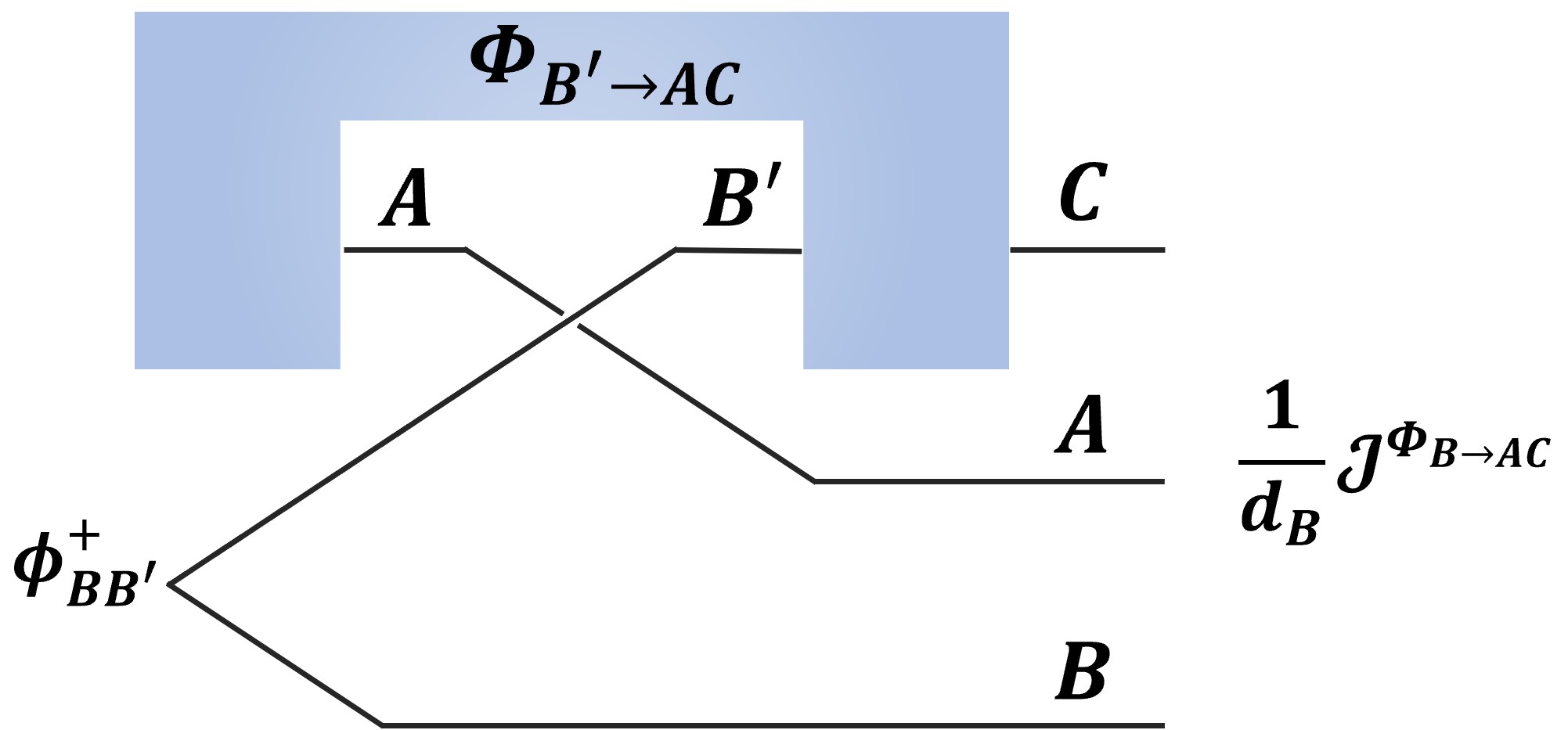}
\caption{(color online) Circuit realization of the CJ state $J^{\Phi_{B\to AC}}/d_{B}$, where $\phi^{+}_{BB^{'}}:= \ketbra{I}{I}_{BB^{'}}/d_{B}$ stands for the maximally entangled state on systems $BB^{'}$.
}
\label{fig:dynamics}
\end{figure}

From an experimental viewpoint, the normalized CJ operator (also called CJ state in this work) $\Phi_{B\to AC}/d_{B}$ can be obtained directly by feeding maximally entangled state to the dynamics, which is given by
\begin{align}
\id_{B}\otimes\Phi_{B^{'}\to AC}(\phi^{+}_{BB^{'}}), 
\end{align}
where $\phi^{+}_{BB^{'}}:= \ketbra{I}{I}_{BB^{'}}/d_{B}$ be the maximally entangled state with $B^{'}$ being a replica of system $B$. The circuit realization is demonstrated in Fig.~\ref{fig:dynamics}.

\begin{figure}[h]
\includegraphics[width=0.55\textwidth]{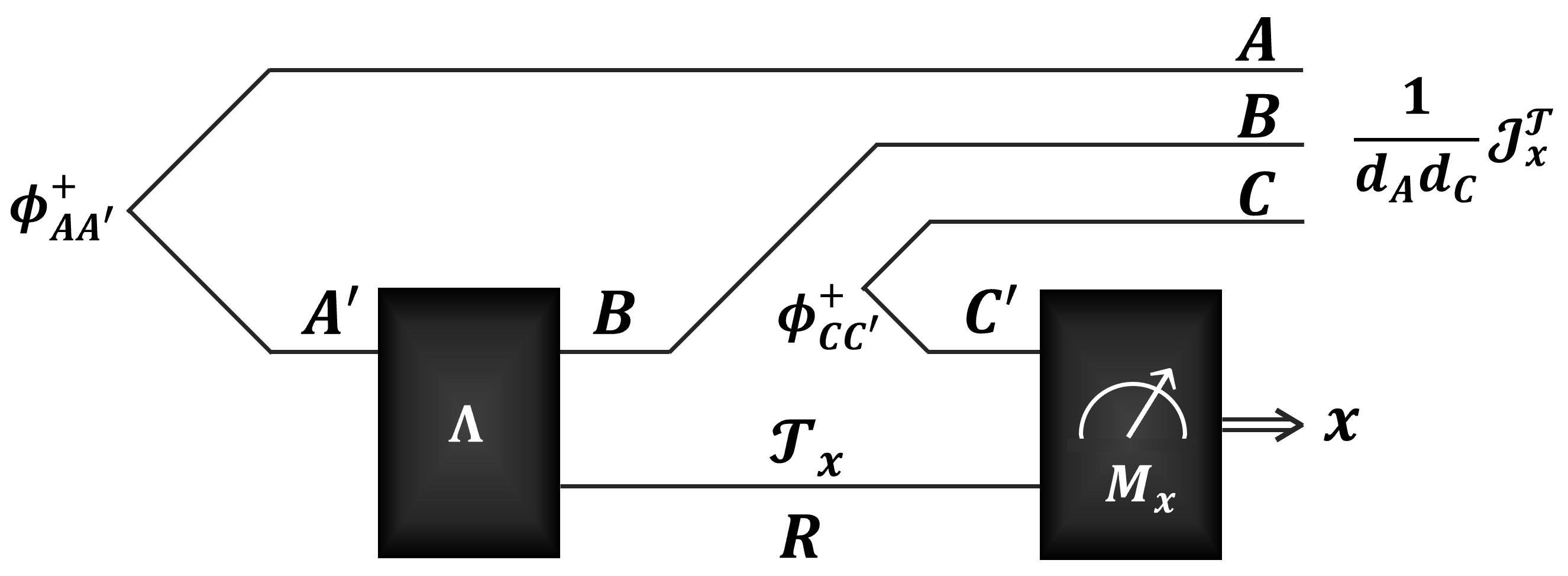}
\caption{(color online) Circuit realization of the CJ state $J_{x}/d_{A}d_{C}$, where $\phi^{+}_{AA^{'}}:= \ketbra{I}{I}_{AA^{'}}/d_{A}$ and $\phi^{+}_{CC^{'}}:= \ketbra{I}{I}_{CC^{'}}/d_{C}$ stand for the maximally entangled states on systems $AA^{'}$ and $CC^{'}$ respectively.}
\label{fig:ao}
\end{figure}

The interactive measurement is essential in inferring the causal structures associated with $\Phi_{B\to AC}$. Such a measuring process $\mathcal{T}$, which is an element of $\mathfrak{T}_2$ (see Def.~\ref{def:IM}), consists of a quantum channel $\Lambda$ from system $A$ to $BR$, a joint POVM $M:=\{M_{x}\}_{x}$ acting on systems $CR$, and an ancillary system $R$ connecting them together (see Fig.~1a of our main text), leading to a set of linear maps
\begin{align}\label{eq:causalmap-T}
\mathcal{T}_{AC\rightarrow B\mathds{C}}(\cdot):=\Big\{\Tr_{CR}[M_{x}\cdot\Lambda_{A\rightarrow BR}(\cdot)]\Big\}_{x},
\end{align}
where each $\mathcal{T}_{x}(\cdot):= \Tr_{CR}[M_{x}\cdot\Lambda_{A\rightarrow BR}(\cdot)]$ forms a CP and trace-non-increasing (TNI) map from $AC$ to $B\mathds{C}$. Since the CJ operator of the positive operator-valued measure (POVM) $M_{x}$ is given by $M_{x}^{\T_{CR}}$, the CJ operator of $\mathcal{T}_{x}$, denoted by $J_{x}$, can be obtained by using link product between $J^{\Lambda}$ and $M_{x}^{\T_{CR}}$. Writing everything out explicitly, we have
\begin{align}
J_{x}= M_{x}^{\T_{CR}}\star J^{\Lambda}.
\end{align}
To realize the CJ state $J_{x}/d_{A}d_{C}$ experimentally, we employ the quantum circuit shown in Fig.~\ref{fig:ao}. For any incompatible interactive measurements $\mathcal{T}_{1}= \{J^{\mathcal{T}_{1}}_x\}_{x=1}^{m}$ and $\mathcal{T}_{2}= \{J^{\mathcal{T}_{2}}_y\}_{y=1}^{n}$ with
\begin{align}
J^{\mathcal{T}_{1}}_x
&:= 
M_{x}^{\T_{CR}}\star J^{\Lambda_{A\rightarrow BR}},\label{eq:tx}\\
J^{\mathcal{T}_{2}}_y
&:= 
N_{y}^{\T_{CR}}\star J^{\Upsilon_{A\rightarrow BR}},\label{eq:ty}
\end{align}
the probability of obtain classical outcomes $x$ and $y$ from causal map $\Phi_{B\to AC}= \Psi^{\text{Post}}_{BE\to C}\circ \Psi^{\text{Pre}}_{\mathbb{C}\to AE}$ is given by
\begin{align}
p_{x}
&=
\Tr[M_{x}\cdot \Psi^{\text{Post}}_{BE\to C}\circ\Lambda_{A\rightarrow BR}(\Psi^{\text{Pre}}_{\mathbb{C}\to AE})],\label{eq:px}\\
q_{y}
&=
\Tr[N_{y}\cdot \Psi^{\text{Post}}_{BE\to C}\circ\Upsilon_{A\rightarrow BR}(\Psi^{\text{Pre}}_{\mathbb{C}\to AE})].\label{eq:qy}
\end{align}
Using the language of CJ operators and link product, Eqs.~\ref{eq:px} and \ref{eq:qy} can also be simplified as
\begin{align}
p_x
&=
J^{\Phi} \star J^{\mathcal{T}_{1}}_x,\label{eq:pxJ}\\
q_y
&=
J^{\Phi} \star J^{\mathcal{T}_{2}}_y,\label{eq:qyJ}
\end{align} 
where $J^{\Phi}$ stands for the CJ operator of causal map $\Phi_{B\to AC}$. Collect $p_x$ and $q_y$ into two probability vector $\p$ and $\q$, namely $\p:=\{p_{x}\}_{x=1}^{m}$ and $\q:=\{q_{y}\}_{y=1}^{n}$. In upcoming sections, we will show how to analyze the causal structure of $\Phi$ by using the uncertainty trade-off between $\p$ and $\q$, and exhibit the corresponding advantages. 

\section{\label{sec:UUR}Universal Uncertainty Relation for Measurements with Interventions}

In this section, starting from the algebraic structure of majorization lattice, we investigate the connections between universal uncertainty relation and quantum roulette, and provide the optimal bound of universal uncertainty relation for all quantum circuit fragments, leading to the entropic uncertainty relation for measurements with interventions (viz., Theorem~$1$ of the main text). Specifically, in Subsec.~\ref{subsec:maj}, we briefly review the concept of majorization lattice and discuss its completeness. The standard steps of finding least upper bound (LUB) for any set and a useful tool known as flatness process, which was originally introduced to investigate the supermodularity and subadditivity of entropies~\cite{992785}, are also supplied. In Subsec.~\ref{susec:intro}, we offer a short review for the historical developments of uncertainty principle, and introduce the requirement for uncertainty measures, which gives rise to the majorization uncertainty relation (known as universal uncertainty relation in literature). In Subsec.~\ref{subsec:QR}, we formulate an universal uncertainty relation for arbitrary quantum circuit fragments, showing its direct operational meaning in quantum roulette. With different uncertainty measures and different quantum dynamics, such a form leads to an infinite number of uncertainty relations. The generality and optimality of our results have also been analyzed. Lemma~$1$ and Theorem~$1$ of the main text are proved in the last subsection, i.e. Subsec.~\ref{subsec:thm1}. Additionally, we have also introduced the entropic uncertainty relation with multiple interactive measurements, extending the results presented in the main text of our work.


\subsection{\label{subsec:maj}Mathematical Toolkit: Majorization Lattice}
In this subsection, we turn our attention to the mathematical concept of \emph{lattice}, which is crucial in constructing the optimal bound of {\it universal uncertainty relation} (see Subsec.~\ref{susec:intro} and Refs.~\cite{PhysRevLett.111.230401,Pucha_a_2013,PhysRevA.89.052115} for more details). Let us start with the definition of Lattice: 

\begin{mydef}
{Lattice}{Lattice}
A quadruple $(S, \sqsubset, \wedge, \vee)$ is called lattice if $\ \sqsubset$ leads to a partial oder on the set $S$ such that for all $\p$, $\q \in S$ there exists a unique greatest lower bound (GLB) $\p \wedge \q$ and a unique least upper bound (LUB) $\p \vee \q$ satisfying 
\begin{align}
\x \sqsubset \p, \, \x \sqsubset \q 
&\Rightarrow 
\x \sqsubset \p \wedge \q, \notag\\
\p \sqsubset \y, \, \q \sqsubset \y
&\Rightarrow
\p \vee \q \sqsubset \y.
\end{align}
for any $\x$, $\y \in S$.
\end{mydef}
\noindent
A special class of lattices are those which have GLB and LUB for all their subsets, namely {\it Complete Lattice}:

\begin{mydef}
{Complete Lattice}{CLattice}
A lattice $(S, \sqsubset, \wedge, \vee)$ is called complete, if for any nonempty subset $R \subset S$, it has a LUB, denoted by $\vee R$ and a GLB, denoted by $\wedge R$. More precisely, if there exist $\x, \y \in S$ such that $\x \sqsubset R \sqsubset \y$, i.e. 
\begin{align}
\x \sqsubset \p \sqsubset \y, \quad \forall \,\p \in R, 
\end{align}
then we immediately have 
\begin{align}
\x \sqsubset \wedge R \quad\text{and}\quad \vee R \sqsubset \y.
\end{align} 
\end{mydef}

We now move on to introducing the concept of majorization~\cite{marshall2010inequalities}, which plays important roles in quantum resource theory and uncertainty relations. In particular, majorization was originally defined by Hardy, Littlewood, and P\'olya 

\begin{mydef}
{Majorization}{Majorization}
For vectors $\x = (x_{k})_{k}$, $\y = (y_{k})_{k} \in \mathds{R}^{d}$, we say $\x$ is majorized by $\y$, denoted as
\begin{align}
    \x\prec\y,
\end{align}
if the following inequality 
\begin{align}
    \sum_{k=1}^{i} x_{k}^{\downarrow} \leqslant \sum_{k=1}^{i} y_{k}^{\downarrow}
\end{align}
holds for all $1 \leqslant i \leqslant d-1$, and
\begin{align}
    \sum_{k=1}^{d} x_{k}^{\downarrow} = \sum_{k=1}^{d} y_{k}^{\downarrow}.
\end{align} 
Here the down-arrow notation $^{\downarrow}$ means that the components of corresponding vector are arranged in non-increasing order.
\end{mydef}

Let us consider a concrete example of majorization. For any $3$-dimensional probability vector $\p$, we see that
\begin{align}
    (1/3, 1/3, 1/3)\prec\p\prec(1, 0, 0).
\end{align}
To build a lattice structure for probability simplex, we hope that majorization can induce a partial order for all probability vectors. However, this is not the case. Majorization only forms a preorder. It is straightforward to check that majorization satisfies the properties of reflexivity ($\x\prec\x$ holds for any $\x$) and transitivity ($\x\prec\y$ and $\y\prec\z$ imply $\x\prec\z$). However, the property of antisymmetry is absent. Take vectors $(1, 0)$ and $(0, 1)$ for instance, it is clear that $(1, 0)\prec(0, 1)$ and $(0, 1)\prec(1, 0)$, but $(1, 0)\neq(0, 1)$. To remedy this problem, we should work on the ordered set of $\mathds{P}_{n}^{d, \, \downarrow}:= \{ \x \in \mathds{R}^{d} \, \| \, x_{k} \geqslant x_{k+1}\geqslant0, \, \forall 1 \leqslant k \leqslant d-1, \, \sum_{k} x_{k} = n \}$ instead of the probability simplex. Noted that when $n=1$, the set $\mathds{P}_{1}^{d, \, \downarrow}$ stands for the ordered probability simplex.

As proven in Ref.~\cite{PhysRevResearch.3.023077}, the quadruple $(\mathds{P}_{n}^{d, \, \downarrow}, \prec, \wedge, \vee)$ indeed forms a complete lattice under majorization, called majorization lattice. The properties of majorization lattice lead to a standard approach in finding the optimal bounds for any subset $S$ of $\mathds{P}_{n}^{d, \, \downarrow}$ \cite{PhysRevResearch.3.023077}. Formally, given a subset $S \subset \mathds{P}_{n}^{d, \, \downarrow}$, there are two steps in constructing its LUB $\vee S$. The first step is to find the quantities $\bb_{S, k}$, which is defined as
\begin{align}\label{eq:step-one}
\bb_{S, k} 
:= 
\left( \max_{\x \in S} \sum_{i=1}^{k} x_{i} \right) -
\sum_{i=1}^{k-1} \bb_{S, i},
\end{align}
for $1 \leqslant k \leqslant d$. Remark that the resultant vector $\bb_{S} := (\bb_{S, k})_{k}$ might not always belongs to the set $\mathds{P}_{n}^{d, \, \downarrow}$. To gain some intuitions, recall the example constructed in Ref.~\cite{992785}. Take $S = \{ \x, \y \}$ with
\begin{align}
\x &= (0.6, 0.15, 0.15, 0.1),\label{eq:sx}\\
\y &= (0.5, 0.25, 0.2, 0.05).\label{eq:sy}
\end{align}
In this case, the vector
\begin{align}
\bb_{S} = (0.6, 0.15, 0.2, 0.05),
\end{align} 
obtained from (\ref{eq:step-one}) does not belong to the set $\mathds{P}_{1}^{d, \, \downarrow}$, since 
\begin{align}
    \bb_{S, 2} = 0.15 < \bb_{S, 3} = 0.2.
\end{align} 
Actually, even if we rearrange the vector $\bb_{S}$ into non-increasing order 
\begin{align}
\bb_{S}^{\downarrow} = (0.6, 0.2, 0.15, 0.05),
\end{align}
the re-ordered vector $\bb_{S}^{\downarrow}$ is not the optimal upper bound, i.e. $\bb_{S}^{\downarrow} \neq \vee S$. 

In order to achieve the optimal bound $\vee S$ of the subset $S$, an additional process $\mathcal{F}$ on $\bb_{S}$ (not $\bb_{S}^{\downarrow}$) is needed, named {\it flatness process}~\cite{992785}:

\begin{mydef}
{Flatness Process}{FProcess}
Let $\x \in \mathds{R}^{d}_{+}$ be a non-negative $d$-dimensional vector, and $j$ be the smallest integer in $\left\{2, \ldots, d\right\}$ such that $x_{j}>x_{j-1}$, and $i$ be the greatest integer in $\left\{1, \ldots, j-1\right\}$ such that $x_{i-1} \geqslant (\sum_{k=i}^{j} x_{k})/(j-i+1):=a$. Define
%
\begin{align}\label{eq; mj bound t}
\mathcal{T} (\x) := \left(x_{1}^{\prime}, \ldots, x_{n}^{\prime}\right) \, \quad\text{with}\quad \, 
x_{k}^{\prime} = 
     \begin{cases}
       a & \text{for}\quad k = i, \ldots, j \\
       x_{k} & \text{otherwise.} \\ 
     \end{cases}
\end{align}
and $\mathcal{F} (\x) := \mathcal{T}^{d-1} (\x) = \mathcal{T} ( \mathcal{T}^{d-2} (\x) )$, i.e. applying $\mathcal{T}$ on the vector $\x$ successively $d-1$ times. We call $\mathcal{F}$ the flatness process of vector $\x$. If the vector $\x$ is already arranged in non-increasing order, or equivalently there does not exist an integer $j$ in $\left\{2, \ldots, d\right\}$ such that $x_{j}>x_{j-1}$, then we simply have $\mathcal{T} (\x) = \x$. 
\end{mydef}
Based on the definition of flatness process, we will show how to obtain the LUB for $\x$ and $\y$ considered in Eqs.~\ref{eq:sx} and \ref{eq:sy}. In this case, $\bb_{S}= (0.6, 0.15, 0.2, 0.05)$ is not arranged in non-increasing order. Let us define $x_{k}$ ($k= 1, 2, 3, 4$) as
\begin{align}
    x_1&= 0.6,\\
    x_2&= 0.15,\\
    x_3&= 0.2,\\
    x_4&= 0.05.
\end{align}
Then the smallest integer in $\left\{2, 3, 4\right\}$ such that $x_{j}>x_{j-1}$ is $3$ as $x_3= 0.2> x_2= 0.15$. Note that now the quantity $(\sum_{k=2}^{3} x_{k})/2$ is given by
\begin{align}
    \frac{x_2+ x_3}{2}= 0.175
    \leqslant x_1= 0.6.
\end{align}
Thus, we can write $\mathcal{T} (\bb_{S})= (0.6, 0.175, 0.175, 0.05)$. As $\mathcal{T} (\bb_{S})$ is already in non-increasing order, we have $\mathcal{F} (\bb_{S})= \mathcal{T} (\bb_{S})$, which is exactly the optimal bound for $\x = (0.6, 0.15, 0.15, 0.1)$ and $\y = (0.5, 0.25, 0.2, 0.05)$, namely
\begin{align}
\vee S = \x \vee \y = \mathcal{F} (\bb_{S}) = (0.6, 0.175, 0.175, 0.05),
\end{align} 
with
\begin{align}
(0.6, 0.15, 0.15, 0.1) 
&\prec (0.6, 0.175, 0.175, 0.05) \prec (0.6, 0.15, 0.2, 0.05),\\
(0.5, 0.25, 0.2, 0.05) 
&\prec (0.6, 0.175, 0.175, 0.05) \prec (0.6, 0.15, 0.2, 0.05),
\end{align}
and for any probability vector $\z$ satisfying $\x\prec\z$ and $\y\prec\z$, it follows that 
\begin{align}
(0.6, 0.175, 0.175, 0.05)\prec\z.
\end{align}
It is worth mentioning that the flatness process $\mathcal{F}$ introduced in Ref.~\cite{992785} is exactly the second step of formulating the optimal bound for $S$. More precisely, $\vee S = \mathcal{F} (\bb_{S})$ holds in general \cite{PhysRevResearch.3.023077}. 

To summarize, the standard approach in finding the optimal bounds for a subset $S$ of $\mathds{P}_{n}^{d, \, \downarrow}$ contains two steps:

\begin{enumerate}
\item[{\bf Step $1$.}] Formulating $\bb_{S}$ defined in Eq.~(\ref{eq:step-one}),
\item[{\bf Step $2$.}] Applying the flatness process defined in Eq.~(\ref{eq; mj bound t}) to obtain $\mathcal{F}(\bb_{S})$,
\end{enumerate}
leading to the following lemma.

\begin{mylem}
{Least Upper Bound (LUB)}{LUB}
For any subset $S \subset \mathds{P}_{n}^{d, \, \downarrow}$, its least upper bound (LUB) $\vee S$ is given by
\begin{align}
    \vee S = \mathcal{F} (\bb_{S}),
\end{align}
where the vector elements of $\bb_{S}$ are defined in Eq.~\ref{eq:step-one}, and $\mF$ represents the flatness process introduced in Def.~\ref{def:FProcess}.
\end{mylem}

Finally, we note that majorization is naturally connected with entropies. More precisely, all \emph{Schur-concave} functions, including R\'enyi entropies, are order-reversing functions for majorization. The relations between majorization, doubly stochastic matrix (square matrix of non-negative real numbers, whose rows and columns sums to $1$), and Schur-concave functions are detailed in the following lemma.

\begin{mylem}
{Majorization Equivalents~\cite{marshall2010inequalities}}{ME}
For two vectors $\x$ and $\y$, the following conditions are equivalent:
\begin{enumerate}[(i)]
\item $\x$ is majorized by $\y$, i.e. $\x\prec\y$;
\item $x= y\cdot D$ for some doubly stochastic matrix $D$;
\item $f(\x)\geqslant f(\y)$ holds for any Schur-concave function $f$.
\end{enumerate}
\end{mylem}

Let us check some simple applications of Lem.~\ref{lem:ME}. First, Shannon entropy is invariant under permutations of its components since $\p\prec\p\cdot D$ and $\p\cdot D\prec\p$ hold for any probability vector $\p$ and permutation matrix $D$. Second, for any $d$-dimensional probability vector $\p$, we always have $\log d\geqslant H(\p)$ as $(1/d, \ldots, 1/d)\prec\p$, where $H$ stands for the Shannon entropy.


\subsection{\label{susec:intro}Brief Introduction to Uncertainty Relations}

If an observer is well-equipped, then estimating the position and momentum of a moving football simultaneously is not a difficult task. However, if the item of observation has been replaced by a particle, then it is impossible to predict the outcomes for both position and momentum with arbitrary precision, no matter how well the observer is equipped. Quantum mechanics constrains what we can learn about the observables of particle. Such a restriction is known as uncertainty principle, which was first introduced by Heisenberg in Ref.~\cite{Heisenberg1927}. Quantitatively, the fundamental trade-off between position and momentum is characterized by the following inequality~\cite{Kennard1927,weyl1928gruppentheorie}
\begin{align}\label{eq:x-p}
    \Delta\x\cdot\Delta\p\geqslant\frac{\hbar}{2},
\end{align}
where $\x$ and $\p$ stand for the position and momentum operators, and $\Delta$ is the standard deviation. For bounded operators $M$ and $N$, Robertson gave a more general form in terms of the commutator $[M, N]$~\cite{PhysRev.34.163}
\begin{align}
    \Delta M\cdot\Delta N\geqslant\frac{1}{2}|\bra{\psi}[M, N]\ket{\psi}|.
\end{align}

From an information-theoretic perspective, it is more natural to quantify the uncertainty associated with quantum measurements in terms of entropy rather than the statistical tools, such as the standard deviation. Using this insight, the entropic version of position-momentum uncertainty relation can be formulated as
\begin{align}\label{eq:diff-xp}
    h(\x)+ h(\p)\geqslant\log(\pi e\hbar)
\end{align}
Remark that the entropic uncertainty relation of Eq.~\ref{eq:diff-xp} is obtained by Bia\l ynicki-Birula and Mycielski in Ref.~\cite{Bialynicki1975}. Here $e$ is the Euler's number, and $h$ stands for the differential entropy: consider a random variable $X$ with density $f(x)$, then its differential entropy is defined as
\begin{align}
    h(X):= -\int_{\infty}^{\infty} f(x)\log f(x)\,d\,f(x).
\end{align}
The first entropic uncertainty relation for general bounded operators was introduced by Deutsch in Ref.~\cite{PhysRevLett.50.631}, which had later been improved by Maassen and Uffink. In particular, given projective measurements $M:=\{\ket{u_i}\}$ and $N:= \{\ket{v_j}\}$, the Maassen-Uffink entropic uncertainty relation reads~\cite{PhysRevLett.60.1103}
\begin{align}\label{eq:MU}
    H(M)+ H(N)\geqslant-\log c,
\end{align}
where $H(M)$ represents the Shannon entropy of probability distribution obtained by implementing measurement $M$, and $c$ is the maximal overlap between measurements $M$ and $N$. More precisely, here the quantity $c$ is defined by
\begin{align}
    c:= \max_{i, j}|\braket{u_i}{v_j}|^2,
\end{align}
which depends only on the measurements, or equivalently state-independent, and hence captures their inherent incompatibility. Technically, the quantity $c$ is derived from \emph{Riesz theorem} in functional analysis.

In the ensuing decades, quantum uncertainty associated with measurements has been quantified by a variety of entropies. Various forms of entropic uncertainty relations have also been proposed~\cite{RevModPhys.89.015002}. However, none of these forms answer the fundamental question -- which uncertainty measure is the most adequate to use? Shannon entropy, Collision entropy, Min-entropy, or even Tsallis entropy? To fully understand the uncertainty induced by measurements and address this question, the concept of `reasonable measure' for uncertainty has been introduced in Ref.~\cite{PhysRevLett.111.230401}: any reasonable measure $f$ of the quantum uncertainty should be a function only of the probability vector associated with measurement and satisfies monotonicity under random relabeling $\mathfrak{R}$. For example, given a probability vector $\p$, function $f$ is an uncertainty measure if the following condition is satisfied
\begin{align}
    f(\mathfrak{R}(\p))\geqslant f(\p).
\end{align}
In other words, the procedure of random relabeling -- known as forgetting outcome labels in classical world -- turns the results to be more uncertain. Technically speaking, the process of random relabeling $\mathfrak{R}$ is characterized by the convex hull of permutations~\cite{marshall2010inequalities}, which implies that
\begin{align}
    \mathfrak{R}(\p)= \p\cdot D_{\mathfrak{R}},
\end{align}
holds for some doubly stochastic matrix $D_{\mathfrak{R}}$. As a direct consequence of Lem.~\ref{lem:ME}, we know that for two probability vectors $\p$ and $\q$, $\p$ is more uncertain than $\q$ if and only if $\p= \q\cdot D$ for some doubly stochastic matrix $D$, or equivalently $\p\prec\q$. Therefore, all Schur-concave functions (including R\'enyi entropies) are reasonable uncertainty measures. Historically, majorization $\prec$ is initially introduced as a tool to extend known inequalities and unify all inequalities based on convex functions. Here, in the context of uncertainty relation, majorization will generate an infinite family of uncertainty relations. Thus, the uncertainty relation in the forms of majorization is also known as \emph{universal uncertainty relation}. A typical example is the direct-sum majorization uncertainty relation~\cite{PhysRevA.89.052115}. Consider the probability vectors $\p:= (p_{i})_{i=1}^{m}$ and $\q:= (q_{j})_{j=1}^{n}$ obtained by measuring a quantum state $\rho$ with respect to a pair of incompatible measurements $M$ and $N$, their direct-sum $\oplus$ is defined as the union of vectors. Formally, 
\begin{align}
    \p\oplus\q:= (p_1, \ldots, p_m, q_1, \ldots, q_n).
\end{align}
For example, given probability vectors $\p= (1, 0)$ and $\q= (1/2, 1/2)$, their direct-sum $\p\oplus\q$ is simply $(1, 0, 1/2, 1/2)$. The goal is find the optimal upper bound $\mathbf{w}_{M, N}$ such that
\begin{align}\label{eq:ds-state}
    \p\oplus\q\prec\mathbf{w}_{M, N},
\end{align}
holds for all quantum state $\rho$, namely $\mathbf{w}_{M, N}$ is a vector independent of the state $\rho$. For any Schur-concave function $f$, uncertainty relation~\ref{eq:ds-state} leads to
\begin{align}
    f(\p\oplus\q)\geqslant f(\mathbf{w}_{M, N}),
\end{align}
which includes all entropic uncertainty relations as special cases. Take Shannon entropy for instance, we immediately obtain the entropic uncertainty relation in the form of $H(\p\oplus\q)\geqslant H(\mathbf{w}_{M, N})$. Before the end of this subsection, we make two remarks. First, direct-sum majorization uncertainty relation is not the only form of universal uncertainty relations. There also exists a direct-product majorization uncertainty relation, which was formulated by Friedland, Gheorghiu, Gour in Ref.~\cite{PhysRevLett.111.230401} and Pucha\l a, Rudnicki, \.{Z}yczkowski in Ref.~\cite{Pucha_a_2013}. Specifically, they are trying to bound the uncertainty associated with $\p\otimes\q:= (p_i\cdot q_j)_{i, j}$. In this work, we mainly focus on the direct-sum form since it shows a provable advantage over the direct-product form (see Ref.~\cite{PhysRevA.89.052115} for more details). Second, all the uncertainty relations considered in this work belong to the category of preparational uncertainty relation. The noise-disturbance uncertainty relation of interactive measurements is beyond the scope of this work, and merits future investigation.


\subsection{\label{subsec:QR}Operational Interpretation of Universal Uncertainty Relation: Quantum Roulette}

Universal uncertainty relations capture the essential trade-off between incompatible measurements in terms of their probability vectors, leading to a family of uncertainty relations. Although it can outperform previous well-known results, such as the Maassen-Uffink entropic uncertainty relation (see Eq.~\ref{eq:MU}) to some extent, its operational interpretation remains unknown. In this subsection, we will connect the direct-sum majorization uncertainty relation with a guessing game which we refer to as the quantum roulette.

In particular, quantum roulette is a game played between two parties -- Alice and Bob. In this game, Alice is an agent that is capable of probing the quantum circuit fragment $\Phi$ (see Def.~\ref{def:QCF}) supplied by Bob with two possible interactive measurements, $\mathcal{T}_{1}:= \{\mT_{1, x_1}\}_{x_1=1}^{m_1}$ and $\mathcal{T}_{2}:= \{\mT_{2, x_2}\}_{x_2=2}^{m_2}$ (see Def.~\ref{def:IM}). Generally, these interactive measurements do not need to have the same number of outcomes, i.e. $m_1\neq m_2$. In each round of the game, Alice and Bob begin with a `roulette table', whose layout consists of all tuples $(b, x_b)$, where $b \in \{1,2\}$ denotes Alice's choice of interactive measurements, and $x_b \in \{ 1, \ldots, m_{b}\}$ represents the corresponding measurement outcome by implementing $\mT_b$. Bob starts the game with $k$ chips, which he can use to place bets on $k$ of the possible tuples and supplies Alice with any $\Phi\in\mathfrak{F}_{a}$ of his choosing. Alice will then select some $b$ at random and probe $\Phi$ with interactive measurement $\mT_b\in\mathfrak{T}_{a}$. She finally announces both $b$ and the resulting measurement outcome $x_b$. Bob
wins if one of his chips is on $(b, x_b)$. We denote Bob's maximum winning probability as $p_{\text{win}, k}$, and our goal is to characterize this maximum winning probability for any quantum circuit fragment $\mathfrak{F}_{a}$.

To gain some intuition, let us first consider the simplest case in which Bob can only provide quantum causal map $\Phi_{B\to AC}\in\mathfrak{F}_{2}$ (see Subsec.~\ref{subsec:CMaps} for more details). Meanwhile, assume the interactive measurements chosen by Alice are $\mT_1$ of Eq.~\ref{eq:tx} and $\mT_2$ of Eq.~\ref{eq:ty}, where $x_1=x \in\{1, \ldots, m\}$ and $x_2=y\in\{1, \ldots, n\}$. In this case, the outcomes $x$ of $\mT_1$ and $y$ of $\mT_2$ happen with the following probabilities (see Eqs.~\ref{eq:pxJ} and \ref{eq:qyJ})
\begin{align}
    \frac{1}{2}p_x
    &=
    \frac{1}{2}J^{\Phi}\star J^{\mT_1}_x,\\
    \frac{1}{2}q_y
    &=
    \frac{1}{2}J^{\Phi}\star J^{\mT_2}_y.\\
\end{align}
Here the coefficient $1/2$ comes from the fact that interactive measurements $\mT_1$ and $\mT_2$ have the same probability of occurring. Now Bob's guessing strategy can be described by two sets $\mS_1\subset\{1,\ldots, m\}$ and $\mS_2\subset\{1,\ldots, n\}$ satisfying
\begin{align}
    |\mathcal{S}_{1}|+ |\mathcal{S}_{2}|=k,
\end{align}
where the set $\mS_b$ ($b\in\{1, 2\}$) demonstrates the collection of his guesses in the form of $(b, \cdot)$. For example, when Bob has $2$ chips, $\mS_1= \{2, 3\}$ means that he will put his chips on $(1, 2)$ and $(1, 3)$. When Bob has $3$ chips, with $\mS_1$ and $\mS_2$ are given by the sets $\{2, 3\}$ and $\{2\}$ respectively, then he will place the chips on $(1, 2)$, $(1, 3)$ and $(2, 2)$. Given $k$ chips and a quantum causal map $\Phi$, Bob’s maximum winning probability is characterized by 
\begin{align}
    p_{\text{win}, k}(\Phi)
    &=
    \max_{|\mathcal{S}_{1}|+ |\mathcal{S}_{2}|=k}
    \sum_{\substack{x\in\mathcal{S}_{1}\subset\{1,\ldots, m\}\\ y\in\mathcal{S}_{2}\subset\{1,\ldots, n\}}}
    \left(
    \frac{1}{2}p_x+\frac{1}{2}q_y
    \right)\label{eq:causal-win-1}\\
    &=
    \max_{|\mathcal{S}_{1}|+ |\mathcal{S}_{2}|=k}
    \sum_{\substack{x\in\mathcal{S}_{1}\subset\{1,\ldots, m\}\\ y\in\mathcal{S}_{2}\subset\{1,\ldots, n\}}}
    \frac{1}{2}
    \left(
    J^{\Phi}\star 
    (J^{\mT_1}_x+ J^{\mT_2}_y)
    \right)\label{eq:causal-win-2}.
\end{align}
The second equation comes from Eqs.~\ref{eq:pxJ} and \ref{eq:qyJ}. Note that the largest probability of $k$  possible tuples is given by
\begin{align}
    \sum_{i=1}^{k}(\frac{1}{2}\p\oplus\frac{1}{2}\q)_{i}^{\downarrow},
\end{align}
where the superscript $\downarrow$ indicates that the corresponding vector is arranged in non-increasing order, and the subscript $i$ means it is the $i$-th element of the vector. Take $(0.1, 0.1, 0.6, 0.2)$ for instance, we have
\begin{align}
    (0.1, 0.1, 0.6, 0.2)_{1}^{\downarrow}
    &=
    0.6,\\
    (0.1, 0.1, 0.6, 0.2)_{2}^{\downarrow}
    &=
    0.2,\\
    (0.1, 0.1, 0.6, 0.2)_{3}^{\downarrow}
    &=
    0.1,\\
    (0.1, 0.1, 0.6, 0.2)_{4}^{\downarrow}
    &=
    0.1.
\end{align}
It now follows immediately that
\begin{align}
    \sum_{i=1}^{k}(\frac{1}{2}\p\oplus\frac{1}{2}\q)_{i}^{\downarrow}
    =
    p_{\text{win}, k}(\Phi)
    =
    \max_{|\mathcal{S}_{1}|+ |\mathcal{S}_{2}|=k}
    \sum_{\substack{x\in\mathcal{S}_{1}\subset\{1,\ldots, m\}\\ y\in\mathcal{S}_{2}\subset\{1,\ldots, n\}}}
    \frac{1}{2}
    \left(
    J^{\Phi}\star 
    (J^{\mT_1}_x+ J^{\mT_2}_y)
    \right).
\end{align}
There are two trivial case of quantum roulette: (\romannumeral1) $k=0$, which means Bob does not have any chips on hand. Thus, his winning probability is simply zero. In this case, we can write $p_{\text{win}, 0}(\Phi)= 0$; (\romannumeral1) $k= m+ n$. In this case, Bob has enough chips to select all tuples, namely $p_{\text{win}, m+ n}(\Phi)= 1$. Our goal here is to find the fundamental limitations of winning the game for all causal maps, which is defined as
\begin{align}
    p_{\text{win}, k}
    &:=
    \max_{\Phi\in\mathfrak{F}_{2}}p_{\text{win}, k}(\Phi)\label{eq:pwin1}\\
    &=
    \max_{|\mathcal{S}_{1}|+ |\mathcal{S}_{2}|=k}
    \sum_{\substack{x\in\mathcal{S}_{1}\subset\{1,\ldots, m\}\\ y\in\mathcal{S}_{2}\subset\{1,\ldots, n\}}}
    \frac{1}{2}
    \left(
    \max_{\Phi\in\mathfrak{F}_{2}}
    J^{\Phi}\star 
    (J^{\mT_1}_x+ J^{\mT_2}_y)
    \right).\label{eq:pwin2}
\end{align}
Thus, for any k ($1\leqslant k\leqslant m+n$) and quantum causal map supplied by Bob, the probability vectors $\p$ and $\q$ obtained from interactive measurements $\mT_1$ and $\mT_2$ satisfy 
\begin{align}
    \sum_{i=1}^{k}(\frac{1}{2}\p\oplus\frac{1}{2}\q)_{i}^{\downarrow}
    \leqslant
    p_{\text{win}, k},
    \quad
    1\leqslant k\leqslant m+n,
\end{align}
leading to
\begin{align}
  \frac{1}{2}\p\oplus\frac{1}{2}\q
  \prec
  \left(p_{\text{win}, 1}, (p_{\text{win}, 2}- p_{\text{win}, 1}), \ldots, (p_{\text{win}, m+n}- p_{\text{win}, m+n-1})
  \right).
\end{align}
Above majorization inequality connects the direct-sum majorization uncertainty relation with Bob's maximum winning
probability in quantum roulette, showing that the largest summation of $k$ elements of $\frac{1}{2}\p\oplus\frac{1}{2}\q$ is completely determined by Bob's maximum winning probability with $k$ chips in quantum roulette. Denote the increment of Bob's maximum winning probability with $k$ rather than $k-1$ chips as $w_{k}$, i.e.
\begin{align}
    w_{k}:= 
    p_{\text{win}, k}
    - 
    p_{\text{win}, k-1},
    \quad
    1\leqslant k\leqslant m+n,
\end{align}
with $p_{\text{win}, 0}:= 0$, and set $\ww:=(w_{k})_{k=1}^{m+n}$, we see that
\begin{align}
    \frac{1}{2}\p\oplus\frac{1}{2}\q
    \prec
    \ww.
\end{align}
To further simplify the representation, we take the union of CJ operators $J^{\mathcal{T}_{1}}_x$ and $J^{\mathcal{T}_{2}}_y$, and denote it by $\{J_z\}_z$, where
\begin{equation}
    J_{z}=\left\{
    \begin{array}{rcl}
        J^{\mathcal{T}_{1}}_{z} 
        && 
        1 \leqslant z \leqslant m, \\
        J^{\mathcal{T}_{2}}_{z-m} 
        &&
        m+1 \leqslant z \leqslant m+n.
    \end{array}
    \right.
\end{equation}
We adopt the convention $J_{\mathcal{S}}$ for 
\begin{align}
    J_{\mS}:=
    \sum_{z\in\mS} J_{z},
\end{align}
where $\mS\subset\{1, \ldots, m+n\}$, e.g. for $\mS= \{2, 5\}$, we immediately have $J_{\{2, 5\}}= J_2+ J_5$. To more precise, if $m= n= 3$, then we then see that
\begin{align}
    J_{\{2, 5\}}= J_2+ J_5 = J^{\mT_1}_2+ J^{\mT_2}_2.
\end{align}
Based on CJ operators, now the universal uncertainty relation in the form of direct-sum can be formulated as
\begin{mythm}
{Universal Uncertainty Relation for $\mathfrak{F}_{2}$}{UUR2}
For any quantum causal map $\Phi\in\mathfrak{F}_{2}$, the probability vectors $\p$ and $\q$ obtained by measuring $\Phi$ with respect to interactive measurements $\mT_{1}$ and  $\mT_{2}\in\mathfrak{T}_{2}$ satisfy the following trade-off
\begin{align}\label{eq:pwink}
    \frac{1}{2}\p\oplus\frac{1}{2}\q
    \prec
    \ww
    =
    (p_{\text{win}, k}
    - 
    p_{\text{win}, k-1})
    _{k=1}^{m+n},
\end{align}
where $p_{\text{win}, 0}:= 0$, and for $1\leqslant k\leqslant m+n$, the quantity $p_{\text{win}, k}$ is characterized by the following optimization problem.
\begin{align}
p_{\text{win}, k}
=
\frac{1}{2}
\max_{|\mathcal{S}|=k}\max\quad &\Tr [J_{\mathcal{S}} \cdot J ]\notag\\
{\rm s.t.}  \quad
&\, J\geqslant0,\quad
\Tr [\rho_{A}] = 1
,\quad
\Tr_{C} [J] = \rho_{A} \otimes \mathds{1}_{B}
.\label{eq:pwin5}
\end{align}
In particular, here the optimal solution of $p_{\text{win}, k}$ forms a semidefinite program (SDP), and hence
can be computed efficiently in polynomial time (up to desired accuracy) by using the ellipsoid method or an interior point method~\cite{doi:10.1137/1038003,boyd_vandenberghe_2004}.
\end{mythm}

\begin{proof}
From Eq.~\ref{eq:pwin2}, we can write
\begin{align}
    p_{\text{win}, k}
    &=
    \frac{1}{2}\max_{|\mS|=k}
    \max_{\Phi\in\mathfrak{F}_{2}}
    J^{\Phi}\star 
    J_{\mS}
    \label{eq:pwin3}\\
    &=
    \frac{1}{2}\max_{|\mS|=k}
    \max_{\Phi\in\mathfrak{F}_{2}}
    \Tr
    [
    J^{\Phi}\cdot
    J_{\mS}
    ]
    \label{eq:pwin4},
\end{align}
where in the second equation, namely Eq.~\ref{eq:pwin4}, we have used the fact that the set of all CJ operators associated with causal map is algebraically closed under the transpose over all of its systems. Remark that, given a CJ operator $J^{\Phi}_{ABC}$ for some causal map $\Phi_{B\to AC}$, the operator $(J^{\Phi}_{ABC})^{\T}$ is still a CJ operator for some causal map. However, this may not be the case for partial transpose, such as $(J^{\Phi}_{ABC})^{\T_{B}}$. As discussed in Subsec.~\ref{subsec:CMaps}, a quantum dynamics $\Phi_{B\to AC}$ is a causal map if and only if its CJ operator $J^{\Phi}$ satisfies the requirements of (\romannumeral1) CP: $J^{\Phi}\geqslant0$, (\romannumeral2) TP: $\Tr_{AC} [J^{\Phi}] = \mathds{1}_{B}$ and (\romannumeral3) NS from $B\to C$ to $A$: $\Tr_{C} [J^{\Phi}] = \Tr_{BC} [J^{\Phi}] \otimes \mathds{1}_{B}/d_{B}$. It hence follows that
\begin{align}
    p_{\text{win}, k}
    =
    \frac{1}{2}
    \max_{|\mathcal{S}|=k}\max\quad &\Tr [J_{\mathcal{S}} \cdot J ]\notag\\ {\rm s.t.}  \quad &\, J\geqslant0,\quad \Tr_{AC} [J] = \mathds{1}_{B},\notag\\
    &
    \Tr_{C} [J] = \Tr_{BC} [J] \otimes \mathds{1}_{B}/d_{B},
\end{align}
which is equivalent to Eq.~\ref{eq:pwin5}, as required.
\end{proof}

In general, for different quantum causal map $\Phi$, we will have different probability vectors $\p$ and $\q$ with respect to interactive measurements $\mT_1$ and $\mT_2$. To indicate their dependencies, we re-express them as $\p(\Phi)$ and $\q(\Phi)$, then the set 
$
   \{\frac{1}{2}\p(\Phi)\oplus\frac{1}{2}\q(\Phi)\}_{\Phi} 
$
forms a subset of the probability simplex. Constructing the probability vector $\ww$ is exactly the first step of finding the optimal bounds for $\{\frac{1}{2}\p(\Phi)\oplus\frac{1}{2}\q(\Phi)\}_{\Phi}$ (see Eq.~\ref{eq:step-one} of Subsec.~\ref{subsec:maj}), and the next step is flatness process $\mF$ (see Def.~\ref{def:FProcess}), which implies that
\begin{align}
    \vee
    \left\{
    \frac{1}{2}\p(\Phi)\oplus\frac{1}{2}\q(\Phi)
    \right\}_{\Phi}^{\downarrow}
    =
    \mF(\ww).
\end{align}
Writing above result formally, we have
\begin{mythm}
{Optimal Universal Uncertainty Relation for $\mathfrak{F}_{2}$}{OUUR2}
For any quantum causal map $\Phi\in\mathfrak{F}_{2}$, the probability vectors $\p$ and $\q$ obtained by measuring $\Phi$ with respect to interactive measurements $\mT_{1}$ and  $\mT_{2}\in\mathfrak{T}_{2}$ satisfy the following trade-off
\begin{align}
    \frac{1}{2}\p\oplus\frac{1}{2}\q
    \prec
    \mF(\ww),
\end{align}
where $\mF$ is the flatness process defined in Def.~\ref{def:FProcess}. The vector $\ww$ is defined as
\begin{align}
    \ww
    :=
    (p_{\text{win}, k}- p_{\text{win}, k-1})_{k=1}^{m+n},
\end{align}
with $p_{\text{win}, 0}:=0$ and $p_{\text{win}, k}$ is solved by SDPs in the form of Eq.~\ref{eq:pwink} for $1\leqslant k\leqslant m+n$. Here the bound $\mF(\ww)$ is optimal, due to the structure of majorization lattice, namely for any probability vector $\x$ satisfying 
\begin{align}
    \left\{
    \frac{1}{2}\p\oplus\frac{1}{2}\q
    \right\}_{\Phi}
    \prec
    \x,
\end{align}
then we also have
\begin{align}
    \mF(\ww)\prec\x.
\end{align}
Physically, the causal map independence of $\x$ leads to an universal uncertainty relation $\frac{1}{2}\p\oplus\frac{1}{2}\q\prec\x$, and $\mF(\ww)$ constructed here is the optimal one. Mathematically, $\mF(\ww)$ is the GLB (see Def.~\ref{def:Lattice}) for all such $\x$. In particular, taken any Schur-concave function $f$, we have
\begin{align}
    f(\frac{1}{2}\p\oplus\frac{1}{2}\q)\geqslant f(\mF(\ww))\geqslant f(\x).
\end{align}
\end{mythm}

Here we would like to remark that permutations will induce an equivalence relation, denoted as $\sim$, on the probability simplex. Given two vectors $\x$ and $\y$, we say that they are equivalent, written as 
\begin{align}
    \x\sim\y,
\end{align} 
if there exists a permutation matrix $D$ such that
\begin{align}\label{eq:equ}
    \x= \y\cdot D.
\end{align}
To illustrate $\sim$, let us consider $(0.2, 0.1, 0.7)$ and $(0.7, 0.2, 0.1)$. Clearly, there exists a permutation matrix satisfying Eq.~\ref{eq:equ}, then we have 
\begin{align}
(0.2, 0.1, 0.7)\sim(0.7, 0.2, 0.1).
\end{align}
The equivalence class of $\x$ under $\sim$, denoted $[\x^{\downarrow}]$, is defined as
\begin{align}
    [\x^{\downarrow}]
    :=
    \{\y\,|\,\y\sim\x\}.
\end{align}
Thus, as a corollary of Thm.~\ref{thm:OUUR2}, we obtain the following result about universal uncertainty relation

\begin{mycor}
{Optimal Universal Uncertainty Relation for $\mathfrak{F}_{2}$ under Permutations}{OUUR2P}
Given a quantum causal map $\Phi\in\mathfrak{F}_{2}$, let us denote the probability vectors obtained by measuring $\Phi$ with respect to interactive measurements $\mT_{1}$ and  $\mT_{2}\in\mathfrak{T}_{2}$ as $\p$ and $\q$. Define the vector $\ww$ as $(p_{\text{win}, k}- p_{\text{win}, k-1})_{k=1}^{m+n}$ with $p_{\text{win}, 0}:=0$ and $p_{\text{win}, k}$ in the form of Eq.~\ref{eq:pwink} for $1\leqslant k\leqslant m+n$. Then for any $\p^{'}\in[\p^{\downarrow}]$, $\q^{'}\in[\q^{\downarrow}]$, and $\bf{w}^{'}\in[\mF(\ww)]$, we have the following universal uncertainty relation
\begin{align}
    \frac{1}{2}\p^{'}\oplus\frac{1}{2}\q^{'}
    \prec
    {\bf w}^{'}.
\end{align}
In particular, given any permutation matrices $D_1$, $D_2$, and $D_3$, we also have
\begin{align}
    \frac{1}{2}(\p^{'}\cdot D_1)
    \oplus
    \frac{1}{2}(\q^{'}\cdot D_2)
    \prec
    {\bf w}^{'}\cdot D_3.
\end{align}
Moreover, if there exists a causal map independent probability vector $\x$, such that
\begin{align}
    \left\{
    \frac{1}{2}\p^{'}\oplus\frac{1}{2}\q^{'}
    \right\}_{\Phi}
    \prec
    \x,
\end{align}
then it is straightforward to see that
\begin{align}
    {\bf w}^{'}\cdot D\prec\x,
\end{align}
holds for any permutation matrix $D$.
\end{mycor}

We now move on to investigating the most general form of quantum roulette, where the player Bob can engineer any quantum circuit fragment $\Phi\in\mathfrak{F}_{a}$ with finite integer $a$. On top of that, the agent Alice will select an interactive measurement $\mT_b$ with  $b\in\{1, 2, \ldots, c\}$ at random and probe $\Phi$ with her choice of interactive measurement $\mT_b$. Without loss of generality, let us assume the quantum circuit fragment (see Def.~\ref{def:QCF}) provided by Bob is in the form of
\begin{align}\label{eq:QCFUUR}
    \Phi
    = 
    \Psi^{a}\circ\Psi^{a-1}
    \circ
    \cdots
    \circ
    \Psi^{2}(\rho)
    \in
    \mathfrak{F}_a.
\end{align}
Meanwhile, the interactive measurement $\mT_{b}:=\{\mT_{b, x_{b}}\}_{x_{b}=1}^{m_{b}}\in\mathfrak{T}_{a}$ ($b=1, 2, \ldots, c$) (see Def.~\ref{def:IM}) is constituted of
\begin{align}
    \mT_{b, x_b}(\cdot)
    := 
    \Tr[
    M_{b, x_b}\cdot\Lambda^{a-1}_{b}
    \circ
    \Lambda^{a-2}_{b}
    \circ
    \cdots
    \circ
    \Lambda^{1}_{b}(\cdot)
    ].
\end{align} 
Here $\{M_{b, x_b}\}_{x_b=1}^{m_{b}}$ forms a POVM for each $b\in\{1, 2, \ldots, c\}$. In this case, the layout of roulette table consists of $\sum_{b=1}^{c}m_{b}$ tuples $(b, x_{b})$, where $b\in\{1, 2, \ldots, c\}$ and $x_{b}\in\{1, 2, \ldots, m_{b}\}$. In particular, for Bob's quantum circuit fragment $\Phi$, the tuple $(b, x_{b})$ happens with probability $p_{x_b}(\Phi, \mT_b)/c$, where the quantity $p_{x_b}(\Phi, \mT_b)$ is given by
\begin{align}
    p_{x_b}(\Phi, \mT_b)
    :=
    \Tr[
    M_{b, x_b}\cdot
    \Psi^{a}
    \circ
    \Lambda^{a-1}_b
    \circ
    \Psi^{a-1}
    \circ
    \Lambda^{a-2}_b
    \circ
    \cdots
    \circ
    \Psi^{2}
    \circ
    \Lambda^{1}_b(\rho)
    ].
\end{align}
Similar to the case of quantum causal maps, when starts with $k$ chips, Bob's guessing strategy is characterized by $c$ sets $\mS_{b}\subset\{1, \ldots, m_b\}$ such that
\begin{align}
    \sum_{b=1}^{c} |\mS_{b}|= k,
\end{align}
and his maximum winning probability is given by
\begin{align}
    p_{\text{win}, k}(\Phi)
    &=
    \max_{\sum_{b=1}^{c} |\mS_{b}|= k}
    \sum_{x_{b}\in\mS_{b}\subset\{1, \ldots, m_b\}}
    \frac{1}{c}\cdot
    p_{x_b}(\Phi, \mT_b)
    \label{eq:multi-win-1}\\
    &=
    \max_{\sum_{b=1}^{c} |\mS_{b}|= k}
    \sum_{x_{b}\in\mS_{b}\subset\{1, \ldots, m_b\}}
    \frac{1}{c}
    \left(
    J^{\Phi}\star 
    J^{\mT_b}_{x_b}
    \right)\label{eq:multi-win-2},
\end{align}
where the operator $J^{\Phi}$ and $J^{\mT_b}_{x_b}$ stand for the CJ operators of quantum circuit fragment $\Phi$ and the component $\mT_{b, x_{b}}$ of interactive measurement $\mT_b$ respectively. To simplify our representation, we take the union of CJ operators $J^{\mT_b}_{x_b}$, and denote it by $\{J_z\}_z$, which is formally defined as
\begin{align}
    J_z
    :=
    J^{\mT_{b}}_{z-\sum_{j=1}^{b-1}m_{j}},
    \quad
    \text{for}
    \,\,\,
    \sum_{j=1}^{b-1}m_{j}+1\leqslant z\leqslant \sum_{j=1}^{b}m_{j},
\end{align}
and set $J_{\mS}$ as
\begin{align}
    J_{\mS}:=
    \sum_{z\in\mS} J_{z},
\end{align}
where $\mS$ is a subset of $\{1, \ldots, \sum_{b=1}^{c}m_{b}\}$. Before going further, let us consider an example of $J_{\mS}$. Assume $m_1= m_2= m_3= 3$, then $J_{\{2, 5, 8\}}$ is an abbreviation of the following summation,
\begin{align}
    J_{\{2, 5, 8\}}
    =
    J_2+ J_5+ J_8
    = 
    J^{\mT_1}_2+ J^{\mT_2}_2+ J^{\mT_3}_2.
\end{align}
Thus, given quantum circuit fragment $\Phi$, Bob’s maximum winning probability, i.e. Eq.~\ref{eq:multi-win-2}, can be simplified to the following form in terms of the CJ operators and link product,
\begin{align}
    p_{\text{win}, k}(\Phi)
    =
    \frac{1}{c}\max_{|\mS|=k}
    J^{\Phi}\star 
    J_{\mS}.
\end{align}
Denote the probability vector obtained by measuring $\Phi$ with respect to interactive measurements $\mT_{b}$ as $\p_{b}$, then the largest summation of $k$ elements of $\oplus_{b=1}^{c} \p_{b}/c$ is exactly $p_{\text{win}, k}(\Phi)$. Thus, written out explicitly, we have
\begin{align}
    \sum_{i=1}^{k} (\bigoplus_{b=1}^{c} \frac{1}{c} \p_{b})_{i}^{\downarrow}
    =
    p_{\text{win}, k}(\Phi)
    =
    \frac{1}{c}\max_{|\mS|=k}
    J^{\Phi}\star 
    J_{\mS},
\end{align}
where the superscript $\downarrow$ indicates that the corresponding vector is arranged in non-increasing order, and the subscript $i$ means it is the $i$-th element of the vector. By maximizing over all quantum circuit fragment $\Phi\in\mathfrak{F}_{a}$, we have the following fundamental limitation $p_{\text{win}, k}$ for Bob's winning probability, namely 
\begin{align}
    p_{\text{win}, k}
    :=
    \max_{\Phi\in\mathfrak{F}_{a}}p_{\text{win}, k}(\Phi).
\end{align}
It hence follows that for any quantum circuit fragment $\Phi\in\mathfrak{F}_{a}$ prepared by Bob, his winning probability with $k$ chips is upper-bounded by $p_{\text{win}, k}$,
\begin{align}\label{eq:multi-upper-bound}
    \sum_{i=1}^{k} (\bigoplus_{b=1}^{c} \frac{1}{c} \p_{b})_{i}^{\downarrow}
    =
    p_{\text{win}, k}(\Phi)
    \leqslant
    p_{\text{win}, k},
\end{align}
for $1\leqslant k\leqslant \sum_{b=1}^{c}m_{b}$. It is worth mentioning that the maximum winning probability $p_{\text{win}, k}$ for Bob is achievable.
Then, by the definition of majorization (see Def.~\ref{def:Majorization}), we can unify $\sum_{b=1}^{c}m_{b}$ inequalities in the form of Eq.~\ref{eq:multi-upper-bound} into a single majorization inequality; that is
\begin{align}
    \bigoplus_{b=1}^{c} \frac{1}{c} \p_{b}
    \prec
    \www,
\end{align}
where the probability vector $\www$ only relies on the set of interactive measurements, i.e. $\{\mT_1, \mT_2, \ldots, \mT_c\}$, and independent of the quantum circuit fragment $\Phi$ chosen by Bob. Formally, we define $\www$ as
\begin{align}
    \www
    :=
    (w_{k})_{k=1}^{\sum_{b=1}^{c}m_{b}}
    =
    (w_1, \ldots, w_{\sum_{b=1}^{c}m_{b}}),
\end{align}
with each component $w_{k}$ is defined as 
\begin{align}
w_{k}:= p_{\text{win}, k}- p_{\text{win}, k-1},    
\end{align} 
for $1\leqslant k\leqslant \sum_{b=1}^{c}m_{b}$, and $p_{\text{win}, 0}:= 0$. Using the language of CJ operators, the maximum winning probability $p_{\text{win}, k}$ can be solved by using SDPs~\cite{PhysRevLett.101.060401,PhysRevA.80.022339},
\begin{align}\label{eq:pwink-multi}
    p_{\text{win}, k}
    =
    \frac{1}{c}\max_{|\mS|=k}
    \max_{\Phi\in\mathfrak{F}_{a}}
    J^{\Phi}\star 
    J_{\mS}
    =
    \frac{1}{c}
    \max_{|\mS|=k}
    \max\quad&\Tr [ J(a) \cdot J_{\mathcal{S}} ]\notag\\
    {\rm s.t.}  \quad &\,J(a)\geqslant0,\quad
    \Tr_{\mathcal{H}_{1}} [J(1)] = 1
    ,\notag\\
    &
    \Tr_{\mathcal{H}_{2i-1}} [J(i)] = \Tr_{\mathcal{H}_{2i-1}\mathcal{H}_{2i-2}} [J(i)] \otimes \frac{\mathds{1}_{\mathcal{H}_{2i-2}}}{d_{\mH_{2i-2}}}, \quad\text{for}\,\,\, 2\leqslant i\leqslant a,
\end{align}
where $d_{\mH_{2i-2}}:= \dim\mH_{2i-2}$. From a causal perspective, the last restriction of Eq.~\ref{eq:pwink-multi} indicates that the quantum circuit fragment $\Phi$ is NS from $\Psi^{i}$ to $\Psi^{i-1}$ for $2\leqslant i\leqslant a$ with $\Psi^{1}:= \rho$ (see Eq.~\ref{eq:QCFUUR}). Now the universal uncertainty relation for multiple interactive measurements can be formulated as
\begin{mythm}
{Multiple Universal Uncertainty Relation for $\mathfrak{F}_{a}$}{UURa}
For any quantum circuit fragment $\Phi\in\mathfrak{F}_{a}$, the probability vector $\p_{b}$ obtained by measuring $\Phi$ with respect to interactive measurements $\mT_{b}\in\mathfrak{T}_{a}$
($b\in\{1, 2, \ldots, c\}$)
satisfies the following trade-off
\begin{align}\label{eq:mpwink}
    \bigoplus_{b=1}^{c} \frac{1}{c} \p_{b}
    \prec
    \www
    =
    (p_{\text{win}, k}
    - 
    p_{\text{win}, k-1})
    _{k=1}^{\sum_{b=1}^{c}m_{b}},
\end{align}
where $p_{\text{win}, 0}:= 0$, and for $1\leqslant k\leqslant \sum_{b=1}^{c}m_{b}$, the quantity $p_{\text{win}, k}$ is characterized by Eq.~\ref{eq:pwink-multi}.
Similar to the case of causal maps (see Eq.~\ref{eq:pwin5}), the optimal solution of $p_{\text{win}, k}$ forms a semidefinite program (SDP), and hence
can be computed efficiently in polynomial time (up to desired accuracy) by using the ellipsoid method or an interior point method~\cite{doi:10.1137/1038003,boyd_vandenberghe_2004}.
\end{mythm}

As discussed in Subsec.~\ref{subsec:maj}, constructing $\www$ is the first step of formulating the optimal bound of $\left\{\oplus_{b=1}^{c}\frac{1}{c} \p_{b}\right\}_{\Phi}$, and the next step is the flatness process $\mF$, which implies that

\begin{mythm}
{Optimal Multiple Universal Uncertainty Relation for $\mathfrak{F}_{a}$}{OUURa}
For any quantum circuit fragment $\Phi\in\mathfrak{F}_{a}$, the probability vector $\p_{b}$ obtained by measuring $\Phi$ with respect to interactive measurements $\mT_{b}\in\mathfrak{T}_{a}$
($b\in\{1, 2, \ldots, c\}$)
satisfies the following trade-off
\begin{align}\label{eq:multiple-main}
     \bigoplus_{b=1}^{c} \frac{1}{c} \p_{b}
    \prec
    \mF(\www)
    ,
\end{align}
where $\mF$ is the flatness process defined in Def.~\ref{def:FProcess}. The vector $\www$ is defined as
\begin{align}
    \www
    :=
    (p_{\text{win}, k}- p_{\text{win}, k-1})_{k=1}^{\sum_{b=1}^{c}m_{b}},
\end{align} 
with $p_{\text{win}, 0}:=0$ and $p_{\text{win}, k}$ is solved by SDPs in the form of Eq.~\ref{eq:pwink-multi} for $1\leqslant k\leqslant \sum_{b=1}^{c}m_{b}$. Here the bound $\mF(\www)$ is optimal, due to the structure of majorization lattice, namely for any probability vector $\x$ satisfying 
\begin{align}
    \left\{
    \bigoplus_{b=1}^{c} \frac{1}{c} \p_{b}
    \right\}_{\Phi}
    \prec
    \x,
\end{align}
then we also have
\begin{align}
    \mF(\www)\prec\x.
\end{align}
Physically, the quantum circuit fragment independence of $\x$ leads to a universal uncertainty relation $\bigoplus_{b=1}^{c} \frac{1}{c} \p_{b}\prec\x$, and $\mF(\www)$ constructed here is the optimal one. Mathematically, $\mF(\www)$ is the GLB (see Def.~\ref{def:Lattice}) for all such $\x$. In particular, taken any Schur-concave function $f$, we have
\begin{align}
    f(\bigoplus_{b=1}^{c} \frac{1}{c} \p_{b})\geqslant f(\mF(\www))\geqslant f(\x).
\end{align}
\end{mythm}

From the equivalence relation $\sim$ based on permutations, we know that permutations will not change the order of majorization. As an illustration, let us consider two equivalence class $[\x]$ and $[\y]$. If $\x\prec\y$, then it is straightforward to check that $\x\cdot D_1\prec\y\cdot D_2$ holds for any permutation matrices $D_1$ and $D_2$. Using this insight, we obtain the following corollary.

\begin{mycor}
{Optimal Universal Uncertainty Relation for $\mathfrak{F}_{a}$ under Permutations}{OUURaP}
Given a quantum circuit fragment $\Phi\in\mathfrak{F}_{a}$, let us denote the probability vectors obtained by measuring $\Phi$ with respect to interactive measurement $\mT_{b}\in\mathfrak{T}_{a}$ ($b\in\{1, 2, \ldots, c\}$) as $\p_b$. Define the vector $\www$ as $(p_{\text{win}, k}- p_{\text{win}, k-1})_{k=1}^{\sum_{b=1}^{c}m_{b}}$ with $p_{\text{win}, 0}:=0$ and $p_{\text{win}, k}$ in the form of Eq.~\ref{eq:pwink-multi} for $1\leqslant k\leqslant \sum_{b=1}^{c}m_{b}$. Then for any $\p^{'}_b\in[\p^{\downarrow}_b]$, and $\bf{w}^{'}\in[\mF(\www)]$, we have the following universal uncertainty relation
\begin{align}
    \bigoplus_{b=1}^{c} \frac{1}{c} \p_{b}{'}
    \prec
    {\bf w}^{'}.
\end{align}
In particular, given any permutation matrices $D_1, D_2, \ldots, D_{c+1}$, we also have
\begin{align}
    \bigoplus_{b=1}^{c} \frac{1}{c} 
    (\p_{b}{'}\cdot D_b)
    \prec
    {\bf w}^{'}\cdot D_{c+1}.
\end{align}
Moreover, if there exists a quantum circuit fragment independent probability vector $\x$, such that
\begin{align}
    \left\{
    \bigoplus_{b=1}^{c} \frac{1}{c} \p_{b}{'}
    \right\}_{\Phi}
    \prec
    \x,
\end{align}
then it is straightforward to see that
\begin{align}
    {\bf w}^{'}\cdot D\prec\x,
\end{align}
holds for any permutation matrix $D$.
\end{mycor}

So far we've been talking about the formulation of universal uncertainty relation and its intrinsic connection with quantum roulette, but the advantages of using this form are just as important. One of the greatest advantages of formulating Eq.~\ref{eq:multiple-main} is its universality. Here we addressed the question of how universal our framework is, and highlight the generality of our uncertainty relations.

Uncertainty principle in quantum mechanics has revolutionized our view of the physical world. Roughly speaking, it captures the uncertainty trade-off between observables associated with static system -- quantum states. However, systems in quantum mechanics evolve. Hence, quantum dynamics with definite causal order, known as quantum circuit fragment in this work, emerge as a central ingredient in state-of-the-art quantum technologies. Our framework extends uncertainty principle to incorporate quantum dynamics, allowing for any quantum circuit fragment $\Phi\in\mathfrak{F}_a$. In particular, our framework, including theorems (Thms.~\ref{thm:UURa} and \ref{thm:OUURa}) and corollary (Cor.~\ref{cor:OUURaP}), does not focus on a particular choice of quantum circuit fragment. By employing our theory, one can build uncertainty relations for any quantum dynamics with definite causal order, e.g., based upon the universal uncertainty relation for quantum causal maps, i.e. $\Phi\in\mathfrak{F}_2$ (see Thms.~\ref{thm:UUR2} and \ref{thm:OUUR2}), we can derived the causal uncertainty relation, which characterizes the fundamental trade-off between causal structures, such as direct-cause and common-cause in causal inference. Besides causality, it would also be interesting to apply our framework to other physical properties, such as completely positive divisibility, non-Markovianity, and so forth, of quantum dynamics, and identify the corresponding trade-off. This could be a good issue for future works.

Finally, we ask if the uncertainty relation $f(\bigoplus_{b=1}^{c} \frac{1}{c} \p_{b})\geqslant f(\mF(\www))$ induced by some Schur-concave function $f$ is optimal? Unfortunately, we cannot guarantee the optimality of $f(\mF(\www))$ with specific uncertainty measure $f$. There might exist a quantum circuit fragment independent vector $\x$ such that
\begin{align}\label{eq:sc-f}
    f(\bigoplus_{b=1}^{c} \frac{1}{c} \p_{b})
    \geqslant
    f(\x)
    \geqslant
    f(\mF(\www)).
\end{align}
Remark that Eq.~\ref{eq:sc-f} only holds for a specific $f$, and is not equivalent to $\oplus_{b=1}^{c} \frac{1}{c} \p_{b}\prec\x\prec\mF(\www)$. 
In this case, there must exist another uncertainty measure $g$, leading to the following uncertainty relation
\begin{align}
    g(\bigoplus_{b=1}^{c} \frac{1}{c} \p_{b})
    \geqslant
    g(\mF(\www))
    \geqslant
    g(\x).
\end{align}
This in turn implies that, as a lower bound of uncertainty relation, the quantity $g(\x)$ is weaker than $g(\mF(\www))$. From above discussion, it is clear that there does not exist another circuit fragment independent vector $\x$ that always outperform our bound $\mF(\www)$. Actually, if there exist a quantum circuit fragment independent vector $\x$ such that Eq.~\ref{eq:sc-f} holds for any uncertainty measure $f$, i.e. Schur-concave function, then Lem.~\ref{lem:ME} immediately implies that
\begin{align}
    \bigoplus_{b=1}^{c} \frac{1}{c} \p_{b}
    \prec
    \x
    \prec
    \mF(\www),
\end{align}
holds for any $\Phi\in\mathfrak{F}_a$. Note that vector $\mF(\www)$ is the LUB of set $\left\{\oplus_{b=1}^{c} \frac{1}{c} \p_{b}\right\}_{\Phi}$, we then find that the reverse is also true, i.e. $\mF(\www)\prec\x$. Therefore, they are equivalent under permutations, namely 
\begin{align}
    \mF(\www)\sim\x,
\end{align}
or there exists a permutation matrix $D$ such that
\begin{align}
    \mF(\www)=\x\cdot D,
\end{align}
which implies 
\begin{align}
    f(\mF(\www))= f(\x),
\end{align} 
holds for any Schur-concave function $f$. Put simply, the completeness of majorization lattice guarantee the optimality of Eq.~\ref{eq:multiple-main}, and there does exist another circuit fragment independent vector that outperforms $\mF(\www)$.


\subsection{\label{subsec:thm1}Lemma~$1$ and Theorem~$1$ of the Main Text: Their Proofs, Improvements, and Generalizations}

In this subsection, we prove the results introduced in the main text of our work, including Lemma~$1$ and Theorem~$1$. Let us begin with Lemma~$1$ of the main text, namely

\begin{mycor}
{Lemma~$1$ of the Main Text}{UUR-Lem}
Collect the probability distributions obtained by implementing interactive measurements $\mathcal{T}_{1}$ and $\mathcal{T}_{2}$ on some dynamical process $\Phi$ as $\p$ and $\q$, then there exists a probability vector $\mathbf{v}(\mathcal{T}_{1}, \mathcal{T}_{2})$ such that
\begin{align}\label{eq:lem-eur}
\frac{1}{2}\p\oplus\frac{1}{2}\q
\prec
\mathbf{v}(\mathcal{T}_{1}, \mathcal{T}_{2}).
\end{align}
The vector-type bound
\begin{align}
    \mathbf{v}(\mathcal{T}_{1}, \mathcal{T}_{2})
    :=
    \ww,
\end{align} 
is independent of the dynamical process $\Phi$ being measured, and hence captures the essential incompatibility between 
$\mathcal{T}_{1}$ and $\mathcal{T}_{2}$.
\end{mycor}

\begin{proof}
By restricting Thm.~\ref{thm:UURa} to the case of two interactive measurements, i.e. $\mT_1$ and $\mT_2$, we immediately obtain Eq.~\ref{eq:lem-eur}, where the upper bound under majorization is completely characterized by the winning probability of quantum roulette.
\end{proof}

From the algebraic structure of majorization lattice (see Subsec.~\ref{subsec:maj}), we know that the upper bound $\ww$ provided by Cor.~\ref{cor:UUR-Lem} (or originally by Thm.~\ref{thm:UURa}) may not be optimal. Thanks to Thm.~\ref{thm:OUURa}, the optimal bound for $\p/2\oplus\q/2$ under majorization can be obtained by further applying the flatness process $\mF$ (see Def.~\ref{def:FProcess}); that is

\begin{mycor}
{Improved Lemma~$1$ of the Main Text}{OUUR-Lem}
Collect the probability distributions obtained by implementing interactive measurements $\mathcal{T}_{1}$ and $\mathcal{T}_{2}$ on some dynamical process $\Phi$ as $\p$ and $\q$, then there exists a probability vector $\mathbf{v}(\mathcal{T}_{1}, \mathcal{T}_{2})$ such that
\begin{align}\label{eq:lem-eur}
\frac{1}{2}\p\oplus\frac{1}{2}\q
\prec
\mathbf{v}(\mathcal{T}_{1}, \mathcal{T}_{2}).
\end{align}
The vector-type bound
\begin{align}
    \mathbf{v}(\mathcal{T}_{1}, \mathcal{T}_{2})
    :=
    \mF(\ww),
\end{align} 
is independent of the dynamical process $\Phi$ being measured, and hence captures the essential incompatibility between 
$\mathcal{T}_{1}$ and $\mathcal{T}_{2}$. Here $\mF$ is the flatness process defined in Def.~\ref{def:FProcess}. Remark that, the bound $\mF(\ww)$ is optimal: for any probability vector $\x$ satisfying
\begin{align}
    \left\{
    \frac{1}{2}\p\oplus\frac{1}{2}\q
    \right\}_{\Phi}
    \prec
    \x,
\end{align}
then we also have
\begin{align}
    \mF(\ww)\prec\x.
\end{align}
Physically, the quantum circuit fragment independence of $\x$ leads to a universal uncertainty relation $\p/2\oplus\q/\prec\x$, and $\mF(\ww)$ constructed here is the optimal one. Mathematically, $\mF(\ww)$ is the GLB (see Def.~\ref{def:Lattice}) for all such $\x$. In particular, taken any Schur-concave function $f$, we have
\begin{align}
    f(\frac{1}{2}\p\oplus\frac{1}{2}\q)\geqslant f(\mF(\ww))\geqslant f(\x).
\end{align}
\end{mycor}

We now move on to demonstrating a generalized entropic uncertainty relation with multiple interactive measurements, which includes the Theorem~$1$ in the main text as a special case. Moreover, compare with the bound $C(\mT_1, \mT_2):= 2 H(\ww)- 2$ offered in the main text, an improved entropic uncertainty relation has also been provided. We start by writing down a direct consequence of Thm.~\ref{thm:UURa}:

\begin{mythm}
{Entropic Uncertainty Relation for Measurements with Interventions}{EURa}
Given $c$ interactive measurements $\mathcal{T}_{b}\in\mathfrak{T}_a$ ($b\in\{1, 2, \ldots, c\}$) acting on some quantum circuit fragment $\Phi\in\mathfrak{F}_a$. The entropy of their measurement outcomes, when summed, satisfies
\begin{align}\label{eq:xeura}
\sum_{b=1}^{c}H(\mT_{b})_{\Phi} 
\geqslant 
C(\mT_1, \mT_2, \ldots, \mT_c)
:=
c H(\www)- c\log c
,
\end{align}
where $H(\mathcal{T}_{b})_{\Phi}:= H(\p_b)$, with $\p_b$ denotes the probability vectors obtained by measuring $\Phi$ with respect to interactive measurement $\mT_{b}$. The bound $C(\mT_1, \mT_2, \ldots, \mT_c)$ -- measuring incompatibility between interactive measurements $\mathcal{T}_{b}$ -- is non-negative and independent of $\Phi$. $C(\mT_1, \mT_2, \ldots, \mT_c)$ can be explicitly computed and is strictly non-zero whenever these interactive measurements have no common eigencircuit. 
\end{mythm}

\begin{proof}
From Eq.~\ref{eq:mpwink} of Thm.~\ref{thm:UURa}, we see that
\begin{align}\label{eq:mpwinks}
    \bigoplus_{b=1}^{c} \frac{1}{c} \p_{b}
    \prec
    \www.
\end{align}
Then, by applying Shannon entropy $H$, it now follows immediately that
\begin{align}
    H(\bigoplus_{b=1}^{c} \frac{1}{c} \p_{b})
    &=
    -\sum_{b, x_b} \frac{p_{x_b}(\Phi, \mT_b)}{c}
    \log \frac{p_{x_b}(\Phi, \mT_b)}{c}\\
    &=
    -\frac{1}{c}\sum_{b, x_b} p_{x_b}(\Phi, \mT_b)
    \log p_{x_b}(\Phi, \mT_b)
    +
    \frac{\sum_{b, x_b} p_{x_b}(\Phi, \mT_b)}
    {c}
    \log c\\
    &=
    \frac{1}{c}
    \sum_{b=1}^{c}H(\mT_{b})_{\Phi}
    +
    \log c\\
    &\geqslant
    H(\www)
    ,
\end{align}
where the second equation comes from the fact that, for each $b\in\{1, 2, \ldots, c\}$, we have $\sum_{x_b} p_{x_b}(\Phi, \mT_b)=1$, and thus $\sum_b \sum_{x_b} p_{x_b}(\Phi, \mT_b)= c$. The third equation follows from the definition, i.e. \begin{align}
H(\mT_{b})_{\Phi}
:= 
H(\p_b)
= 
-\sum_{x_b}p_{x_b}(\Phi, \mT_b)
\log p_{x_b}(\Phi, \mT_b).
\end{align}
Note that throughout this supplemental material, all logarithms are base $2$. Now we have
\begin{align}
   \sum_{b=1}^{c}H(\mT_{b})_{\Phi}
   \geqslant
   c
   \left(
   H(\www)
   -
   \log c
   \right)
   =
   c H(\www)
   -
   c\log c.
\end{align}
Bu further define 
\begin{align}
    C(\mT_1, \mT_2, \ldots, \mT_c)
    :=
    c H(\www)
   -
   c\log c,
\end{align}
we obtain the entropic uncertainty relation of Eq.~\ref{eq:xeura}, as required. Remark that, as a probability vector with $\sum_{b=1}^{c}m_{b}$ components, our bound $\www$ is bounded by
\begin{align}\label{eq:bound-w}
    (\underbrace{\frac{1}{\sum_{b=1}^{c}m_{b}}, \ldots, \frac{1}{\sum_{b=1}^{c}m_{b}}}_{\sum_{b=1}^{c}m_{b}})
    \prec
    \www
    \prec
    (\underbrace{\underbrace{\frac{1}{c}, \ldots, \frac{1}{c}}_{c}, \underbrace{0, \ldots, 0}_{\sum_{b=1}^{c}m_{b}- c}}_{\sum_{b=1}^{c}m_{b}}),
\end{align}
where the captions in Eq.~\ref{eq:bound-w} indicate the number of elements inside the vector. Hence 
\begin{align}
    H(\www)\geqslant\log c,
\end{align}
and the bound $C(\mT_1, \mT_2, \ldots, \mT_c)= c H(\www)-c\log c$ is  non-negative. The majorization upper bound of Eq.~\ref{eq:bound-w} is achieved, i.e.
$\www= (1/c, \ldots, 1/c, 0, \ldots, 0)$, if and only if $(\p_b)^{\downarrow}_1= 1$ for all $b\in\{1, 2, \ldots, c\}$, which is equivalent to say that the bound $C(\mT_1, \mT_2, \ldots, \mT_c)$ is zero whenever these interactive measurements $\{\mT_1, \mT_2, \ldots, \mT_c\}$ have a common eigencircuit $\Phi\in\mathfrak{F}_a$ (see Def.~\ref{def:EC}). According to the definition of $\www$, we know that $C(\mT_1, \mT_2, \ldots, \mT_c)$ is independent of the quantum circuit fragment $\Phi$, and hence quantifies the inherent incompatibility between interactive measurements. Eq.~\ref{eq:pwink-multi} further ensures that $\www$ (and thus $C(\mT_1, \mT_2, \ldots, \mT_c)$) can be explicitly computed.
\end{proof}

Thm.~\ref{thm:EURa} forms a fundamental limitation of incompatible interactive measurements in terms of Shannon entropy, covering all quantum dynamics with definite causal order (called quantum circtuit fragment in this work, see Def.~\ref{def:QCF} for more details). In fact, Eq.~\ref{eq:xeura} works for quantum circuit fragment $\Phi\in\mathfrak{F}_{a}$ with arbitrary $a$ and arbitrary number of interactive measurements, and hence generates an infinite number of entropic uncertainty relations. As a special case, we consider the situation where $c=2$, i.e. a pair of interactive measurements, and obtain the following corollary straightforwardly, which is exactly Theorem~$1$ of the main text.

\begin{mycor}
{Theorem~$1$ of the Main Text}{EUR2}
Given two interactive measurements $\mT_1$ and $\mT_2$ acting on some quantum circuit fragment $\Phi\in\mathfrak{F}_a$. The entropy of their measurement outcomes, when summed, satisfies
\begin{align}\label{eq:xeur2}
H(\mT_{1})_{\Phi}
+
H(\mT_{2})_{\Phi}
\geqslant 
C(\mT_1, \mT_2),
\end{align}
where $H(\mathcal{T}_{b})_{\Phi}:= H(\p_b)$, with $\p_b$ denotes the probability vectors obtained by measuring $\Phi$ with respect to interactive measurement $\mT_{b}$ ($b=1, 2$). The bound 
\begin{align}
    C(\mT_1, \mT_2)
    :=
    2 H(\ww)- 2,
\end{align}
which measures incompatibility between $\mT_1$ and $\mT_2$, is non-negative and independent of $\Phi$. $C(\mT_1, \mT_2)$ can be explicitly computed and is strictly non-zero whenever these interactive measurements have no common eigencircuit. 
\end{mycor}

\noindent
Here we have two remarks: First, when the quantum circuit fragment $\Phi$ degenerates to the case of a quantum channel, we can simply adopt the framework introduced in \cite{PhysRevResearch.3.023077} to construct $C(\mathcal{T}_{1}, \mathcal{T}_{2})$, which includes a Maassen-Uffink type uncertainty relation. However, in such a case, the intervention is absent. Second, till now we have a variety of different forms of entropic uncertainty relations~\cite{RevModPhys.89.015002}, but none of them can surpass all the others. Among them, the form derived from universal uncertainty relation
outperforms other forms of entropic uncertainty relations in a large number of instances.
For readers who are unfamiliar with entropic uncertainty relation and majorization, recent works, and overviews exist (e.g., Refs.~\cite{PhysRevLett.111.230401,PhysRevA.89.052115,RevModPhys.89.015002}).

If the vector $\www$ is not arranged in non-increasing order, then the flatness process $\mF$ will further improve the result~\cite{992785}. Recall the example of $S=\{\x = (0.6, 0.15, 0.15, 0.1), \y = (0.5, 0.25, 0.2, 0.05)\}$ considered in Subsec.~\ref{subsec:maj}, where
\begin{align}
\mF(\bb_{S})
=
(0.6, 0.175, 0.175, 0.05)
\prec
\bb_{S} 
= 
(0.6, 0.15, 0.2, 0.05).
\end{align}
Using this insight, we can employ Eq.~\ref{eq:multiple-main} instead of Eq.~\ref{eq:mpwink} to formulate entropic uncertainty relation. Thus, written out explicitly, we have

\begin{mythm}
{Improved Entropic Uncertainty Relation for Measurements with Interventions}{IEURa}
Given $c$ interactive measurements $\mathcal{T}_{b}\in\mathfrak{T}_a$ ($b\in\{1, 2, \ldots, c\}$) acting on some quantum circuit fragment $\Phi\in\mathfrak{F}_a$. The entropy of their measurement outcomes, when summed, satisfies
\begin{align}\label{eq:ixeura}
\sum_{b=1}^{c}H(\mT_{b})_{\Phi} 
\geqslant 
C(\mT_1, \mT_2, \ldots, \mT_c)
:=
c H(\mF(\www))- c\log c
,
\end{align}
where $H(\mathcal{T}_{b})_{\Phi}:= H(\p_b)$, with $\p_b$ denotes the probability vectors obtained by measuring $\Phi$ with respect to interactive measurement $\mT_{b}$, and $\mF$ stands for the flatness process defined in Def.~\ref{def:FProcess}. The bound $C(\mT_1, \mT_2, \ldots, \mT_c)$ -- measuring incompatibility between interactive measurements $\mathcal{T}_{b}$ -- is non-negative and independent of $\Phi$. $C(\mT_1, \mT_2, \ldots, \mT_c)$ can be explicitly computed and is strictly non-zero whenever these interactive measurements have no common eigencircuit. 
\end{mythm}

Thm.~\ref{thm:IEURa} follows directly by replacing $\www$ with $\mF(\www)$ in the proof of Thm.~\ref{thm:EURa}. Thanks to the property of flatness process $\mF$, now we have
\begin{align}
    \mF(\www)\prec\www,
\end{align}
so that
\begin{align}
    H(\mF(\www))\geqslant H(\www).
\end{align}
Written in full, that is
\begin{align}\label{eq:EURa-com}
    \sum_{b=1}^{c}H(\mT_{b})_{\Phi}
    \geqslant
    c H(\mF(\www))- c\log c
    \geqslant
    c H(\www)- c\log c.
\end{align}
In particular, given two interactive measurements $\mT_1$ and $\mT_2$, we would like to show that

\begin{mycor}
{Improved Theorem~$1$ of the Main Text}{IEUR2}
Given two interactive measurements $\mT_1$ and $\mT_2$ acting on some quantum circuit fragment $\Phi\in\mathfrak{F}_a$. The entropy of their measurement outcomes, when summed, satisfies
\begin{align}\label{eq:ixeur2}
H(\mT_{1})_{\Phi}
+
H(\mT_{2})_{\Phi}
\geqslant 
C(\mT_1, \mT_2),
\end{align}
where $H(\mathcal{T}_{b})_{\Phi}:= H(\p_b)$, with $\p_b$ denotes the probability vectors obtained by measuring $\Phi$ with respect to interactive measurement $\mT_{b}$ ($b=1, 2$). The bound 
\begin{align}
    C(\mT_1, \mT_2)
    :=
    2 H(\mF(\ww))- 2\log 2,
\end{align}
which measures incompatibility between $\mT_1$ and $\mT_2$, is non-negative and independent of $\Phi$. $C(\mT_1, \mT_2)$ can be explicitly computed and is strictly non-zero whenever these interactive measurements have no common eigencircuit. 
\end{mycor}
Similar to Eq.~\ref{eq:EURa-com}, we have the following relation between Cor.~\ref{cor:IEUR2} and Cor.~\ref{cor:EUR2} (i.e. Theorem~$1$ of the main text),
\begin{align}\label{eq:EUR2-com}
H(\mT_{1})_{\Phi}
+
H(\mT_{2})_{\Phi}
\geqslant
2 H(\mF(\ww))- 2\log 2
\geqslant
2 H(\ww)- 2\log 2.
\end{align}

To summarize, in this subsection, Thm.~\ref{thm:EURa} introduces the entropic uncertainty relation for arbitrary quantum circuit fragments with multiple interactive measurements, which includes Cor.~\ref{cor:EUR2} (namely, Theorem~$1$ of the main text) as a special case. By applying the flatness process, Thm.~\ref{thm:IEURa} and Cor.~\ref{cor:IEUR2} further improve the results presented in Thm.~\ref{thm:EURa} and Cor.~\ref{cor:EUR2} respectively.


\section{\label{sec:CUR}Causal Uncertainty Relation}

In this section, we extend the uncertainty principle from studying observables for quantum states to inferring causal structures of quantum dynamics, providing a lower bound for the fundamental trade-off between maximal common-cause indicator and the maximal direct-cause indicator.
Such a trade-off is called causal uncertainty relation in this work. Specifically, the causal uncertainty relations in terms of both Shannon entropy and majorization are provided. We further infer the causality associated with system-environment unitary dynamics by utilizing entropic causal uncertainty relation. In Subsec.~\ref{subsec:CC-DC}, we introduce the concepts of maximal common-cause indicator and maximal direct-cause indicator, and show that it is possible to formulate a universal uncertainty relation for these two families of interactive measurements. As a special case, we obtain the entropic uncertainty relation for maximal common-cause indicators and maximal direct-cause indicators, characterizing the incompatibility between common-cause and direct-cause in quantum causal inference. We then discuss the necessary and sufficient conditions for system-environment unitary dynamics to be purely common-cause and purely direct-cause, and further detail the application of our causal uncertainty relation to inferring causality in Subsec.~\ref{subsec:NS}. Finally, in Subsec.~\ref{subsec:coh}, we demonstrate that the parameterized quantum circuit considered in our main text contains a coherent superpositions of common-cause and direct-cause as a special case.


\subsection{\label{subsec:CC-DC}Uncertainty Relation for Common-Cause and Direct-Cause: Eq.~$3$ of the Main Text and Its Extension}

Physicists have studied the fundamental limitations of observable pairs for quantum states -- like the position and momentum of a particle, the phase and excitation number of a harmonic oscillator, the orthogonal components of spin angular momentum, and etc. Even with the complete description of the quantum state, it is impossible to predict the outcomes of these observable pairs. Such a restriction leads to a key principle in quantum mechanics -- known as Heisenberg's uncertainty principle. However, less is known about the intrinsic constraints on underlying physical properties of quantum dynamics. In this subsection, we give a complete characterization of the uncertainty trade-off between common-cause and direct-cause, establishing uncertainty relations for the maximal common-cause indicator and the maximal direct-cause indicator. Unlike previous studies with static object (e.g. quantum states), here we focus on the dynamical process of quantum causal maps, i.e. $\mathfrak{F}_2$ (see Subsec.~\ref{subsec:CMaps}).

There are so many types of interactive measurements that can be used to get information about the dynamical properties of quantum causal maps. For the purpose of causal inference, which of them are likely to be most useful to us? For these useful interactive measurements, how to characterize their incompatibility and formulate  the corresponding trade-off between them? To address these questions and 
explain our uncertainty relation for common-cause and direct-cause, let us first recall two families of interactive measurements introduced in the main text -- the set of all maximal common-cause indicators $\mM_{\text{CC}}\subset\mathfrak{T}_2$ and the set of all maximal direct-cause indicators $\mM_{\text{DC}}\subset\mathfrak{T}_2$ (see Fig.~$4$ of the main text for an illustration). For consistency purposes, we use the same notation as in Subsec.~\ref{subsec:CMaps}, in particular that a causal map $\Phi\in\mathfrak{F}_2$ is a linear map from system $B$ to systems $A$ and $C$. Without loss of generality, we assume $d_{A}= d_{B}= d_{C}= d$ in this work. Formally, the set $\mM_{\text{CC}}$ is defined as

\begin{mydef}
{Maximal Common-Cause Indicator}{MCCI}
An interactive measurement
$\mT_{\text{CC}}(\mU_1, \mU_2):=\{\mT_{\text{CC}, i}(\mU_1, \mU_2)\}_{i}\in\mathfrak{T}_2$ is called maximal common-cause indicator, if it has the form of 
\begin{align}
    \mT_{\text{CC}, i}(\mU_1, \mU_2)(\cdot)
    =
    \Tr[
    \Phi_i
    \cdot
    \mU_{2}(\cdot)
    \otimes
    \frac{\1_{B}}{d}
    \otimes
    \mU_{1}(\cdot)
    ],
\end{align}
where $\mU_1$ and $\mU_2$ are some local unitary channels acting on systems $A$ and $C$ respectively, namely $\mU_b(\cdot)= U_b (\cdot) U_b^{\dag}$ ($b=1, 2$), with $\ket{\Phi_1}:= \ket{\phi^{+}}= \sum_{k=0}^{d-1}\ket{kk}/\sqrt{d}$ being the maximally entangled state. Measurements are done with respect to a maximally entangling basis $\{\Phi_i\}_i$ with $d^2$ possible outcomes. Denote the collection of all maximal common-cause indicators as $\mM_{\text{CC}}$.
\end{mydef}

Remark that, in the main text of our work, we simply denote the elements of $\mM_{\text{CC}}$ as $\mT_1$. But, to identify the dependence and causation, we keep all the subscripts and arguments of interactive measurements in this Supplemental Material. On the other hand, the maximal direct-cause indicator is defined as

\begin{mydef}
{Maximal Direct-Cause Indicator}{MDCI}
An interactive measurement
$\mT_{\text{DC}}(\mU_3, \mU_4):=\{\mT_{\text{DC}, j}(\mU_3, \mU_4)\}_{j}\in\mathfrak{T}_2$ is called maximal direct-cause indicator, if it has the form of 
\begin{align}
    \mT_{\text{DC}, j}(\mU_3, \mU_4)(\cdot)
    =
    \Tr[
    \Phi_j
    \cdot
    \mU_{4}(\cdot)
    \otimes
    \mU_{3}(\Phi_{1})
    \otimes
    \Tr_{A}(\cdot)
    ],
\end{align}
where $\mU_3$ and $\mU_4$ are some local unitary channels acting on systems $B$ and $C$ respectively, namely $\mU_b(\cdot)= U_b (\cdot) U_b^{\dag}$ ($b=1, 2$).  $\ket{\Phi_1}:= \ket{\phi^{+}}= \sum_{k=0}^{d-1}\ket{kk}/\sqrt{d}$ is the maximally entangled state acting on systems $BR$. Measurements are done with respect to a maximally entangling basis $\{\Phi_j\}_j$ with $d^2$ possible outcomes. Denote the collection of all maximal direct-cause indicators as $\mM_{\text{DC}}$.
\end{mydef}

The circuit illustrations of Def.~\ref{def:MCCI} and Def.~\ref{def:MDCI} are given by Figs.~4a and 4b of the main text. The principal objective of this subsection is to characterize the fundamental trade-off between $\mT_{\text{CC}}(\mU_1, \mU_2)\in\mM_{\text{CC}}$ and $\mT_{\text{DC}}(\mU_3, \mU_4)\in\mM_{\text{DC}}$, and formulate their uncertainties in terms of Shannon entropy. To simplify our discussion and gain some intuitions, here we mainly focus on the qubit case, where $d_{A}= d_{B}= d_{C}= 2$. The general qudit case can be obtained by a straightforward extension of the procedure.

Before stating our uncertainty relation for common-cause and direct-cause, let us first introduce some notations related with maximal common-cause indicator and maximal direct-cause indicator. Denote the probability vectors obtained by measuring causal map $\Phi$ with respect to interactive measurements
$\mT_{\text{CC}}(\mU_1, \mU_2):=\{\mT_{\text{CC}, i}(\mU_1, \mU_2)\}_{i}\in\mathfrak{T}_2$ and $\mT_{\text{DC}}(\mU_3, \mU_4):=\{\mT_{\text{DC}, j}(\mU_3, \mU_4)\}_{j}\in\mathfrak{T}_2$ as $\p(\mU_1, \mU_2)_{\Phi}= (p_{i}(\mU_1, \mU_2)_{\Phi})_i$ and $\q(\mU_3, \mU_4)_{\Phi}= (q_{j}(\mU_3, \mU_4)_{\Phi})_{j}$, where each $p_{i}(\mU_1, \mU_2)_{\Phi}$ and $q_{j}(\mU_3, \mU_4)_{\Phi}$ are given by 
\begin{align}
    p_{i}(\mU_1, \mU_2)_{\Phi}
    &=
    \mT_{\text{CC}, i}(\mU_1, \mU_2)(\Phi)
    =
    J^{\Phi}\star J^{\text{CC}}_{i}(\mU_1, \mU_2),\label{eq:pi-u}\\
    q_{j}(\mU_3, \mU_4)_{\Phi}
    &=
    \mT_{\text{DC}, j}(\mU_3, \mU_4)(\Phi)
    =
    J^{\Phi}\star J^{\text{DC}}_{j}(\mU_3, \mU_4)
    .\label{eq:qj-u}
\end{align}
Here $J^{\text{CC}}_{i}(\mU_1, \mU_2)$ and $J^{\text{DC}}_{j}(\mU_3, \mU_4)$ are CJ operators of the measuring processes $\mT_{\text{CC}, i}(\mU_1, \mU_2)$ and $\mT_{\text{DC}, j}(\mU_3, \mU_4)$ respectively. The probability $p_{i}(\mU_1, \mU_2)_{\Phi}$ is a functional of unitary channels $\mU_1$, $\mU_2$, and the causal map $\Phi$. Similarly, the probability $q_{j}(\mU_3, \mU_4)_{\Phi}$ is a functional of unitary channels $\mU_3$, $\mU_4$, and the causal map $\Phi$. In the case of qubit systems, the CJ operators of $\mT_{\text{CC}, i}(\mU_1, \mU_2)$ and $\mT_{\text{DC}, j}(\mU_3, \mU_4)$ are written as
\begin{align}
    J^{\text{CC}}_{i}(\mU_1, \mU_2)
    &=
    \frac{\mathds{1}_B}{2}\otimes (U_{1}^{\dagger}\otimes U_{2}^{\dagger} \ketbra{\Phi_i}{\Phi_i} U_1 \otimes U_2)^{\T},\quad i\in \{1, 2, 3, 4\}.\label{eq:cc-i}\\
    J^{\text{DC}}_{j}(\mU_3, \mU_4)
    &=
    \mathds{1}_A \otimes (U_3 \ketbra{\Phi_1}{\Phi_1}_{BR} U_{3}^{\dagger}) \star (U_4^{\dagger} \ketbra{\Phi_j}{\Phi_j}_{CR} U_{4})^{\T},\quad j\in \{1, 2, 3, 4\}.\label{eq:dc-j}
\end{align}
The coefficient of $1/2$ in Eq.~\ref{eq:cc-i} comes from the fact that $d_{B}= 2$, ensuring $\1_{B}/2$ is a maximally mixed state on system $B$. In Eq.~\ref{eq:dc-j}, $\1_A$ is the CJ operator of $\Tr_{A}$ (partial trace over system $A$), i.e. $J^{\Tr_{A}}= \1_A$. For example, given a bipartite quantum state $\rho_{AB}$, it reduced state on $B$ system can be rewritten as $\rho_{B}:= \Tr_{A}[\rho_{AB}]= \rho_{AB}\star\1_{A}$.

Now the quantum uncertainty associated with the maximal common-cause indicator $\mT_{\text{CC}}(\mU_1, \mU_2)$ and the maximal direct-cause indicator $\mT_{\text{DC}}(\mU_3, \mU_4)$ can be quantified through the Shannon entropy of probability vectors $\p(\mU_1, \mU_2)_{\Phi}$ and $\q(\mU_3, \mU_4)_{\Phi}$. Mathematically, they are given by
\begin{align}
    H(\mT_{\text{CC}}(\mU_1, \mU_2))_{\Phi}
    &:=
    H(\p(\mU_1, \mU_2)_{\Phi})
    =
    -\sum_{i}
    p_{i}(\mU_1, \mU_2)_{\Phi}
    \log p_{i}(\mU_1, \mU_2)_{\Phi},\label{eq:H-CC}\\
    H(\mT_{\text{DC}}(\mU_3, \mU_4))_{\Phi}
    &:=
    H(\q(\mU_3, \mU_4)_{\Phi})
    =
    -\sum_{j}
    q_{j}(\mU_3, \mU_4)_{\Phi}
    \log q_{j}(\mU_3, \mU_4)_{\Phi}.\label{eq:H-DC}
\end{align}
Particularly, we are interested in the minimization of their joint uncertainty over all causal maps $\Phi\in\mathfrak{F}_2$, maximal common-cause indicators $\mT_{\text{CC}}(\mU_1, \mU_2)\in\mM_{\text{CC}}$, and maximal direct-cause indicators $\mT_{\text{DC}}(\mU_3, \mU_4)\in\mM_{\text{DC}}$, which is
\begin{align}
    \mB
    :&=
    \min_{\substack{
    \mT_{\text{CC}}(\mU_1, \mU_2)\in\mM_{\text{CC}} 
    \\ 
    \mT_{\text{DC}}(\mU_3, \mU_4)\in\mM_{\text{DC}}
    }}
    \min_{\Phi}
    \Big\{
    H(\mT_{\text{CC}}(\mU_1, \mU_2))_{\Phi}
    +
    H(\mT_{\text{CC}}(\mU_1, \mU_2))_{\Phi}
    \Big\}\label{eq:B}\\
    &=
    \min_{
    \mU_1, \mU_2, \mU_3, \mU_4
    }
    \min_{\Phi}
    \Big\{
    H(\mT_{\text{CC}}(\mU_1, \mU_2))_{\Phi}
    +
    H(\mT_{\text{CC}}(\mU_1, \mU_2))_{\Phi}
    \Big\}.
\end{align}
We note that the characterization of $\mB$ identifies the uncertainty trade-off between two families of interactive measurements, leading to a quantitative connection between different causal structures -- common-cause and direct-cause -- in quantum theories.

Roughly speaking, the universal uncertainty relation investigated in Subsec.~\ref{subsec:QR} is the key to solving $\mB$. In the first stage, we collect all the measurement data into a set $\Q$, and study its algebraic properties. Formally, the set $\Q$ is defined as
\begin{align}
    \Q:=
    \{
    \p(\mU_1, \mU_2)_{\Phi}
    \oplus
    \q(\mU_3, \mU_4)_{\Phi}
    \}
    _{\mU_1, \ldots, \mU_4, \Phi},
\end{align}
and we denote the ordered version of $\Q$ as $\Q_1$, namely
\begin{align}\label{eq:Q_1}
    \Q_1:= \Q^{\downarrow}
    =
    \{
    \p(\mU_1, \mU_2)_{\Phi}
    \oplus
    \q(\mU_3, \mU_4)_{\Phi}
    \}^{\downarrow}
    _{\mU_1, \ldots, \mU_4, \Phi}
    ,
\end{align}
where the down-arrow $^{\downarrow}$ indicates that all the elements of $\Q$ are arranged in non-increasing order. In qubit case, $\Q_1$ forms a subset of $\mathds{P}_{2}^{8, \, \downarrow}$, i.e. $\Q_1\subset\mathds{P}_{2}^{8, \, \downarrow}$. Here we have three remarks: First, instead of investigating $\frac{1}{2}\p(\mU_1, \mU_2)_{\Phi}\oplus\frac{1}{2}\q(\mU_3, \mU_4)_{\Phi}$, we work on $\p(\mU_1, \mU_2)_{\Phi}\oplus\q(\mU_3, \mU_4)_{\Phi}$ directly to simplify our representation. Second, according to the definition of majorization, the least upper bound (LUB) of $\Q_1$ is also a LUB of $\Q$ itself. Thanks to the completeness of $(\mathds{P}_{2}^{8, \, \downarrow}, \prec, \wedge, \vee)$ (see Subsec.~\ref{subsec:maj}), there exist a unique LUB, denoted by $\vee\Q_1$, in $\mathds{P}_{2}^{8, \, \downarrow}$; that is
\begin{align}
    (\vee\Q_1)_k
    \geqslant
    (\vee\Q_1)_{k+1}, 
    \quad
    \text{for}
    \,\,
    1\leqslant k\leqslant 7,
\end{align}
where $(\vee\Q_1)_k$ denotes the $k$-th element of $\vee\Q_1$. Third, the LUB of $\Q$ is not unique in the probability simplex. To be more precise, for any permutation matrix $D$, vector $(\vee\Q_1)\cdot D$ forms a LUB for set $\Q$. Another useful set is $\Q_2$, which is derived by taking all unitary channels in maximal common-cause indicator and maximal direct-cause indicator as noiseless channel, i.e. $\id$.
\begin{align}\label{eq:Q_2}
    \Q_2:=
    \{\p(\id, \id)_{\Phi}\oplus\q(\id, \id)_{\Phi}\}^{\downarrow}_{\Phi}.
\end{align}
Clearly, the newly constructed $\Q_2$ forms a subset of $\Q_1$, satisfying 
\begin{align}
\Q_2\subset\Q_1\subset\mathds{P}_{2}^{8, \, \downarrow}.    
\end{align}
In addition to inclusion, a key observation about the sets $\Q_1$ and $\Q_2$ is given in the following lemma. 

\begin{mylem}
{}{Second}
Given maximal common-cause indicators $\mT_{\text{CC}}(\mU_1, \mU_2)\in\mM_{\text{CC}}$ (see Def.~\ref{def:MCCI}), and maximal direct-cause indicators $\mT_{\text{DC}}(\mU_3, \mU_4)\in\mM_{\text{DC}}$ (see Def.~\ref{def:MDCI}), the largest sum of any two elements in $\vee\Q_1$ (see Eq.~\ref{eq:Q_1}) is achieved by only considering noiseless channels, namely
\begin{align}
    \sum_{k=1}^{2}
    (\vee\Q_1)_k
    =
    \sum_{k=1}^{2}
    (\vee\Q_2)_k,
\end{align}
with
\begin{align}
    \sum_{k=1}^{2}
    (\vee\Q_1)_k
    &:=
    \sum_{k=1}^{2}
    (\vee
    \{
    \p(\mU_1, \mU_2)_{\Phi}\oplus\q(\mU_3, \mU_4)_{\Phi}
    \}
    ^{\downarrow}
    _{\mU_1, \ldots, \mU_4, \Phi})_{k},\\
    \sum_{k=1}^{2}
    (\vee\Q_2)_k
    &:=
    \sum_{k=1}^{2}
    (\vee
    \{
    \p(\id, \id)_{\Phi}\oplus\q(\id, \id)_{\Phi}
    \}
    ^{\downarrow}
    _{\Phi})_{k}.
\end{align}
Here the down-arrow notation $^{\downarrow}$ means that the components of corresponding vector are arranged in non-increasing order. The probability vectors are defined by $\p(\mU_1, \mU_2)_{\Phi}:= (p_{i}(\mU_1, \mU_2)_{\Phi})_i$ and $\q(\mU_3, \mU_4)_{\Phi}:= (q_{j}(\mU_3, \mU_4)_{\Phi})_{j}$, with each $p_{i}(\mU_1, \mU_2)_{\Phi}$ and $q_{j}(\mU_3, \mU_4)_{\Phi}$ being obtained from Eqs.~\ref{eq:pi-u} and \ref{eq:qj-u} respectively. 
\end{mylem}

\begin{figure}[h]
\includegraphics[width=1\textwidth]{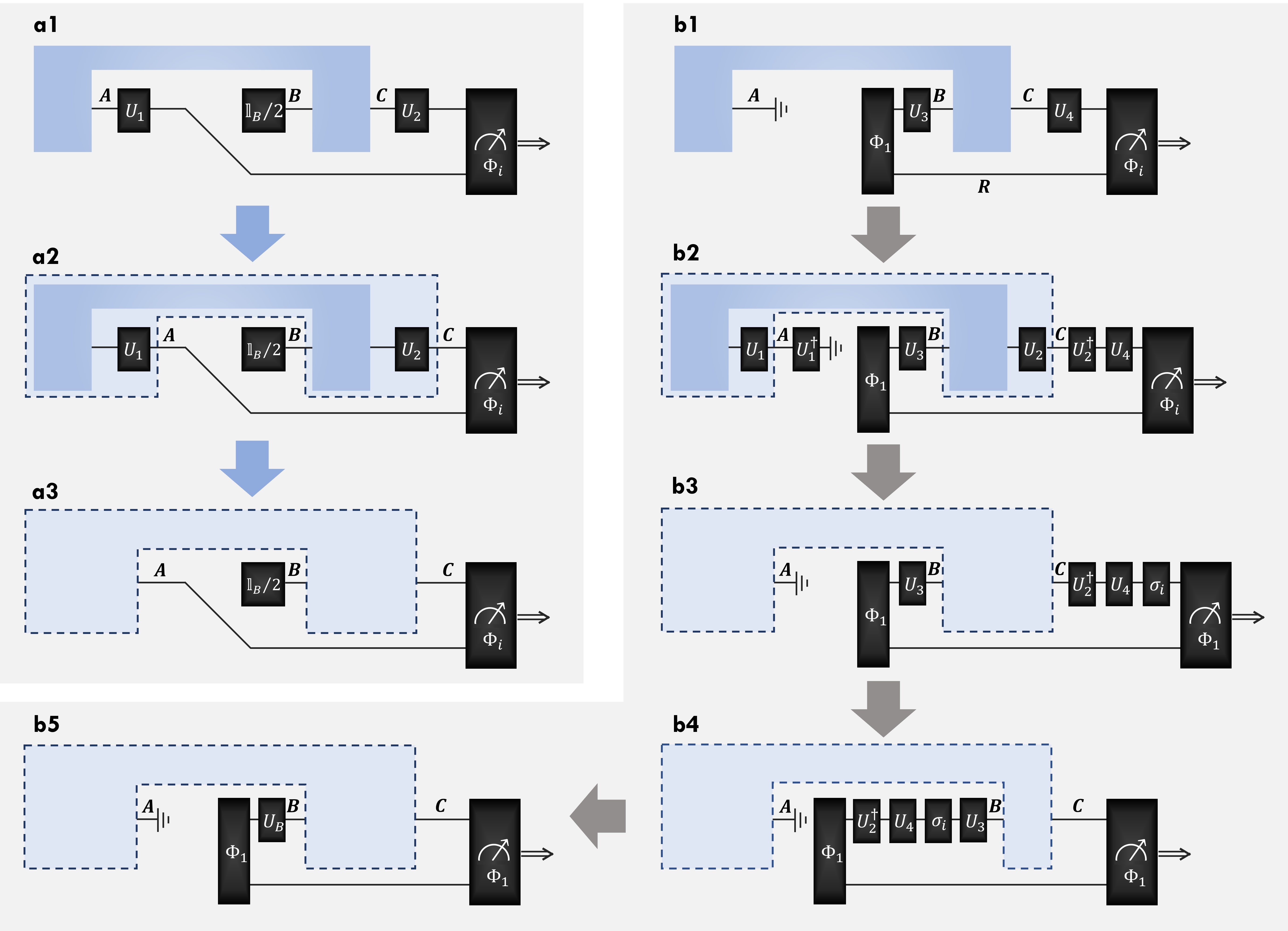}
\caption{(color online) Visualization of the process transforming Eq.~(\ref{eq:four-uni}) to Eq.~(\ref{eq:one-uni}). By absorbing $\mU_1$ and $\mU_2$ into the quantum causal map, we transform the uncertainty relation associated with interactive measurements, illustrated by (a1) and (b1), into the uncertainty relation for (a2) and (b2). As the optimization is taken over all causal maps, we can simply package the quantum dynamics inside the blue dashed box of (a2) and (b2), and reformulate them into (a3) and (b3). The transformation of (b3)$\rightarrow$(b4)$\rightarrow$(b5) follows directly from the properties of Bell state. Noted that from above visualization, it is clear that Eq.~(\ref{eq:one-uni}) can be further simplified to the following semidefinite programming (SDP) $\max_{i}\max_{\Phi\in\mathfrak{F}_2}
    \Tr[((
    \frac{\mathds{1}_B}{2}\otimes 
    (\ketbra{\Phi_i}{\Phi_i} )^{\T}
    +
    \mathds{1}_A \otimes \ketbra{\Phi_1}{\Phi_1}_{BR}  \star \ketbra{\Phi_1}{\Phi_1}_{CR}^{\T}
    )
    \cdot J^{\Phi}
    ]$, whose numerical value is $5/4$.}
\label{fig:equivalent}
\end{figure}

\begin{proof}
Remark that for both $\Q_1$
and $\Q_2$, their largest element is $1$; that is
\begin{align}
    (\vee\Q_1)_1
    =
    \bb_{\Q_1, 1}
    =
    (\vee\Q_2)_1
    =
    \bb_{\Q_2, 1}
    =
    1,
\end{align}
where $\bb_{S}$ is defined in Eq.~\ref{eq:step-one}, with $S= \Q_b$ ($b=1, 2$). 

If we only pick up two elements from $\p(\mU_1, \mU_2)_{\Phi}$ or $\q(\mU_3, \mU_4)_{\Phi}$, their summation is still upper-bounded by $1$. Thus, to investigate the largest sum of any two elements in $\p(\mU_1, \mU_2)_{\Phi}\oplus\q(\mU_3, \mU_4)_{\Phi}$, we should select one element from $\p(\mU_1, \mU_2)_{\Phi}$, and choose another one from $\q(\mU_3, \mU_4)_{\Phi}$. Similar situation applies to $\p(\id, \id)_{\Phi}$ and $\q(\id, \id)_{\Phi}$. Writing everything out explicitly, we have
\begin{align}
    &\sum_{k=1}^{2} \bb_{\Q_1, k}\\
    =
    &\max_{\mU_1, \ldots, \mU_4}\max_{i, j}\max_{\Phi\in\mathfrak{F}_2}
    \Tr[(
    J^{\text{CC}}_{i}(\mU_1, \mU_2)
    + 
    J^{\text{DC}}_{j}(\mU_3, \mU_4)
    ) \cdot J^{\Phi}]\\
    =
    &\max_{\mU_1, \ldots, \mU_4}\max_{i, j}\max_{\Phi\in\mathfrak{F}_2}
    \Tr[(
    \frac{\mathds{1}_B}{2}\otimes (U_{1}^{\dagger}\otimes U_{2}^{\dagger} \ketbra{\Phi_i}{\Phi_i} U_1 \otimes U_2)^{\T} + 
    \mathds{1}_A \otimes (U_3 \ketbra{\Phi_1}{\Phi_1}_{BR} U_{3}^{\dagger}) \star (U_4^{\dagger} \ketbra{\Phi_j}{\Phi_j}_{CR} U_{4})^{\T}
    ) \cdot J^{\Phi}]\label{eq:four-uni}\\
    =
    &\max_{\mU_B}\max_{i, j}\max_{\Phi\in\mathfrak{F}_2}
    \Tr[(
    \frac{\mathds{1}_B}{2}\otimes 
    (\ketbra{\Phi_i}{\Phi_i} )^{\T}
    +
    \mathds{1}_A \otimes (U_B \ketbra{\Phi_1}{\Phi_1}_{BR} U_B^{\dagger} ) \star \ketbra{\Phi_j}{\Phi_j}_{CR}^{\T}
    )
    \cdot J^{\Phi}
    ]\label{eq:one-uni}\\
    =
    &\max_{i, j}\max_{\Phi\in\mathfrak{F}_2}
    \Tr[(
    \frac{\mathds{1}_B}{2}\otimes 
    (\ketbra{\Phi_i}{\Phi_i} )^{\T}
    +
    \mathds{1}_A \otimes \ketbra{\Phi_1}{\Phi_1}_{BR}  \star \ketbra{\Phi_j}{\Phi_j}_{CR}^{\T}
    )
    \cdot J^{\Phi}
    ]\label{eq:sdp54}\\
    =
    &\sum_{k=1}^{2} \bb_{\Q_2, k}\\
    =
    &\frac{5}{4},
\end{align}
where the third equation is obtained by absorbing unitaries into the causal map (visualized in Fig.~\ref{fig:equivalent}), and the fourth equation is a direct consequence of $U_B\,U_B^{\dagger}=\mathds{1}$. Finally, the last equation is the result of SDP formulated in Eq.~(\ref{eq:sdp54}). This immediately implies that for the set $\Q_1$, we have
\begin{align}\label{q1-q2-geq}
    \sum_{k=1}^{2}
    (\vee\Q_1)_k
    \leqslant
    \sum_{k=1}^{2} \bb_{\Q_1, k}
    =
    \sum_{k=1}^{2} \bb_{\Q_2, k}
    =
    \frac{5}{4}.
\end{align}
On the other hand, by taking $\Phi= \Tr_{E}[\ketbra{\Phi_1}{\Phi_1}_{AE}] \otimes \id_{B\to C}$, maximal common-cause indicators $\mT_{\text{CC}}(\id, \id)\in\mM_{\text{CC}}$, and maximal direct-cause indicators $\mT_{\text{DC}}(\id, \id)\in\mM_{\text{DC}}$, we obtain
\begin{align}
    \p(\id, \id)_{\Tr_{E}[\ketbra{\Phi_1}{\Phi_1}_{AE}] \otimes \id_{B\to C}}
    &=
    (1/4, 1/4, 1/4, 1/4),\label{eq:unitary-p}\\
    \q(\id, \id)_{\Tr_{E}[\ketbra{\Phi_1}{\Phi_1}_{AE}] \otimes \id_{B\to C}} 
    &= 
    (1, 0, 0, 0).\label{eq:unitary-q}
\end{align}
In this case, we have
\begin{align}
    (1, 1/4, 1/4, 1/4, 1/4, 0, 0, 0)
    \in
    \Q_2
    \subset
    \Q_1,
\end{align}
and the largest sum of any two elements in both $\vee\Q_1$ and $\vee\Q_2$ are lower bounded by $5/4$, i.e.
\begin{align}\label{q1-q2-leq}
    \frac{5}{4}
    \leqslant
    \sum_{k=1}^{2}
    (\vee\Q_2)_k
    \leqslant
    \sum_{k=1}^{2}
    (\vee\Q_1)_k.
\end{align}
Combining Eq.~\ref{q1-q2-geq} with Eq.~\ref{q1-q2-leq} leads to the following equation,
\begin{align}\label{eq:vee-sum-2}
    \sum_{k=1}^{2}
    (\vee\Q_1)_k
    =
    \sum_{k=1}^{2}
    (\vee\Q_2)_k
    =
    \frac{5}{4},
\end{align}
what is to be shown.

\end{proof}

As a by-product of Eq.~\ref{eq:vee-sum-2}, it is now straightforward to see that
\begin{align}
    (\vee\Q_1)_2
    =
    \sum_{k=1}^{2}
    (\vee\Q_1)_k
    -
    (\vee\Q_1)_1
    =
    \frac{5}{4}- 1
    =
    \frac{1}{4}.
\end{align}
Then, it can be shown that

\begin{mythm}
{}{all-vector}
Given maximal common-cause indicators $\mT_{\text{CC}}(\mU_1, \mU_2)\in\mM_{\text{CC}}$ (see Def.~\ref{def:MCCI}), and maximal direct-cause indicators $\mT_{\text{DC}}(\mU_3, \mU_4)\in\mM_{\text{DC}}$ (see Def.~\ref{def:MDCI}), the LUB $\vee\Q_1$ (see Eq.~\ref{eq:Q_1}) is given by
\begin{align}\label{eq:vq-vector}
    \vee\Q_1
    =
    (1, 1/4, 1/4, 1/4, 1/4, 0, 0, 0).
\end{align}
\end{mythm}

\begin{proof}
From the analysis of Lem.~\ref{lem:Second}, we know that the largest element of $\vee\Q_1$ is $1$ (i.e. $(\vee\Q_1)_1= 1$), and the second largest element of $\vee\Q_1$ is $1/4$ (i.e. $(\vee\Q_1)_2= 1/4$). Consider an eigencircuit $\Phi$ of $\mT_{\text{DC}}(\mU_3, \mU_4)$. Without loss of generality, we assume that $q_{1}(\mU_3, \mU_4)_{\Phi}= 1$. For example, see Eqs.~\ref{eq:unitary-p} and \ref{eq:unitary-q}. In this case, it follows immediately that
\begin{align}
    2=
    \sum_{i}p_{i}(\mU_1, \mU_2)_{\Phi}
    +
    q_{1}(\mU_3, \mU_4)_{\Phi}
    \leqslant
    \sum_{k=1}^{5}
    (\vee\Q_1)_k
    \leqslant
    \sum_{k=1}^{8}
    (\vee\Q_1)_k
    =
    2,
\end{align}
and hence we have
\begin{align}
    (\vee\Q_1)_6
    =
    (\vee\Q_1)_7
    =
    (\vee\Q_1)_8
    =
    0.
\end{align}
Let us now come back to $(\vee\Q_1)_k$ ($k= 3, 4, 5$). Since the elements of $\vee\Q_1$ are arranged in non-increasing order, it follows that
\begin{align}
    (\vee\Q_1)_5
    \leqslant
    (\vee\Q_1)_4
    \leqslant
    (\vee\Q_1)_3
    \leqslant
    (\vee\Q_1)_2
    =
    \frac{1}{4}.
\end{align}
If any of them is smaller than $1/4$, then $\sum_{k=1}^{8}(\vee\Q_1)_k< 2$, which is a contradiction. Therefore, the LUB $\vee\Q_1$ of $\Q_1$ is given by Eq.~\ref{eq:vq-vector} as required.
\end{proof}

In the second stage, equipped with Thm.~\ref{thm:all-vector}, we now introduce the universal uncertainty relation for maximal common-cause indicators and maximal direct-cause indicators, which is characterized by the following theorem.

\begin{mythm}
{Universal Causal Uncertainty Relation}{universal-CUR}
Given a quantum causal map $\Phi:B\to AC$ with $d_A= d_B= d_C= 2$, let us denote the probability vectors obtained by measuring $\Phi$ with respect to 
maximal common-cause indicators $\mT_{\text{CC}}(\mU_1, \mU_2)\in\mM_{\text{CC}}$ (see Def.~\ref{def:MCCI}) and maximal direct-cause indicators $\mT_{\text{DC}}(\mU_3, \mU_4)\in\mM_{\text{DC}}$ (see Def.~\ref{def:MDCI}) as $\p(\mU_1, \mU_2)_{\Phi}$ and $\q(\mU_3, \mU_4)_{\Phi}$ respectively. Then for any causal map $\Phi\in\mathfrak{F}_2$ and unitary channels $\mU_{b}$ ($b\in\{1, 2, 3, 4\}$), we have the following universal causal uncertainty relation
\begin{align}\label{eq:uni-causal}
    \frac{1}{2}
    \p(\mU_1, \mU_2)_{\Phi}
    \oplus
    \frac{1}{2}
    \q(\mU_3, \mU_4)_{\Phi}
    \prec
    \frac{1}{2}
    \cdot
    (\vee\Q_1)
    =
    (1/2, 1/8, 1/8, 1/8, 1/8, 0, 0, 0).
\end{align}
In particular, by applying Schur-concave functions, we can generate a family of infinite causal uncertainty relations from Eq.~\ref{eq:uni-causal}.
\end{mythm}

Now, direct calculation implies that
\begin{align}
    2H(\frac{1}{2}
    \cdot
    (\vee\Q_1))- 2
    =
    2.
\end{align}
Since the Shannon entropy $H$ is a Schur-concave function, we hence obtain the following entropic uncertainty relation for maximal common-cause indicators and maximal direct-cause indicators

\begin{mycor}
{Entropic Causal Uncertainty Relation}{entropic-CUR}
Given maximal common-cause indicators $\mT_{\text{CC}}(\mU_1, \mU_2)\in\mM_{\text{CC}}$ (see Def.~\ref{def:MCCI}) and maximal direct-cause indicators $\mT_{\text{DC}}(\mU_3, \mU_4)\in\mM_{\text{DC}}$ (see Def.~\ref{def:MDCI}) acting on some qubit quantum causal map $\Phi\in\mathfrak{F}_2$. The entropy of their measurement outcomes, when summed, satisfies
\begin{align}\label{eq:entropic-CUR}
    H(\mT_{\text{CC}}(\mU_1, \mU_2))_{\Phi}
    +
    H(\mT_{\text{CC}}(\mU_1, \mU_2))_{\Phi}
    \geqslant
    2,
\end{align}
where the uncertainties $H(\p(\mU_1, \mU_2)_{\Phi})$ and $H(\q(\mU_3, \mU_4))_{\Phi}$ are defined by Eqs.~\ref{eq:H-CC} and \ref{eq:H-DC} respectively. 
\end{mycor}

In particular, for the case with $\Phi= \Tr_{E}[\ketbra{\Phi_1}{\Phi_1}_{AE}] \otimes \id_{B\to C}$ and $\mU_1= \mU_2= \mU_3= \mU_4= \id$, the joint uncertainty is $2$, i.e.
\begin{align}
    H(\mT_{\text{CC}}(\id, \id))_{\Tr_{E}[\ketbra{\Phi_1}{\Phi_1}_{AE}] \otimes \id_{B\to C}}
    +
    H(\mT_{\text{CC}}(\id, \id))_{\Tr_{E}[\ketbra{\Phi_1}{\Phi_1}_{AE}] \otimes \id_{B\to C}}
    =
    2.
\end{align}
Thus, in Eq.~\ref{eq:entropic-CUR} of Cor.~\ref{cor:entropic-CUR}, the lower bound $2$ is optimal.
In other words, here $\mB= 2$ (see Eq.~\ref{eq:B}). All results presented in this subsection, including Lem.~\ref{lem:Second}, Thm.~\ref{thm:all-vector}, Thm.~\ref{thm:universal-CUR}, and Cor.~\ref{cor:entropic-CUR}, for qubit systems can be extended to the case of qudit systems straightforwardly by replacing the Pauli operator appeared in Bell measurements and Eq.~(\ref{eq:unitary-p}) with Heisenberg-Weyl operators and $(1/d^2,\ldots,1/d^2)$ respectively. In particular that, for the case with $d_A= d_B= d_C= d$, we have
\begin{align}
   H( \mathcal{T}_{\text{CC}}(\mU_1, \mU_2) )_{\Phi_{B\rightarrow AC}}
+
H( \mathcal{T}_{\text{DC}}(\mU_3, \mU_4) )_{\Phi_{B\rightarrow AC}}
\geqslant
2\log d,
\end{align}
where $\mU_1$, $\mU_2$, $\mU_3$, and $\mU_4$ are now $d$-dimensional unitary channels.


\subsection{\label{subsec:NS}Necessary and Sufficient Conditions for Common-Cause and Direct-Cause}

In this subsection, we give a complete characterization of when the quantum causal map $\Phi\in\mathfrak{F}_2$ obtained from system-environment unitary dynamics in open quantum
systems (see Eq.~\ref{eq:phi'decom}) indicates a common-cause, direct-cause, or even their mixture. The method of our causal inference is based on  the entropic causal uncertainty relation, namely Eq.~\ref{eq:entropic-CUR}. 

\begin{figure}[h]
\includegraphics[width=0.5\textwidth]{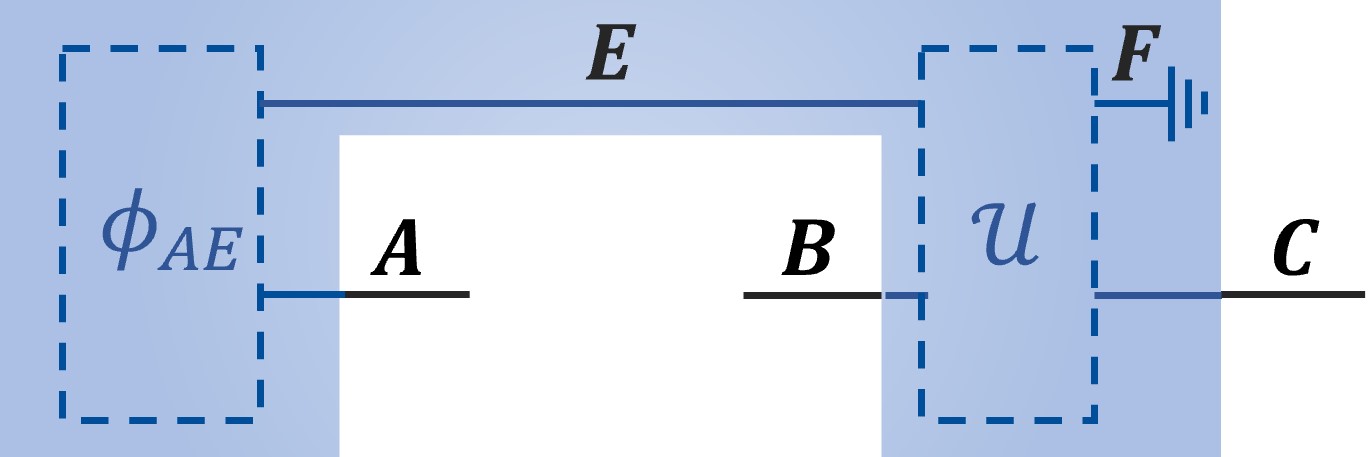}
\caption{(color online) Quantum causal map $\Phi_{B\to AC}$, where the initial state is $\phi_{AE}$, and the systems $BE$ and $CF$ are connected through the bipartite unitary channel $\mathcal{U}_{BE\to CF}$.}
\label{fig:causal-map-without}
\end{figure}

In particular, given a quantum causal map $\Phi\in\mathfrak{F}_2$, its quantum dynamics is given by a initial pure state $\phi_{AE}$ and a system-environment unitary evolution $\mU_{BE\to CF}$ (see Fig.~\ref{fig:causal-map-without} for an illustration). Practically, we do not have access to the environmental systems. Hence, $\Phi\in\mathfrak{F}_2$ is characterized by
\begin{align}
    \Phi_{B\to Ac}
    =
    \Tr_{F}
    [\mU_{BE\to CF}(\phi_{AE})],
\end{align}
where $E$ and $F$ stand for the environment before and after the  unitary evolution $\mU$. Meanwhile, $A$, $B$, and $C$ represent the system of interest at different time points $t_A$, $t_B$, and $t_C$ respectively, with $\dim A= \dim B= \dim C$.

Causal understanding of the quantum dynamics, especially the system-environment unitary dynamics , enables us to reason about the environment and its interactions with the system of interest at multiple time points. Here, our primary goal is to discover the structural dependencies among environmental and object systems, namely $\mU_{BE\to CF}(\phi_{AE})$. This is a particularly challenging task -- Typically, quantum tomography of $\mU_{BE\to CF}(\phi_{AE})$ offers us a general solution. But it requires lots of experimental effort, including a large number of measurements, an extensive analysis of data, and etc. What is even worse is that usually we do not have complete access to all information in the environment. A main contribution of this work lies in the quantum causal inference without tomography. To do do, let us first introduce the definitions of purely common-cause, purely direct-cause, and their mixture for $\mU_{BE\to CF}(\phi_{AE})$; that are

\begin{mydef}
{Common-Cause}{CC}
For a system-environment unitary dynamics $\mU_{BE\to CF}(\phi_{AE})$, its causal structure is purely common-cause, if the following condition holds
\begin{align}\label{eq:cc-decom}
    \mU_{BE\to CF}(\phi_{AE})
    =
    \rho_{AC}\otimes\mE_{B\to F},
\end{align}
where $\rho_{AC}$ is an entangled state acting on systems $AC$, and $\mE_{B\to F}$ is a quantum channel from system $B$ to $F$.
\end{mydef}

In the case where the quantum dynamics is constituted of pure initial state and unitary channel, like $\mU_{BE\to CF}(\phi_{AE})$, its CJ operator is a rank-$1$ operator, implying that the operator $\rho_{AC}\otimes J^{\mE}_{BF}$ is also rank-$1$. In other words, in Eq.~\ref{eq:cc-decom}, $\rho_{AC}$ must be a pure state and meanwhile $\mE_{B\to F}$ must be a unitary channel. Thus, written out explicitly, we have the following lemma.

\begin{mylem}
{Common-Cause}{CC}
For a system-environment unitary dynamics $\mU_{BE\to CF}(\phi_{AE})$, its causal structure is purely common-cause, if the following condition holds
\begin{align}\label{eq:cc-decom}
    \mU_{BE\to CF}(\phi_{AE})
    =
    \psi_{AC}\otimes\mU_{B\to F},
\end{align}
where $\psi_{AC}$ is a pure entangled state acting on systems $AC$, and $\mU_{B\to F}$ is a unitary channel from system $B$ to $F$.
\end{mylem}
When measuring $\mU_{BE\to CF}(\phi_{AE})$, if an outcome (without loss of generality, we take $1$ for instance) of $\mT_{\text{CC}}(\mU_1, \mU_2)\otimes\Tr_{F}$ happens with certainty, namely
\begin{align}
    \Tr[
    \Phi_1\otimes\1_{F}
    \cdot
    \mU_2
    \circ
    \mU_{BE\to CF}
    \left(
    \mU_{1}(\psi_{AE})
    \otimes\frac{\1_{B}}{d_{B}}
    \right)
    ]
    =
    1.
\end{align}
Then, according to the definition of maximal common-cause indicator $\mT_{\text{CC}}(\mU_1, \mU_2)$ (see Def.~\ref{def:MCCI}), it is straightforward to check that in this case we have $\mU_{BE\to CF}(\phi_{AE})= \mU_1^{\dag}\otimes\mU_2^{\dag}(\phi^{+}_{AC})\otimes\mU_{B\to F}$, where
$\phi^{+}$ is the maximally entangled state, and $\mE^{\dag}$ stands for the dual channel of $\mE$. Conversely, if system-environment unitary dynamics $\mU_{BE\to CF}(\phi^{+}_{AE})$ exhibits common-cause, where $\phi^{+}$ represents the maximally entangled state, then we can always find one maximal common-cause indicator $\mT_{\text{CC}}(\mU_1, \mU_2)\in\mM_{\text{CC}}$, such that a measurement outcome of $\mT_{\text{CC}}(\mU_1, \mU_2)\otimes\Tr_{F}$ will occur with certainty. Thus, in the case of $\mU_{BE\to CF}(\phi^{+}_{AE})$, we have the following necessary and sufficient condition for quantum dynamics to be purely common-cause; that is

\begin{mythm}
{Common-Cause}{CC}
For a system-environment unitary dynamics $\mU_{BE\to CF}(\phi^{+}_{AE})$ with $\phi^{+}_{AE}$ being the maximally entangled state, its causal structure is purely common-cause, if and only if there exist unitary channels $\mU_1$ and $\mU_2$ such that
\begin{align}
    H(\mT_{\text{CC}}(\mU_1, \mU_2)\otimes\Tr_F)_{\mU_{BE\to CF}(\phi^{+}_{AE})}
    =
    H(\mT_{\text{CC}}(\mU_1, \mU_2))_{\Tr_F[\mU_{BE\to CF}(\phi^{+}_{AE})]}
    =
    0.
\end{align}
That is, an outcome of the maximal common-cause indicator $\mT_{\text{CC}}(\mU_1, \mU_2)\in\mM_{\text{CC}}$ happens with certainty for some unitary channels $\mU_1$ and $\mU_2$.
\end{mythm}

On the other hand, the definition of purely direct-cause for $\mU_{BE\to CF}(\phi_{AE})$ is given by

\begin{mydef}
{Direct-Cause}{DC}
For a system-environment unitary dynamics $\mU_{BE\to CF}(\phi_{AE})$, its causal structure is purely direct-cause, if the following condition holds
\begin{align}\label{eq:dc-decom-o}
    \mU_{BE\to CF}(\phi_{AE})
    =
    \rho_{AF}\otimes\mE_{B\to C},
\end{align}
where $\rho_{AF}$ is a quantum state acting on systems $AF$, and $\mE_{B\to C}$ is a quantum channel from system $B$ to $C$. Here the channel $\mE$ can not be the one constituted of tracing $B$ and then preparing a state on system $C$.
\end{mydef}

It is worth mentioning that if the system-environment unitary dynamics $\mU_{BE\to CF}(\phi_{AE})$ is decomposed as $\rho_A\otimes\mathcal{E}_{B\to CF}$, the causal structure associated with system-environment dynamics can also be defined as purely direct-cause. However, the unitarity of evolution $\mU_{BE\to CF}$ and the dimensional restriction on the systems (i.e. $\dim A= \dim B= \dim C$) will force the environment to be a trivial system, i.e. $F=\mathds{C}$. Hence, the system-environment unitary dynamics can still be described by Eq.~\ref{eq:dc-decom-o}. Similar to Lem.~\ref{lem:CC}, now the rank-$1$ property of $\mU_{BE\to CF}(\phi_{AE})$ implies that 

\begin{mylem}
{Direct-Cause}{DC}
For a system-environment unitary dynamics $\mU_{BE\to CF}(\phi_{AE})$, its causal structure is purely direct-cause, if the following condition holds
\begin{align}\label{eq:dc-decom}
    \mU_{BE\to CF}(\phi_{AE})
    =
    \psi_{AF}\otimes\mU_{B\to C},
\end{align}
where $\psi_{AF}$ is a pure quantum state acting on systems $AF$, and $\mU_{B\to C}$ is a unitary channel from system $B$ to $C$.
\end{mylem}

When measuring $\mU_{BE\to CF}(\phi_{AE})$, if an outcome (without loss of generality, we take $1$ for instance) of $\mT_{\text{DC}}(\mU_3, \mU_4)\otimes\Tr_{F}$ happens with certainty, namely
\begin{align}
    \Tr[
    \1_{AF}
    \otimes
    \Phi_{1, CR}
    \cdot
    \mU_4
    \circ
    \mU_{BE\to CF}\left(\phi_{AE}\otimes\mU_3(\Phi_{1, BR})\right)
    ]
    =
    1.
\end{align}
Then it is straightforward to conclude that $\mU_{BE\to CF}(\phi_{AE})$ can be decomposed as $\psi_{AF}\otimes(\mU_4^{\dag}\circ\mU_3^{\dag})$, hence reveals a direct-cause. On the other hand, if system-environment unitary dynamics $\mU_{BE\to CF}(\phi_{AE})$ exhibits a direct-cause, i.e. $\mU_{BE\to CF}(\phi_{AE})= \psi_{AF}\otimes\mU_{B\to C}$, then we have $H(\mT_{\text{DC}}(\id, (\mU_{B\to C})^\dag)\otimes\Tr_F)_{\mU_{BE\to CF}(\phi_{AE})}=0$, leading to the following theorem.

\begin{mythm}
{Direct-Cause}{DC}
For a system-environment unitary dynamics $\mU_{BE\to CF}(\phi_{AE})$, its causal structure is purely direct-cause, if and only if there exist unitary channels $\mU_3$ and $\mU_4$ such that
\begin{align}
    H(\mT_{\text{DC}}(\mU_3, \mU_4)\otimes\Tr_F)_{\mU_{BE\to CF}(\phi_{AE})}
    =
    H(\mT_{\text{DC}}(\mU_3, \mU_4))_{\Tr_F[\mU_{BE\to CF}(\phi_{AE})]}
    =
    0.
\end{align}
That is, an outcome of the maximal direct-cause indicator $\mT_{\text{DC}}(\mU_3, \mU_4)\in\mM_{\text{DC}}$ happens with certainty for some unitary channels $\mU_3$ and $\mU_4$.
\end{mythm}

After we obtain the necessary and sufficient condition for system-environment unitary dynamics $\mU_{BE\to CF}(\phi^{+}_{AE})$ to be purely common-cause, we can now infer the causal model associated with $\mU_{BE\to CF}(\phi^{+}_{AE})$ without using tomography. More specifically, we would like to show that the causal uncertainty relation (see Eq.~\ref{eq:entropic-CUR}) implies causality. Here we take quantum dynamics with qubit system as an example, that is $d_A= d_B= d_C= 2$. In particular, for $\mU_{BE\to CF}(\phi^{+}_{AE})$, we have 

\begin{mycor}
{Non-Markovianity}{NM}
For a system-environment unitary dynamics $\mU_{BE\to CF}(\phi^{+}_{AE})$, if there exist some unitary channels $\mU_1$ and $\mU_2$, such that
\begin{align}
    2>
    H(\mT_{\text{CC}}(\mU_1, \mU_2)\otimes\Tr_F)_{\mU_{BE\to CF}(\phi^{+}_{AE})}
    >0.
\end{align}
Then the system-environment unitary dynamics $\mU_{BE\to CF}(\phi^{+}_{AE})$ is non-Markovian.
\end{mycor}

\begin{proof}
According to entropic uncertainty relation presented in Cor.~\ref{cor:entropic-CUR}, it follows that
\begin{align}
    H(\mT_{\text{DC}}(\mU_3, \mU_4))_{\Tr_F[\mU_{BE\to CF}(\phi_{AE})]}
    &\geqslant
    2- 
    H(\mT_{\text{CC}}(\mU_1, \mU_2))_{\Tr_F[\mU_{BE\to CF}(\phi^{+}_{AE})]}\\
    &=
    2-
    H(\mT_{\text{DC}}(\mU_3, \mU_4)\otimes\Tr_F)_{\mU_{BE\to CF}(\phi_{AE})}\\
    &>0,
\end{align}
holds for arbitrary unitary channels $\mU_3$ and $\mU_4$. Hence, Thm.~\ref{thm:DC} immediately implies that the system-environment unitary dynamics $\mU_{BE\to CF}(\phi^{+}_{AE})$ can either be purely common-cause, or the mixture of common-cause and direct-cause. But in either case, the initial state $\phi^{+}_{AE}$ will influence the quantum dynamics from $B$ to $C$. Thus, in this case, the system-environment unitary dynamics $\mU_{BE\to CF}(\phi^{+}_{AE})$ is non-Markovian.
\end{proof}

Let us further consider how to infer the mixture of common-cause and direct-cause by employing the entropic causal uncertainty relation formulated in Cor.~\ref{cor:entropic-CUR}.

\begin{mycor}
{Mixture of Common-Cause and Direct-Cause}{Mixture}
For a system-environment unitary dynamics $\mU_{BE\to CF}(\phi^{+}_{AE})$, if there exist some unitary channels $\mU_b$ ($b\in\{1, 2, 3, 4\}$), such that
\begin{align}
    2>
    &H(\mT_{\text{CC}}(\mU_1, \mU_2)\otimes\Tr_F)_{\mU_{BE\to CF}(\phi^{+}_{AE})}
    >0,\label{eq:0cc2}\\
    2>
    &H(\mT_{\text{DC}}(\mU_3, \mU_4)\otimes\Tr_F)_{\mU_{BE\to CF}(\phi^{+}_{AE})}
    >0.\label{eq:0dc2}
\end{align}
Then the system-environment unitary dynamics $\mU_{BE\to CF}(\phi^{+}_{AE})$ is a mixture of common-cause and direct-cause.
\end{mycor}

\begin{proof}
On the one hand, Eq.~\ref{eq:0cc2} implies that the causal structure of $\mU_{BE\to CF}(\phi^{+}_{AE})$ cannot be purely direct-cause (see Thm.~\ref{thm:DC}). On the other hand, Eq.~\ref{eq:0dc2} tells us that the causal model of $\mU_{BE\to CF}(\phi^{+}_{AE})$ will not be pure common-cause (see Thm.~\ref{thm:CC}). Thus, the system-environment unitary dynamics $\mU_{BE\to CF}(\phi^{+}_{AE})$ should be a mixture of common-cause and direct-cause, which completes the proof.
\end{proof}

From Cor.~\ref{cor:Mixture}, we know that by using the entropic causal uncertainty relation (see Eq.~\ref{eq:entropic-CUR}), the causal structure of system-environment unitary dynamics $\mU_{BE\to CF}(\phi^{+}_{AE})$ can be determined by implementing only two interactive measurements, namely a maximal common-cause indicator $\mT_{\text{CC}}(\mU_1, \mU_2)\in\mM_{\text{CC}}$ (see Def.~\ref{def:MCCI}) and a maximal direct-cause indicator $\mT_{\text{DC}}(\mU_3, \mU_4)\in\mM_{\text{DC}}$ (see Def.~\ref{def:MDCI}).


\subsection{\label{subsec:coh}Coherent Mixture of Common-Cause and Direct-Cause}

Most things in nature are mixtures. Classically, two or more substances can be combined in a probabilistic way, such as the  convex combination of probability density functions in statistics. Quantumly, operations or states of the systems can be controlled coherently, such as the superpositions of trajectories (i.e. quantum switch~\cite{PhysRevA.88.022318}) in quantum communication~\cite{PhysRevLett.120.120502,PhysRevLett.127.190502,PhysRevResearch.3.013093}. In quantum causal inference, the models of pure common-cause and purely direct-cause can also be combined in a coherent way~\cite{causal-NC}. In this subsection, we further investigate the quantum circuit (which is also known as quantum causal map in this work) $\Phi_{\alpha,\beta}\in\mathfrak{F}_2$ considered in the main text (see also Fig. 5 of the main text), and show that $\Phi_{\alpha= -\pi/4, \beta= -\pi/2}$ is indeed a coherent mixture purely common-cause and purely direct-cause.

In quantum circuit $\Phi_{\alpha,\beta}= \Tr_{F}[\mU(\alpha, \beta)(\phi^{+}_{AE})]\in\mathfrak{F}_2$, the unitary channel $\mU(\alpha, \beta)$ from systems $BE$ to $CF$ is given by $\mU(\alpha, \beta)(\cdot)= U(\alpha,\beta)(\cdot)U^{\dag}(\alpha,\beta)$, where the unitary matrix $U(\alpha,\beta)$ is characterized by the following form
\begin{align}
U(\alpha,\beta) =e^{i\alpha/2}
\begin{pmatrix} 
e^{-i\alpha} 
& 0 & 0 & 0 \\ 
0 &\cos(\beta/2) & -i\sin(\beta/2) & 0 \\ 
0 &-i\sin(\beta/2) & \cos(\beta/2) & 0 \\ 
0 & 0 & 0 & e^{-i\alpha} 
\end{pmatrix}.
\end{align}
When the parameters are taken as $\alpha= -\pi/4$ and $\beta= -\pi/2$, the corresponding unitary matrix turns out to be
\begin{align}
U(\alpha= -\pi/4, \beta= -\pi/2) =e^{-i\pi/8}
\begin{pmatrix} 
e^{i\pi/4} 
& 0 & 0 & 0 \\ 
0 & \cos(\pi/4) & i\sin(\pi/4) & 0 \\ 
0 &i\sin(\pi/4) & \cos(\pi/4) & 0 \\ 
0 & 0 & 0 & e^{i\pi/4} 
\end{pmatrix}.
\end{align}
Thus in this case, written out explicitly, the quantum circuit is given by
\begin{align}
\Phi_{\alpha= -\pi/4, \beta= -\pi/2}
=
\Tr_{F} [\mathcal{U}(\alpha= -\pi/4, \beta= -\pi/2)_{BE\to CF} (\phi^{+}_{AE})].
\end{align}
Meanwhile, the existence of coherent mixture of causal models of purely common-cause and purely direct-cause has been proved by exampling the quantum circuit $\Tr_{F} [\mU_{\ast} (\phi^{+}_{AE})]$ \cite{causal-NC}, where the unitary $\mU_{\ast}$ is defined as
\begin{align}\label{eq:pswap}
\mU_{\ast}= (\cos{\pi/4})\text{id}_{B\to C}\otimes\text{id}_{E\to F}+(i\sin{\pi/4})\cdot\text{SWAP}(BE\to FC).
\end{align}
Noted that here $\mU_{\ast}$ forms a partial SWAP from $BE$ to $CF$. For any input state $\cdot$, we have $\mU_{\ast}(\cdot)= U_{\ast}(\cdot)U_{\ast}^{\dag}$, where
\begin{align}
U_{\ast} =
\begin{pmatrix} 
e^{i\pi/4} 
& 0 & 0 & 0 \\ 
0 & \cos(\pi/4) & i\sin(\pi/4) & 0 \\ 
0 &i\sin(\pi/4) & \cos(\pi/4) & 0 \\ 
0 & 0 & 0 & e^{i\pi/4} 
\end{pmatrix}.
\end{align}
Remark that for any input state $\rho$ acting on systems $BE$, we have 
\begin{align}
U_{\ast} \rho U_{\ast}^{\dagger}
=
U(\alpha= -\pi/4, \beta= -\pi/2) \rho U(\alpha= -\pi/4, \beta= -\pi/2)^{\dagger}.
\end{align}
It now follows immediately that
\begin{align}
\Phi_{\alpha= -\pi/4, \beta= -\pi/2}=\Tr_{F} [\mathcal{U}_{\ast} (\phi^{+}_{AE})]. 
\end{align}
In other words, the quantum circuit $\Phi_{\alpha= -\pi/4, \beta= -\pi/2}$ studied in Fig. 5(c) of the main text is indeed a coherent mixture of both purely common-cause (see Def.~\ref{def:CC}) and purely direct-cause (see Def.~\ref{def:DC}), which goes beyond the classical way of mixing causal models in causal inference.

\begin{figure}[h]
\includegraphics[width=1\textwidth]{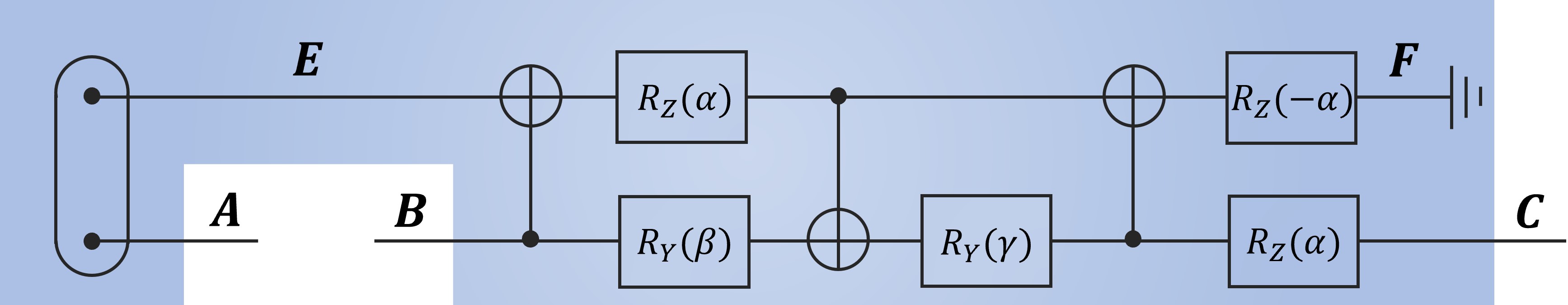}
\caption{(color online) Parameterized quantum circuit $\Phi_{\alpha, \beta, \gamma}= \Tr_{F}[\mU(\alpha, \beta, \gamma)(\phi^{+}_{AE})]\in\mathfrak{F}_2$, with each parameter controls a Pauli rotation. System $A$ and environment $E$ are initiated in the maximally entangled state $\phi^{+}_{AE}$. Meanwhile, the bipartite evolution $\mU(\alpha, \beta, \gamma)$ from systems $BE$ to $CF$ is an universal interaction between two qubits~\cite{PhysRevA.69.062321}.}
\label{fig:para-circuit} 
\end{figure}
\section{\label{sec:NE}Numerical Experiments}

In this section, we focus on a parameterized quantum circuit, and illustrate the uncertainty associated with this parameterized quantum circuit numerically. In Subsec.~\ref{subsec:land}, we demonstrate the entropic causal uncertainty relation with respect to a special pair of maximal common-cause indicator and maximal direct-cause indicator, showing that our lower bound is tight. In Subsec.~\ref{subsec:adv}, we detail the application of our entropic causal uncertainty relation to inferring causality by using few interactive measurements -- one maximal common-cause indicator and one maximal direct-cause indicator.



\begin{figure}[h]
\includegraphics[width=0.8\textwidth]{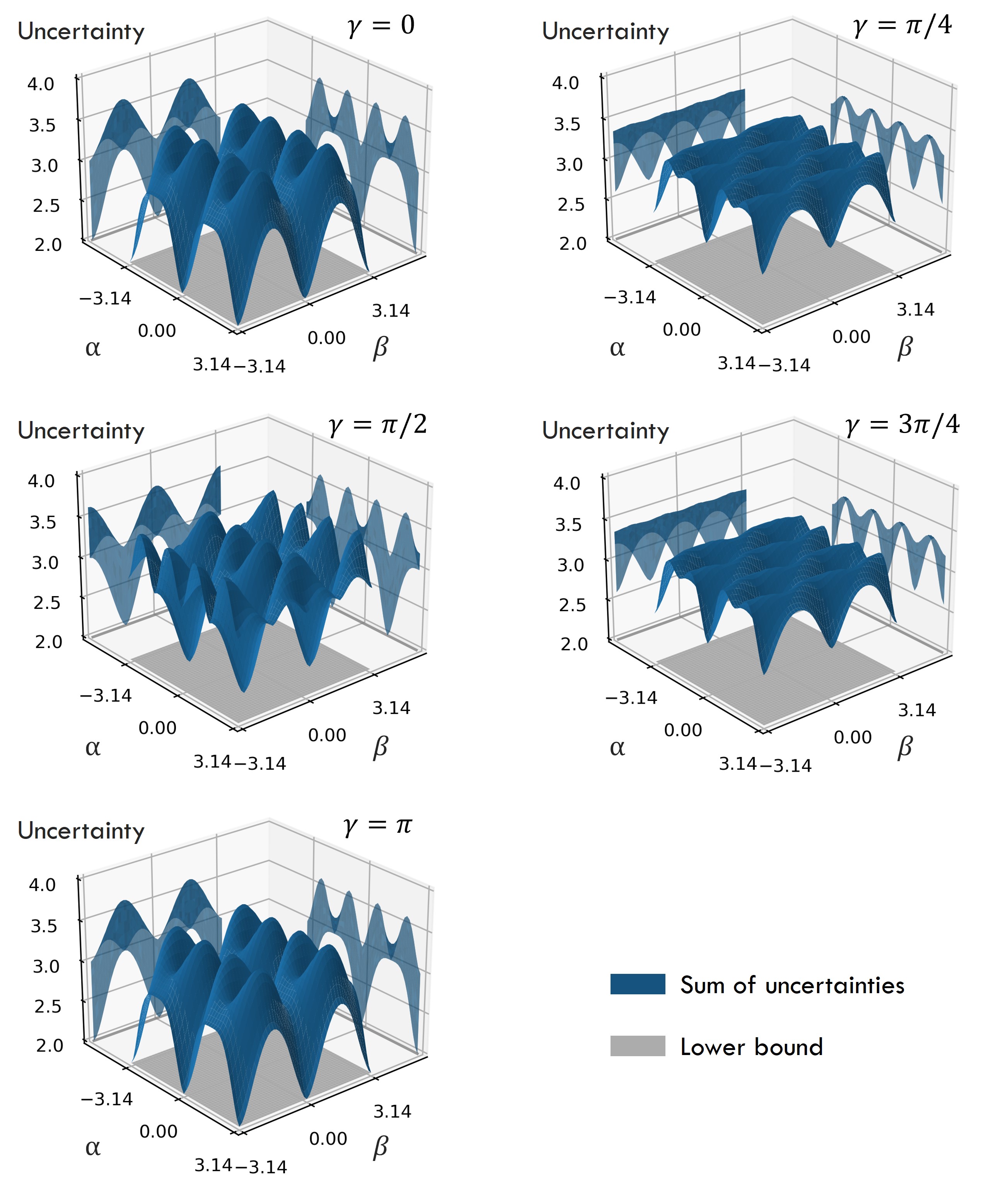}
\caption{(color online) Comparison between the joint uncertainty $H( \mT_{\text{CC}}(\id, \id) )_{\Phi_{\alpha, \beta, \gamma}}
+
H( \mT_{\text{DC}}(\id, \id) )_{\Phi_{\alpha, \beta, \gamma}}$ (blue) and the lower bound $2$ (gray).}
\label{fig:causal-uncertainty}
\end{figure}

\subsection{\label{subsec:land}The Landscape of Joint Uncertainty}

Instead of investigating a special bipartite unitary (see Fig.~$5a$ of the main text), we consider the universal two-qubit unitary gates $\mU(\alpha,\beta,\gamma)$ (see Fig.~\ref{fig:para-circuit} of this Supplemental Material) between system and environment in this subsection. We numerically demonstrate the optimality of entropic uncertainty relation (see Cor.~\ref{cor:entropic-CUR}) by comparing the value of
\begin{align}
    H(\mT_{\text{CC}}(\id, \id))_{\Phi_{\alpha, \beta, \gamma}}
    +
    H(\mT_{\text{DC}}(\id, \id))_{\Phi_{\alpha, \beta, \gamma}}
    ,
\end{align}
with the lower bound $2$. Here the maximal common-cause indicator $\mT_{\text{CC}}(\id, \id)\in\mM_{\text{CC}}$ (see Def.~\ref{def:MCCI}) and the maximal direct-cause indicators $\mT_{\text{DC}}(\id, \id)\in\mM_{\text{DC}}$ (see Def.~\ref{def:MDCI}) are obtained by setting $\mU_b= \id$ for all $b\in\{1, 2, 3, 4\}$. Meanwhile, the parameterized quantum circuit $\Phi_{\alpha, \beta, \gamma}\in\mathfrak{F}_2$ (Fig.~\ref{fig:para-circuit}) is given by
\begin{align}
    \Phi_{\alpha, \beta, \gamma}
    =
    \Tr_{F}
    [\mU(\alpha,\beta,\gamma)_{BE\to CF}(\phi^{+}_{AE})],
\end{align}
where the system $A$ and environment $E$ are initialized in the maximally entangled state $\phi^{+}$. In this case, the bipartite unitary $\mathcal{U}(\alpha,\beta,\gamma)_{BE\to CF}$ constitutes of three CNOT gates and five Pauli rotations. More precisely, the rotations with respect to $Z$ and $Y$ axis are defined as
\begin{equation}
    R_Z(\theta) := \begin{pmatrix}
    e^{-i\theta/2} & 0 \\
    0 & e^{i\theta/2}
    \end{pmatrix}, \quad
    R_Y(\theta) := \begin{pmatrix}
    \cos(\theta/2) & -\sin(\theta/2) \\
    \sin(\theta/2) & \cos(\theta/2)
    \end{pmatrix}.
\end{equation}
Here we investigate $\mathcal{U}(\alpha,\beta,\gamma)_{BE\to CF}$ since it forms a family of universal two-qubit unitary gates, up to some local unitary pre-processing and post-processing~\cite{PhysRevA.69.062321}. On the one hand,
the environment $F$ will be traced out at the end. Hence, the post-processing on $F$ can be ignored. On the other hand,
as the initial system-environment quantum dynamics is prepared in the maximally entangled state $\phi^{+}$, any unitary pre-processing on system $E$ is equivalent to its transpose acting on system $A$.
Consequently, this family of two-qubit unitary operators are universal up to some local unitary channels on systems $A,B$ and $C$.
The two $R_z$ gates at the end are inserted merely for aesthetic purposes. 

\begin{figure}[h]
\includegraphics[width=0.8\textwidth]{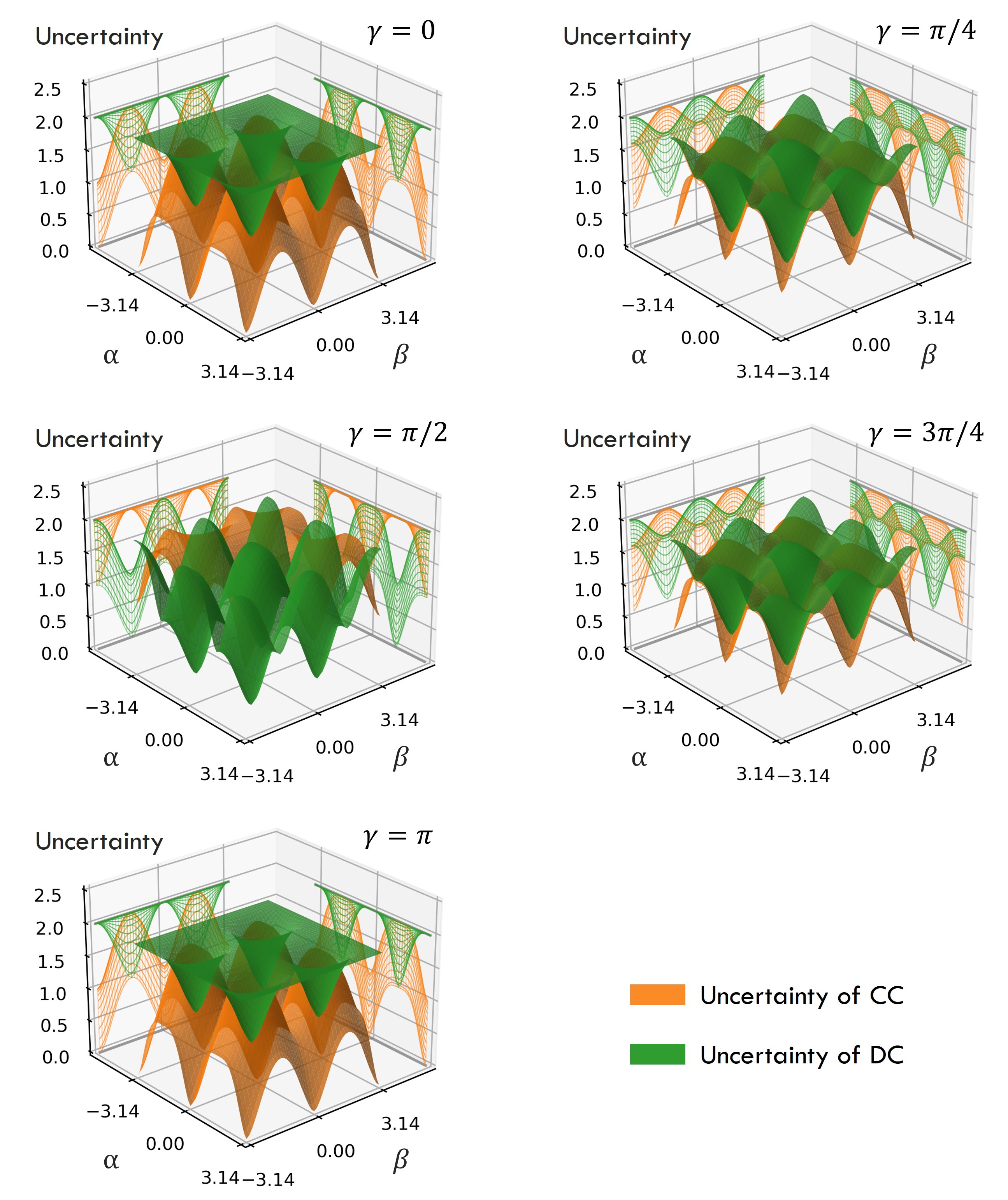}
\caption{(color online) Illustration of the Shannon entropies $H( \mathcal{T}_{\text{CC}}(\id, \id) )_{\Phi_{\alpha, \beta, \gamma}}$ (yellow) and $H( \mathcal{T}_{\text{DC}}(\id, \id) )_{\Phi_{\alpha, \beta, \gamma}}$ (green) with different parameters. If the numerical value of $H( \mathcal{T}_{\text{CC}}(\id, \id{1}) )_{\Phi_{\alpha^\star, \beta^\star, \gamma^\star}}$ falls into the interval of $(0,2)$ for some $(\alpha^\star, \beta^\star, \gamma^\star)$, then the circuit $\Phi_{\alpha^\star, \beta^\star, \gamma^\star}$ cannot reveal a purely direct-cause. Meanwhile, if $H( \mathcal{T}_{\text{DC}}(\id, \id) )_{\Phi_{\alpha^\star, \beta^\star, \gamma^\star}}$ falls into the interval of $(0,2)$ for some $(\alpha^\star, \beta^\star, \gamma^\star)$, then the causal structure of $\Phi_{\alpha^\star, \beta^\star, \gamma^\star}$ will not be purely common-cause. Finally, if both $H( \mathcal{T}_{\text{CC}}(\id, \id) )_{\Phi_{\alpha^\star, \beta^\star, \gamma^\star}}$ and $H( \mathcal{T}_{\text{DC}}(\id, \id) )_{\Phi_{\alpha^\star, \beta^\star, \gamma^\star}}$ belong to $(0,2)$ for some $(\alpha^\star, \beta^\star, \gamma^\star)$, then the causal structure of quantum circuit $\Phi_{\alpha^\star, \beta^\star, \gamma^\star}$ exhibits a mixture of both direct-cause and common-cause.}
\label{fig:CC-DE-Causal}
\end{figure}

Our numerical experiments are plotted at different values of $\gamma$, including the case of $\gamma=0, \pi/4, \pi/2, 3\pi/4, \pi$, and runs over $\alpha$ and $\beta$ from $-\pi$ to $\pi$. Fig.~\ref{fig:causal-uncertainty} illustrates the the tightness of lower bound $2$, demonstrating the optimality of our entropic causal uncertainty relation (see Eq.~\ref{eq:entropic-CUR}).

\subsection{\label{subsec:adv}Advantage in Inferring Causal Structures}

In this subsection, we infer the causal model of parameterized quantum circuit $\Phi_{\alpha, \beta, \gamma}$ (see Fig.~\ref{fig:para-circuit}) by using our entropic causal uncertainty relation (see Eq.~\ref{eq:entropic-CUR}). In this case, the initial state is $\phi^{+}$, and hence we can apply the results of Thms.~\ref{thm:CC} and \ref{thm:DC} directly. To simplify the experimental processes, we simply implement the maximal common-cause indicator $\mT_{\text{CC}}(\id, \id)\in\mM_{\text{CC}}$ (see Def.~\ref{def:MCCI}) and the maximal direct-cause indicators $\mT_{\text{DC}}(\id, \id)\in\mM_{\text{DC}}$ (see Def.~\ref{def:MDCI}), and evaluating the corresponding Shannon entropies of $H(\mT_{\text{CC}}(\id, \id))_{\Phi_{\alpha, \beta, \gamma}}$ and $H(\mT_{\text{DC}}(\id, \id))_{\Phi_{\alpha, \beta, \gamma}}$, which are defined by Eqs.~\ref{eq:H-CC} and \ref{eq:H-DC} respectively.

\begin{figure}[h]
\includegraphics[width=0.6\textwidth]{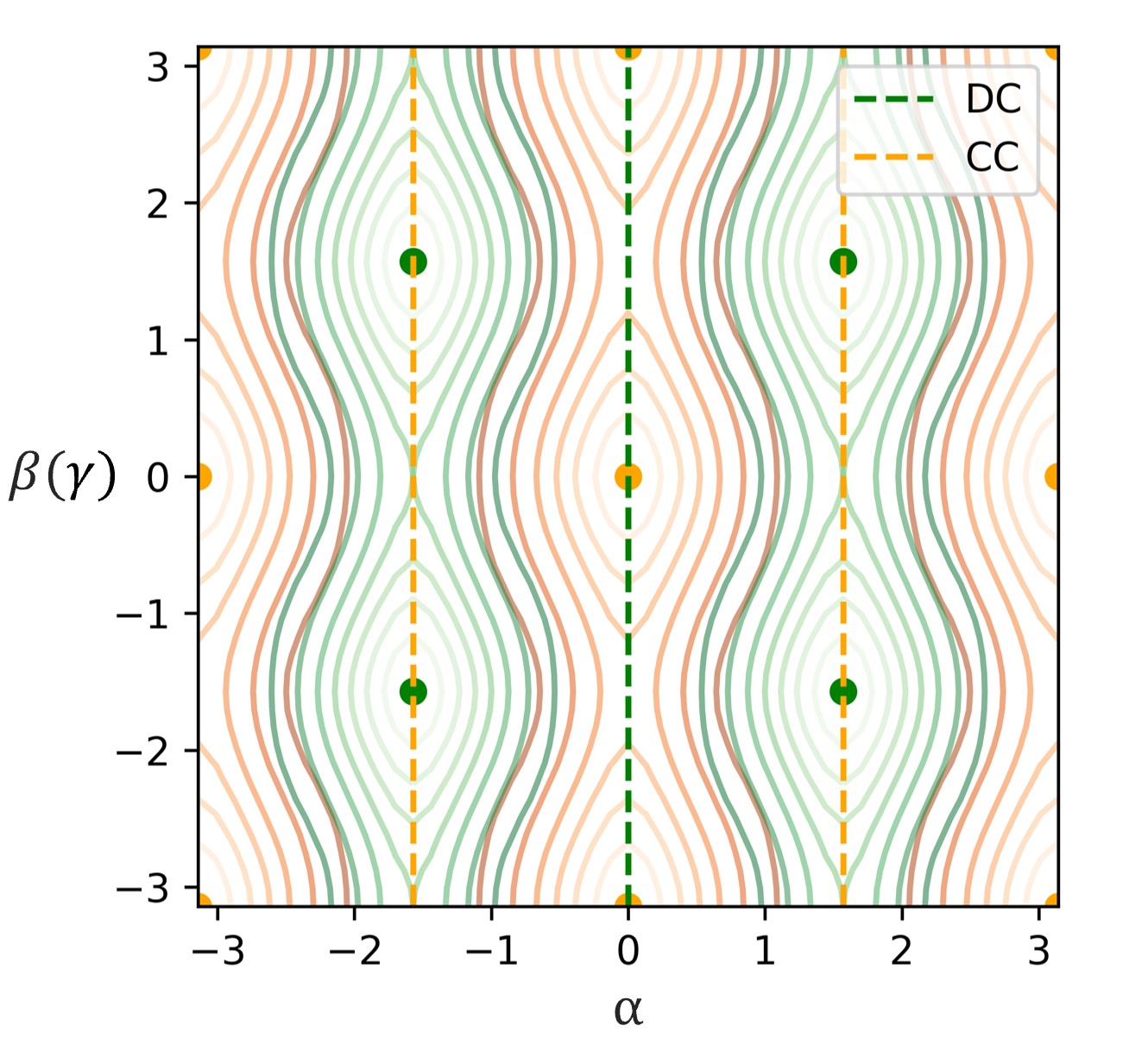}
\caption{(color online) Contour lines for the uncertainty associated with maximal direct-cause indicator $\mathcal{T}_{\text{DC}}(\id, \id)$ and maximal common-cause indicator $\mathcal{T}_{\text{CC}}(\id, \id)$. Here the green dashed line stand for the cases where $H(\mathcal{T}_{\text{DC}}(\id, \id) )_{\Phi_{\alpha, \beta, \beta}}=2$, and the yellow dashed line represents the cases where $H( \mathcal{T}_{\text{CC}}(\id, \id) )_{\Phi_{\alpha, \beta, \beta}}=2$. The green and yellow dots are the tuples $(\alpha,\beta)$ such that $H( \mathcal{T}_{\text{DC}}(\id, \id) )_{\Phi_{\alpha, \beta, \beta}}=0$ and $H( \mathcal{T}_{\text{CC}}(\id, \id) )_{\Phi_{\alpha, \beta, \beta}}=0$. In such cases, the corresponding causal structures are purely direct-cause and purely common-cause respectively. Besides dashed lines and dotted, all other quantum circuits exhibit a mixture of both direct-cause and common-cause.
}
\label{fig:CC-DE-Infer}
\end{figure}

We have depicted the landscape of Shannon entropies $H( \mathcal{T}_{\text{CC}}(\id, \id) )_{\Phi_{\alpha, \beta, \gamma}}$ and $H( \mathcal{T}_{\text{DC}}(\id, \id) )_{\Phi_{\alpha, \beta, \gamma}}$ in Fig.~\ref{fig:CC-DE-Causal}. For parameters $\alpha^\star$, $\beta^\star$ and $\gamma^\star$, if the numerical value of $H( \mathcal{T}_{\text{CC}}(\id, \id) )_{\Phi_{\alpha^\star, \beta^\star, \gamma^\star}}$ belongs to $(0,2)$, then the Shannon entropy associated with maximal direct-cause indicators satisfies
\begin{align}
    H( \mathcal{T}_{\text{DC}}(\mU_3, \mU_4) )_{\Phi_{\alpha^\star, \beta^\star, \gamma^\star}}
    \geqslant
    2- H( \mathcal{T}_{\text{CC}}(\id, \id) )_{\Phi_{\alpha^\star, \beta^\star, \gamma^\star}}
    >0,
\end{align}
for any unitary channels $\mU_3$ and $\mU_4$. Thus, Thm.~\ref{thm:DC} implies that the causal model of quantum circuit $\Phi_{\alpha^\star, \beta^\star, \gamma^\star}$ will never be purely direct-cause. Meanwhile, for some parameters $\alpha^\star$, $\beta^\star$ and $\gamma^\star$, if the numerical value of $H( \mathcal{T}_{\text{DC}}(\id, \id) )_{\Phi_{\alpha^\star, \beta^\star, \gamma^\star}}$ belongs to $(0,2)$, then the Shannon entropy associated with maximal common-cause indicators meets the following inequality 
\begin{align}
    H( \mathcal{T}_{\text{CC}}(\mU_1, \mU_2) )_{\Phi_{\alpha^\star, \beta^\star, \gamma^\star}}
    \geqslant
    2- H( \mathcal{T}_{\text{DC}}(\id, \id) )_{\Phi_{\alpha^\star, \beta^\star, \gamma^\star}}
    >0,
\end{align}
for any unitary channels $\mU_1$ and $\mU_2$. Hence, Thm.~\ref{thm:CC} implies that the causal structure of $\Phi_{\alpha^\star, \beta^\star, \gamma^\star}$ cannot be purely common-cause. At last, if the numerical value of both $H( \mathcal{T}_{\text{CC}}(\id, \id) )_{\Phi_{\alpha^\star, \beta^\star, \gamma^\star}}$ and $H( \mathcal{T}_{\text{DC}}(\id, \id) )_{\Phi_{\alpha^\star, \beta^\star, \gamma^\star}}$ belong to $(0,2)$ for the same $\alpha^\star$, $\beta^\star$ and $\gamma^\star$, then the corresponding causality of $\Phi_{\alpha^\star, \beta^\star, \gamma^\star}$ must be a mixture of both direct-cause and common-cause. 

In particular, for the case $\beta=\gamma$, we have demonstrated all the circuits whose causal structure can be inferred by implementing only two interactive measurements in Fig.~\ref{fig:CC-DE-Infer}, where the green (dashed) line stands for the uncertainty associated with maximal direct-cause indicator $\mathcal{T}_{\text{DC}}(\id, \id)$ and the yellow (dashed) line represents the uncertainty associated with maximal common-cause indicators $\mathcal{T}_{\text{CC}}(\id, \id)$. The dashed lines stand for the case when the uncertainty is equal to $2$, where we cannot get any useful information about the causal structure of $\Phi_{\alpha, \beta, \gamma}$. Meanwhile, the dots stand for the case whose uncertainty is equal to $0$. For these dots, we can get a deterministic statement about its causal structure. Take the green (yellow) dots for instance, then the corresponding causal structure of circuit $\Phi_{\alpha, \beta, \beta}$ is a purely direct-cause (purely common-cause). Compared with the tomography of system-environment unitary dynamics, we have seen at least two advantages in causal inference by using causal uncertainty relation (see Eq.~\ref{eq:entropic-CUR}): First, we can infer the structural dependencies among environmental and object systems without the need of information from the environment. Second, the causality can sometimes be determined by using very few interactive measurements, e.g. for $\Phi_{\alpha, \beta, \beta}$, the causal structure of almost all circuits can be determined by using only two interactive measurements -- $\mT_{\text{CC}}(\id, \id)\in\mM_{\text{CC}}$ (see Def.~\ref{def:MCCI}) and $\mT_{\text{DC}}(\id, \id)\in\mM_{\text{DC}}$ (see Def.~\ref{def:MDCI}).





\bibliography{Bib}